\documentclass[aps,prd,onecolumn,groupedaddress,nofootinbib,preprintnumbers,superscriptaddress]{revtex4-2}  
\usepackage{graphicx}
\usepackage{amsmath}
\usepackage{amsfonts}
\usepackage{bm,bbm}
\usepackage[colorlinks=true,citecolor=red,linkcolor=blue,breaklinks=true]{hyperref}
\usepackage[figure, table]{hypcap}
\usepackage{amssymb}
\usepackage{xcolor}
\usepackage{subfigure}
\usepackage{setspace}
\usepackage{footnote}
\usepackage{multirow}
\usepackage[normalem]{ulem}
\usepackage[utf8]{inputenc}
\usepackage{mathrsfs}
\usepackage{slashed}
\usepackage{longtable}
\usepackage{enumitem}
\usepackage{cancel}
\usepackage{booktabs}

\numberwithin{equation}{section}

\allowdisplaybreaks

\definecolor{tabred}{rgb}{0.8392156862745098, 0.15294117647058825, 0.1568627450980392}
\definecolor{tabblue}{rgb}{0.12156862745098039, 0.4666666666666667, 0.7058823529411765}
\hypersetup{colorlinks=true, linkcolor={tabred}, citecolor={tabblue}, urlcolor={tabblue}} 

\usepackage{orcidlink}

\usepackage[skins,theorems]{tcolorbox}
\tcbset{highlight math style={enhanced,
  colframe=red,colback=white,arc=0pt,boxrule=1pt}}

\newcommand{\orcid}[1]{\,\orcidlink{#1}}

\newcommand{\EMQED}{\mathrm{em}}

\newcommand{\dd}{\mathrm{d}}

\newcommand{\rmi}[1]{{\mbox{\scriptsize #1}}}
\newcommand{\rmii}[1]{{\mbox{\tiny\rm{#1}}}}
\newcommand{\rmiii}[1]{{\mbox{\tiny{$\scriptstyle{\rm#1}$}}}}
\newcommand{\rmO}{{\mathcal{O}}}
\newcommand{\eq}{eq.~}
\newcommand{\eqs}{eqs.~}
\newcommand{\se}{sec.~}

\newcommand{\app}{appendix~}
\newcommand{\fig}{fig.~}

\newcommand{\tabl}{table~}
\newcommand{\nr}[1]{(\ref{#1})}
\newcommand{\aL}{a^{ }_\rmii{L}}
\newcommand{\aR}{a^{ }_\rmii{R}}
\newcommand{\iF}{\rmii{F}}
\newcommand{\iI}{\rmii{I}}
\newcommand{\iL}{\rmii{L}}
\newcommand{\iR}{\rmii{R}}
\newcommand{\iT}{\rmii{$T$}}
\newcommand{\M}{\mathcal{M}}
\newcommand{\s}{\sigma}
\renewcommand{\t}{\tau}
\renewcommand{\vec}[1]{{\bf #1}}
\newcommand{\nn}{\nonumber \\}
\newcommand{\now}{\rmi{0}}
\newcommand{\inow}{\rmii{0}}
\newcommand{\be}{\begin{equation}}
\newcommand{\ee}{\end{equation}}
\newcommand{\ba}{\begin{eqnarray}}
\newcommand{\ea}{\end{eqnarray}}
\newcommand{\la}[1]{\label{#1}}
\newcommand{\bi}{\begin{itemize}}
\newcommand{\ei}{\end{itemize}}
\newcommand{\e}{\rho}
\newcommand{\p}{p} 
\newcommand{\tot}{\rmi{t}}

\newcommand{\msl}[1]{\,\slash\!\!\!{#1}\,}
\newcommand{\nF}{n_\rmii{F}} 
\newcommand{\bmu}{\bar\mu}
\newcommand{\msbar}{{\overline{\mbox{\rm MS}}}}
\newcommand{\mpl}{m_\rmii{pl}}
\newcommand{\N}{\mathcal{N}}
\newcommand{\re}{\mathop{\mathrm{Re}}}

\usepackage{fontawesome5}
\definecolor{color_git}{rgb}{0.098, 0.160, 0.345}
\newcommand{\gitlink}{\href{https://github.com/MiguelEA/nudec_BSM}{\textsc{g}it\textsc{h}ub {\large\color{color_git}\faGithub}}$\,$}

\newcommand{\hiddenappsubsection}[2][]{%
  \begingroup
    \renewcommand{\addcontentsline}[3]{} 
    \refstepcounter{subsection}
    \vspace{3.25ex plus1ex minus.2ex}
    \centering\textbf{\thesubsection.\; #2}\par
    \vspace{1.5ex plus .2ex}
    \label{#1}
  \endgroup
}

\begin{document}

\preprint{LA-UR-25-30442, CERN-TH-2025-225}

\title{Fast and Flexible Neutrino Decoupling Part I: The Standard Model \\[2mm]}

\author{M.~Escudero\orcid{0000-0002-4487-8742}}
\email{miguel.escudero@cern.ch}
\affiliation{Theoretical Physics Department, CERN, 1 Esplanade des Particules, CH-1211 Geneva 23, Switzerland}

\author{G.~Jackson\orcid{0000-0003-1283-3264}}
\email{jackson@subatech.in2p3.fr}
\affiliation{SUBATECH UMR 6457 (IMT Atlantique, Nantes Universit\'e, IN2P3/CNRS), \\ 4 rue Alfred Kastler, 44307 Nantes, France}

\author{M.~Laine\orcid{0000-0002-2680-4213}}
\email{laine@itp.unibe.ch}
\affiliation{AEC, Institute for Theoretical Physics, University of Bern, Sidlerstrasse 5, CH-3012 Bern, Switzerland}

\author{S.~Sandner\orcid{0000-0002-1802-9018}}
\email{stefan.sandner@lanl.gov}
\affiliation{Theoretical Division, Los Alamos National Laboratory, Los Alamos, NM 87545, USA}

\begin{abstract}
\noindent
Cosmological determinations of the number of relativistic  neutrino species, 
$N^{ }_\rmi{eff}$, are becoming increasingly accurate, and further 
improvements are expected both from CMB and BBN data. Given  this context, 
we update the evaluation of $N^{ }_{\rm eff}$ and the current entropy 
density via the momentum-averaged approach. This allows for a numerically 
fast description of neutrino decoupling, easily portable 
to an array of new physics scenarios. We revisit all aspects of this 
approach, including collision terms with full electron mass dependence, 
finite temperature QED corrections to the equation of state, 
neutrino oscillations, and the modelling of neutrino ensembles 
with effective chemical potentials. 
For integrated observables, our results differ by less than 0.04\%  from
the solution of the momentum-dependent evolution equation. We outline 
how to extend the approach to BSM settings, and will highlight 
its power in Part II. To facilitate the practical implementation, 
we release a \texttt{Mathematica} and \texttt{python} code 
within \texttt{nudec\_BSM\_v2} \gitlink, 
easily linkable to BBN codes. 
\end{abstract}

\maketitle

\vspace{-4.5mm}
{
\hypersetup{linkcolor=black}
\setlength{\parskip}{0pt} 
\tableofcontents
}


\vspace{0.5cm}

\section{Introduction}
\label{sec:intro}
\addtocontents{toc}{\vspace{-0.5em}}

The measurement of the effective number of relativistic neutrino species, 
$N_{\rm eff}$, 
from Cosmic Microwave Background (CMB) observations 
has improved at an impressive pace over the last 15 years:
\[
\begin{alignedat}{2}
  N_{\rm eff} &> 2.3 \,\, {\rm at}\,\, 95\%\ \text{CL}  & \quad & (2008)\ \text{WMAP--5}    \,  \mbox{\cite{WMAP5_Cosmology_2009}} \,,\\
  N_{\rm eff} &= 3.66 \pm 0.34 &\quad & (2013)\ \text{Planck + WMAP}                 \, \mbox{\cite{Planck2013_CosmoParams}} \,,\\
  N_{\rm eff} &= 2.92 \pm 0.19 &\quad & (2018)\ \text{Planck}                        \, \mbox{\cite{Planck2018_CosmoParams}} \,,\\
  N_{\rm eff} &= 2.81 \pm 0.12 &\quad & (2025)\ \text{Planck + ACT + SPT-3G}          \,   \,\mbox{\cite{ACT:2025fju,ACT:2025tim,SPT-3G:2025bzu}}\,.
\end{alignedat}
\]
Further improvements are expected from SPT-3G+~\cite{SPT-3G:2024qkd} 
($\pm\, 0.07$) 
and the Simons Observatory~\cite{SimonsObservatory:2018koc} 
($\pm\, 0.05$). 
In parallel, inferences of the expansion rate 
from Big Bang Nucleosynthesis (BBN) have improved in 
the past $\sim 20$ years due to a more robust measurement of 
the primordial helium abundance~\cite{Izotov:2014fga,Aver:2020fon}, 
a $1\%$ determination of the primordial deuterium 
abundance~\cite{Cooke:2017cwo}, 
and better knowledge about nuclear reaction rates 
that control the deuterium 
abundance~\cite{Mossa:2020gjc,Pisanti:2020efz,Pitrou:2020etk,Yeh:2022heq}.

Independently, 
there has been increased interest in particles beyond the 
Standard Model (SM) which are lighter than 
the weak scale and which can affect $N_{\rm eff}$ and BBN. 
Well-motivated 
scenarios include MeV-scale thermal dark matter 
particles~\cite{Boehm:2003hm,Feng:2008ya,Boehm:2013jpa,Knapen:2017xzo,Sabti:2019mhn,Giovanetti:2021izc,Chu:2022xuh}, 
hidden photons~\cite{Pospelov:2007mp,Ilten:2018crw,Bauer:2018onh,Escudero:2019gzq,Esseili:2023ldf}, 
dark scalars~\cite{Bezrukov:2009yw,Krnjaic:2015mbs,Winkler:2018qyg,Fradette:2018hhl}, 
late-decaying inflatons~\cite{Kawasaki:2000en,Hannestad:2004px,Hasegawa:2019jsa},
axions~\cite{Cadamuro:2011fd,Depta:2020wmr,Jain:2024dtw,Bouzoud:2024bom,Escudero:2025avx}, 
sterile neutrinos~\cite{Dolgov:2000jw,Hernandez:2014fha,Boyarsky:2020dzc}, 
and many others, 
see e.g.~refs.~\cite{Alexander:2016aln,Beacham:2019nyx,Antel:2023hkf} 
for extensive discussions.

In this context, it is important to have flexible tools to explore 
the phenomenology of extensions of the Standard Model 
that can modify $N_{\rm eff}$ and BBN. 
This entails solving for the process of neutrino decoupling, 
which occurred when the universe was $t\sim 1\,{\rm s}$ old 
and had a temperature of about $T^{ }_\gamma\sim 1\,{\rm MeV}$. 
At this point the electroweak interactions that kept neutrinos 
in thermal equilibrium stopped being efficient. 
Soon afterwards, at $T^{ }_\gamma \sim m_e$, 
electrons and positrons became non-relativistic 
and 
their decays released more energy to photons than to neutrinos, 
leading to the famous $T_\gamma/T_\nu \simeq 1.4$ 
ratio. 
Solving for this process requires tracking the evolution of neutrinos, 
of the electromagnetic plasma, and of the exchanges of energy, momentum, 
and particle number between them.

The most precise approach to calculate the process of neutrino decoupling 
is to solve for the neutrino density matrix using the momentum-dependent 
Liouville equation that takes into account the neutrino-neutrino 
and neutrino-electron 
interactions~\cite{Hannestad:1995rs,Dolgov:1997mb,Mangano:2001iu,Mangano:2005cc,deSalas:2016ztq}. 
With a comoving momentum grid, this leads to 
$\mathcal{O}(100)$ coupled non-linear integro-differential equations 
which are stiff and thus challenging to track numerically, 
see e.g. refs.~\cite{Dolgov:1997mb,deSalas:2016ztq}. 
The issue becomes worse in many extensions of the Standard Model, 
because new physics states can interact with neutrinos 
at rates much larger than the Hubble rate. 
This implies a scale hierarchy, 
making solving the Liouville equation even harder. 
On the other hand, a large interaction rate leads to rapid equilibration, 
and then we already know the shape 
of the corresponding distribution function, 
suggesting that there is no need to track 
it dynamically.\footnote{%
 We note that the Liouville equation also becomes hard to solve 
 if high-energy neutrinos are injected, 
 as happens for instance in the case of 
 GeV-scale sterile neutrinos~\cite{Boyarsky:2020dzc}. 
 For these types of scenarios, 
 a 
 recently developed method using direct Monte Carlo sampling of 
 the collisions~\cite{Ovchynnikov:2024rfu,Ovchynnikov:2024xyd} 
 has been shown to give very good results, both in the 
 Standard Model~\cite{Ihnatenko:2025kew} 
 and beyond~\cite{Akita:2024nam,Akita:2024ork}.
} 

The idea to insert a thermal shape for a distribution function, 
parametrized by 
a temperature and a chemical potential, has a long history. 
It was already used for pioneering recombination 
calculations~\cite{Peebles:1968ja,Zeldovich:1969ff}, 
and is commonly employed in WIMP literature to calculate the evolution 
of thermal relics~\cite{Lee:1977ua,Gondolo:1990dk}, 
because their kinetic equilibration rate is 
much faster than the chemical one. However, it was also used to 
provide early estimates of $N_{\rm eff}$ in the Standard Model~\cite{Dicus:1982bz,1989ApJ...336..539H,1985PhFl...28.3253H,Rana:1991xk}, 
and was more recently exploited for BSM scenarios in the context of neutrino decoupling~\cite{Escudero:2018mvt,EscuderoAbenza:2020cmq}. 
The approach is fast, flexible, 
and sufficiently accurate for many practical purposes, 
which has 
motivated new BBN codes 
to adopt it~\cite{Burns:2023sgx,Giovanetti:2024zce}. 

While refs.~\cite{Escudero:2018mvt,EscuderoAbenza:2020cmq} 
introduced the essential elements to solve 
for neutrino decoupling in the presence of new thermal relics, 
several approximations and simplifications were adopted. 
The main goal of the present paper and its companion~\cite{Escudero:2025mvt} 
is to scrutinize and significantly improve upon these works. The improvements are:
\begin{enumerate}

    \item \textit{Updated neutrino-electron interaction rates in the Standard Model}. The interaction rates used in refs.~\cite{Escudero:2018mvt,EscuderoAbenza:2020cmq} employed the massless electron limit and assumed Maxwell-Boltzmann statistics for neutrinos and electrons. In this study we calculate the rates exactly at leading order (LO) in $\alpha^{ }_\rmi{em}$, and provide an accurate linear response representation of the full mass dependence and also featuring non-zero chemical potentials. 
    The new rates are more flexible and accurate than the previous ones, 
but remain simple and compact.\\ 
    
    \item \textit{Increased computational speed}. The thermodynamic formulae for the number and energy densities used in~refs.~\cite{Escudero:2018mvt,EscuderoAbenza:2020cmq} were integrated over momentum explicitly, which led to a significant computational cost. Here, inspired by the implementation of some BBN codes~\cite{Pisanti:2007hk,Consiglio:2017pot,Arbey:2011nf,Giovanetti:2024zce}, we make use of accurate low-temperature expansions
in terms of Bessel functions. This yields a speed-up of about a factor of 10, without any loss of precision (provided that effective chemical potentials are small or negative, which is the typical case).

    \item \textit{Exploration of neutrino oscillations}. Given the observed mass differences and mixing angles, neutrinos start to oscillate at $T\sim 10\,{\rm MeV}$ in the early universe. While physically important, this effect is numerically small in the Standard Model calculation for $N_{\rm eff}$, and for simplicity refs.~\cite{Escudero:2018mvt,EscuderoAbenza:2020cmq} did not account for it.  It was rather argued that given the rapid neutrino oscillations, 
the $\nu_e$, $\nu_\mu$ and $\nu_\tau$ neutrinos effectively share a common temperature. Here we do include oscillations, and
demonstrate that they only have a small effect in the Standard Model, 
while improving on the match to the full Liouville solution. 
In our companion paper~\cite{Escudero:2025mvt}, 
we study the issue in BSM settings,  
with particles interacting at different rates with different neutrino species.
    
    \item \textit{Accuracy of the approach and spectral distortions}. At the time of refs.~\cite{Escudero:2018mvt,EscuderoAbenza:2020cmq} there were only a handful of high-accuracy solutions to the Liouville equation, but a resurgence of interest in $N_{\rm eff}$ led to a consensus $N_{\rm eff}$ calculation by several groups~\cite{Akita:2020szl,Froustey:2020mcq,Bennett:2020zkv}. In this work, by comparing against the public code \texttt{FortEPiaNO}~\cite{Gariazzo:2019gyi}, we provide a comparison between the full Liouville solution and that assuming thermally shaped distribution functions. This enables us to understand what are the spectral distortions of the Cosmic Neutrino Background as well as the accuracy of the momentum-averaged approach to solve for neutrino decoupling.
    
    \item \textit{Higher-order QED corrections to the energy density and pressure of the plasma}.
    State-of-the-art calculations of $N_{\rm eff}$ implement the  
    $\rmO(e^2_{ })$~\cite{Heckler:1994tv,Fornengo:1997wa} 
    and $\rmO(e^3_{ })$~\cite{Bennett:2019ewm} 
    QED thermal corrections (cf.\ ref.~\cite{Kapusta:1979fh} and references therein)
    to the electromagnetic energy density of the universe, 
    as this leads to a change $\Delta N_\rmi{eff}^{ }  \sim 0.01$. 
    Here we recalculate and re-implement these corrections, 
    and estimate the $\rmO(e^4_{ })$ and $\rmO(e^5_{ })$ ones.
    We show that though the last ones introduce qualitatively new 
    structures, they are numerically small and can be safely neglected.  

\end{enumerate}

Our study is structured as follows. First, in \se\ref{sec:Evolution}, 
we review the basic ingredients for the solution of neutrino decoupling 
using a momentum-averaged approach. 
We introduce our newly developed collision terms, 
and write simple versions thereof using a linear response approach. 
In \se\ref{ss:osc}, we show how neutrino oscillations can be incorporated
in the momentum-averaged approach. 
In \se\ref{sec:QED_corrections}, 
we present a new calculation of the QED equation of state 
featuring corrections up to $\mathcal{O}(e^5)$. 
In \se\ref{se:expansion}, 
we define the physical observables extracted from the solution.
In \se\ref{se:comparison},
we compare our 
results against the solution of the momentum-dependent Liouville equation.
In \se\ref{se:bsm}, 
we anticipate how the framework can be extended to include 
BSM particles. 
In \se\ref{sec:data}, we 
briefly summarize the codes and data tables
that accompany this work, before concluding in \se\ref{sec:conclusions}. 
Several details are relegated to a number of appendices, 
including a quantification of spectral 
distortions of the Cosmic Neutrino Background.

\section{Basic variables, rate equations and rate coefficients}
\label{sec:Evolution}

Neutrinos interact with electrons, positrons and themselves in the early universe, via the processes $\nu \bar{\nu} \leftrightarrow e^+e^-$, $\nu e^\pm\to \nu e^\pm$, $\nu \bar{\nu} \to \nu \bar{\nu}$, and $\nu \nu \to \nu \nu$. The rates of these interactions drop below the Hubble rate at $T^{ }_\gamma < 2\,{\rm MeV}$. Soon afterwards, at around $T_\gamma \sim m_e$, electrons and positrons become non-relativistic, and the energy released in their decays is mostly taken up by photons. 
In this section, we describe how to track this system in the momentum-averaged approach, which assumes that all particles follow thermally shaped distribution functions.

\subsection{Outline of the momentum-averaged approach}
\label{ss:momave}

In the momentum-averaged approach, we think of the system as consisting of a number of separate subsystems. These include a QED plasma and the neutrino and antineutrino ensembles, but generically also other sectors depending on the new physics scenario considered (see \se\ref{se:bsm} and ref.~\cite{Escudero:2025mvt}). 
Focusing 
on the Standard Model for the moment, energy-momentum conservation equations for the subsystems can be written as
\be
 \frac{{\rm d}\e_i^{ }}{{\rm d}t} + 3 H (\e^{ }_i + \p^{ }_i) 
 \; = \; Q^{ }_i
 \;, \quad \sum_i Q^{ }_i = 0
 \;, \quad i \in \{ \EMQED, \nu^{ }_e, \nu^{ }_\mu, \nu^{ }_\tau \} \;,
 \la{e_i_cons}
\ee
where $H$ is the Hubble rate, $\e^{ }_i$ denote energy densities, 
$\p^{ }_i$ pressures, 
$Q^{ }_i$ energy transfer rates, and ``$\EMQED$'' refers to 
the entire QED ensemble formed of electrons, positrons and photons 
(we disregard baryons as their density is negligible at $T^{ }_\gamma \sim {\rm MeV}$).
We assume that 
there are no substantial lepton asymmetries, so that the neutrino
and antineutrino ensembles can be treated as equivalent; 
and that neutrino oscillations are either very slow or very
fast compared to the interaction rate 
(we return to 
dynamical neutrino oscillations in \se\ref{ss:osc}).
Given that neutrino number densities, $n^{ }_i$, are slowly evolving, we can
also write down the corresponding evolution equations, 
\be
 \frac{{\rm d} n_i^{ }}{{\rm d} t} + 3 H n^{ }_i \; = \; J^{ }_i
 \;, \quad i \; \in \{\nu^{ }_e, \nu^{ }_\mu, \nu^{ }_\tau \} 
 \;, \la{n_i_cons}
\ee
where the source terms, $J^{ }_i$, originate from 
the same Fermi vertices (cf.\ \app\ref{se:couplings}) 
as the energy transfer rates, $Q^{ }_i$. 

As the universe cools down, neutrinos eventually fall out of equilibrium with the QED plasma. The essence of the momentum-averaged approach is to parametrize their distribution functions by two quantities, an effective temperature ($T^{ }_i$) and an effective chemical potential ($\mu^{ }_i$).\footnote{%
 This $\mu^{ }_i$ should not be confused with a chemical potential parametrizing particle-antiparticle asymmetries, but rather refers to a distortion of the equilibrium Fermi-Dirac distribution, as defined in eq.~\eqref{eq:f_nu}.
}
We can then write 
$
 \dot{\e}^{ }_i = (\partial \e^{ }_i/\partial T^{ }_i)
 \dot{T}^{ }_i + (\partial \e^{ }_i/\partial \mu^{ }_i)
 \dot{\mu}^{ }_i
$, and
correspondingly for $\dot{n}^{ }_i$. Inserting these in 
\eqs\nr{e_i_cons} and \nr{n_i_cons}, we can solve for the
time dependence of the effective temperatures and chemical
potentials, 
\begin{subequations}\label{eq:dTdt_dmudt_general}
\ba
  \frac{{\rm d} T_i^{ }}{{\rm d}t}
  & = &
  \frac{
    \raisebox{3mm}{$
    \displaystyle
    - 3 H \bigg(
    ( \e^{ }_i + \p^{ }_i )
    \frac{\partial n_i}{\partial \mu_i}
    \, - \, 
    n_i
    \frac{\partial \e_i}{\partial \mu_i}
    \bigg) 
    \, + \, 
    Q_i \, \frac{\partial n_i}{\partial \mu_i}
    \, - \, 
    J_i \, \frac{\partial \e_i}{\partial \mu_i}
    $}
  }{
    \raisebox{-3mm}{$
    \displaystyle
    \frac{\partial n_i}{\partial \mu_i}
    \,
    \frac{\partial \e_i}{\partial T_i}
    \, - \, 
    \frac{\partial n_i}{\partial T_i }
    \,
    \frac{\partial \e_i}{\partial \mu_i}
    $}
  }
  \; ,
  \la{T dot}
  \\[2mm]
 \frac{{\rm d} \mu_i^{ }}{{\rm d} t}
  & = &
  \frac{
    \raisebox{3mm}{$
    \displaystyle
     3 H \bigg(
    ( \e^{ }_i + \p^{ }_i )
    \frac{\partial n_i}{\partial T_i}
    \, - \, 
    n_i
    \frac{\partial \e_i}{\partial T_i}
    \bigg) 
    \, - \, 
    Q_i \, \frac{\partial n_i}{\partial T_i}
    \, + \, 
    J_i \, \frac{\partial \e_i}{\partial T_i}
    $}
  }{
    \raisebox{-3mm}{$
    \displaystyle
    \frac{\partial n_i}{\partial \mu_i}
    \,
    \frac{\partial \e_i}{\partial T_i}
    \, - \, 
    \frac{\partial n_i}{\partial T_i }
    \,
    \frac{\partial \e_i}{\partial \mu_i}
    $}
  }
  \; .
  \la{mu dot}
\ea
\end{subequations}
For the QED plasma, which has a negligible chemical potential, 
the evolution equation from
\eq\nr{e_i_cons} reads\footnote{%
 To connect with the notation in refs.~\cite{Escudero:2018mvt,EscuderoAbenza:2020cmq}, 
 we recall that it is common to separate the Stefan-Boltzmann contributions from photons
 ($\gamma$) and electrons and positrons ($e$) from the interaction part (${\rm int}$).
 In the last we can employ the thermodynamic identities
  $\e^{ }_\rmii{int} + \p^{ }_\rmii{int} = T^{ }_\gamma {\rm d} \p^{ }_\rmii{int}/{\rm d} T^{ }_\gamma$
  and ${\rm d} \e^{ }_\rmii{int}/{\rm d}T^{ }_\gamma = T^{ }_\gamma {\rm d}^2 \p^{ }_\rmii{int}/{\rm d} T^2_\gamma$,
  valid because the baryon chemical potential is negligible. Then
 \begin{align}
    \frac{{\rm d}T_{\gamma}}{{\rm d}t}  &
    \overset{\rmii{\nr{dot_T_gamma}}}{=} - 
  \frac{  4 H \e_{\gamma} + 3 H \left( \e_{e} + \p_{e}\right)  + 3 H  \, T_\gamma
  \frac{d \p_{\rm int}}{dT_\gamma} - Q^{ }_\EMQED }{\frac{d \rho_{\gamma}}{d T_\gamma}
  + \frac{d \rho_e}{d T_\gamma} + T_\gamma \frac{d^2 \p_{\rm int}}{d T_\gamma^2}  } \,.
 \end{align}
 For neutrinos with $\mu^{ }_i \ll T^{ }_i$, 
 \eqs\nr{n_plus_massless} and \nr{e_plus_massless} imply
 $
  n^{ }_i \approx b T_i^3 + c \mu_i T_i^2
 $
 and
 $
  3 \p^{ }_i = \e^{ }_i \approx d\, T_i^4 + 3 b \mu^{ }_i T_i^3
 $, 
 where $b$,$c$ and $d$ are constants. Then 
 $
 \Delta^{ }_i \equiv \frac{\partial n_i}{\partial \mu_i}
    \,
    \frac{\partial \e_i}{\partial T_i}
    \, - \, 
    \frac{\partial n_i}{\partial T_i }
    \,
    \frac{\partial \e_i}{\partial \mu_i}
    \approx (4 c d- 9 b^2_{ })T_i^5
 $, 
 $ 
 ( \e^{ }_i + \p^{ }_i )
    \frac{\partial n_i}{\partial \mu_i}
    \, - \, 
    n_i
    \frac{\partial \e_i}{\partial \mu_i} \approx T^{ }_i\Delta^{ }_i/3
 $
 and 
 $
  ( \e^{ }_i + \p^{ }_i )
    \frac{\partial n_i}{\partial T_i}
    \, - \, 
    n_i
    \frac{\partial \e_i}{\partial T_i} \approx -\mu^{ }_i\Delta^{ }_i/3
 $, whereby
 \be
 \frac{{\rm d}T^{ }_i}{{\rm d}t} 
 \overset{\rmii{\nr{T dot}}}{\approx}
  -H T^{ }_i + \frac{Q^{ }_i \frac{\partial n_i^{ }}{\partial \mu_i^{ }}
  - J^{ }_i \frac{\partial \e_i^{ }}{\partial \mu_i^{ }} }{\Delta^{ }_i}
  \; \underset{\rmii{\nr{e_i_cons}}}
     {\overset{{\rm no}\;n^{ }_i,\mu^{ }_i}{\approx}}\;
 -H T^{ }_i + \frac{Q^{ }_i}{{{\rm d}\e^{ }_i} / {{\rm d}T^{ }_i}}
 \;, \quad
 \frac{{\rm d}\mu^{ }_i}{{\rm d}t} 
 \overset{\rmii{\nr{mu dot}}}{\approx}
 -H \mu^{ }_i - \frac{Q^{ }_i \frac{\partial n_i^{ }}{\partial T_i^{ }}
  - J^{ }_i \frac{\partial \e_i^{ }}{\partial T_i^{ }} }{\Delta^{ }_i}
 \;.
\ee
 } 
\ba
  \frac{{\rm d} {T}^{ }_{\gamma} }{{\rm d} t}
  & = &
  \frac{
    - 3 H ( \e^{ }_\EMQED + \p^{ }_\EMQED )
    + Q^{ }_\EMQED
  }{
    \displaystyle
    \frac{ d \e_\EMQED }{d T_\gamma}
  }
  \;. \la{dot_T_gamma}
\ea
The system of equations is completed by an equation for the Hubble rate, 
\be
 H^2_{ } 
 \; \equiv \; \frac{\dot{a}^2_{ }}{a^2_{ }}
 \; = \; \frac{8\pi}{3} \frac{\e^{ }_\tot}{\mpl^2}
 \;, \quad
 \e^{ }_\tot 
 \; \equiv \; \sum_i \e^{ }_i
 \;, \la{hubble}
\ee
where $a$ is the scale factor, $\rho_{\rm t}$ the overall energy density, and 
$\mpl^{ } = 1.2209 \times 10^{19}_{ }\,$GeV the Planck mass. 

\pagebreak

Apart from the temperatures and chemical potentials, 
it is also interesting to determine how many $e$-folds, 
$\N \equiv \ln a$, take place during the decoupling period
(cf.\ \se\ref{se:expansion}). For this we need to integrate
$\dot{\N} \; = \; H $. However, the evolution of $a$ correlates
strongly with the redshift of the temperatures, 
so it is simpler to consider
\be
 z \; \equiv \; \frac{a T^{ }_\gamma}{(a T^{ }_\gamma)^{ }_\rmii{ini}}
 \;, \quad 
 \dot{z} \; = \; z \, \biggl( H +
 \frac{\dot{T}^{ }_\gamma}{T^{ }_\gamma}\biggr)
 \;, \la{e-folds}
\ee
where $(...)^{ }_\rmi{ini}$ denotes some initial value well before the decoupling process starts
(cf.\ \eq\nr{int_boundaries}). We remark that $z$ does not influence the solution 
of \eqs\nr{eq:dTdt_dmudt_general}--\nr{dot_T_gamma}, but it can be obtained as their by-product. 

In order to compute the coefficients $Q^{ }_i$ and $J^{ }_i$
that appear in \eqs\nr{eq:dTdt_dmudt_general}--\nr{dot_T_gamma},
we need
to parametrize the underlying neutrino and antineutrino density matrices, 
$\varrho^{ }_\nu$, in terms of $T^{ }_i$ and $\mu^{ }_i$. Assuming
the system to be homogeneous, and denoting by $\vec q$ a physical
momentum, with $q \equiv |\vec q|$, the density matrix 
(in the flavour basis) 
is assumed to take the form 
\be
 \varrho^{ }_\nu(t,\vec q)
 \; = \; 
  \left\{
  \begin{array}{ll}
   \left( 
     \begin{array}{ccc}
       f^{ }_{\nu^{ }_e}(t,q) & & \\
        & f^{ }_{\nu^{ }_\mu}(t,q) & \\
        & & f^{ }_{\nu^{ }_\tau}(t,q)
     \end{array}
   \right) & 
   \hbox{[``no oscillations'']\;,}
   \\ & \\
      f^{ }_{\nu^{ }}(t,q)\left( 
     \begin{array}{ccc}
      \,\, 1\,\,\,\,\,\,\,\, & & \\
        & 1\,\,\,\,\,\,\,\, & \\
        & & 1\,\,
     \end{array}
   \right) & 
   \hbox{[``fast oscillations'']\;.}
  \end{array}
  \right.
  \la{varrho}
\ee
Here the diagonal entries are assumed to take the  {\em equilibrium} Fermi-Dirac shape for {\em massless} neutrinos, 
\be
  f^{ }_{ \nu^{ }_\alpha }
  \; \equiv \; 
  f^{ }_{ \bar \nu^{ }_\alpha }
  \; \equiv \; 
  \big[ e^{(q-\mu_{\nu^{ }_\alpha})/T_{\nu^{ }_\alpha}} + 1 \big]^{-1}
  \;.
  \la{eq:f_nu}
\ee
In practice, the $\nu^{ }_\mu$ and 
$\nu^{ }_\tau$ flavours are treated as degenerate, 
and they are denoted by $\nu^{ }_\mu$.
The generalization of \eq\nr{varrho} to include neutrino oscillations
is given in \eq\nr{factor}.

Given the form in \eq\nr{varrho}, the neutrino energy densities, pressures, 
and number densities that appear in \eqs\nr{T dot} and \nr{mu dot} read
\be
 \e^{ }_{\nu^{ }_\alpha} \; = \; 2 \int_\vec{q} q\, f^{ }_{\nu^{ }_\alpha}
 \;, \quad
 \p^{ }_{\nu^{ }_\alpha} \; = \; 2 \int_\vec{q}
 \frac{ q\, f^{ }_{\nu^{ }_\alpha} }{3}
 \; = \; -2 T^{ }_{\nu^{ }_\alpha}
 \int_\vec{q} \ln (1 - f^{ }_{\nu^{ }_\alpha})
 \;, \quad
 n^{ }_{\nu^{ }_\alpha} \; = \; 2 \int_\vec{q} f^{ }_{\nu^{ }_\alpha}
 \;, \la{e_p_n}
\ee
where the factor 2 counts 
the left-handed neutrinos and right-handed antineutrinos; 
$\int_\vec{q} = (2 \pi^2_{ })^{-1}_{ }\int_0^\infty \! {\rm d}q \, q^2_{ }$;
and for $\p^{ }_{\nu^{ }_a}$ we have shown two versions related by partial
integration. 
The integrals in \eq\nr{e_p_n} can be carried out in terms of polylogarithms, 
as summarized in \app\ref{sec:Bessel}. 
We remark that neutrinos have a vanishing trace anomaly in this massless and 
non-interacting limit, $\e^{ }_{\nu^{ }_a} - 3 \p^{ }_{\nu^{ }_a} = 0$.
The QED energy density and pressure, $\e^{ }_\EMQED$ and $\p^{ }_\EMQED$, 
have more structure,
and we return to them in \se\ref{sec:QED_corrections} 
and in \app\ref{sec:qed_eos}. 

\subsection{Determination of rate coefficients via matching to linear-response regime}
\label{ss:matching}

In the momentum-averaged approach of \se\ref{ss:momave}, 
the interactions of the neutrino 
and QED ensembles are modelled by energy density 
transfer rates ($Q^{ }_i$) and number density transfer rates ($J^{ }_i$). 
These rates refer to the given density going to the destination ensemble, 
but they are a sum over all sources. 
In the following, in order to streamline the notation, 
we first consider the partial contributions 
$Q \equiv Q^{ }_{\nu^{ }_\alpha \leftarrow \EMQED}$ 
and $J \equiv J^{ }_{\nu^{ }_\alpha \leftarrow \EMQED}$. 
We return to the neutrino-neutrino rates 
at the end of this section (cf.\ \eqs\nr{Q_matrix_gamma}--\nr{J_matrix_nu}). 

If two ensembles 
are in thermal equilibrium with each other, the rate coefficients vanish, 
as required by detailed balance. Therefore, quite generally, 
the values of the coefficients are of {\em first order} in 
the differences of the characteristics of the neutrino 
and QED ensembles. However, as the universe 
cools down, in the end the two ensembles differ from each other
by effects of $\rmO(1)$. To describe the whole decoupling 
process, we need approximate expressions for $Q$ and $J$ 
also {\em beyond the linear order}. In the following, we describe how
such expressions can be obtained, by combining a strict linear response
philosophy with a phenomenological Maxwell-Boltzmann (MB) approximation. 

If we introduce the kinematic Maxwell-Boltzmann approximation for
describing the phase-space distributions of electrons, and furthermore
treat the electrons as massless ($m^{ }_e = 0$) for a moment, then $Q$ and $J$ can
be computed exactly at LO 
in $\alpha^{ }_\rmi{em}$. The expressions 
read~\cite{EscuderoAbenza:2020cmq}
\begin{subequations}\label{eqs:MB_rates}
\begin{eqnarray}
 Q^\rmii{MB}_{ } 
 & 
 \equiv 
 & 
 \frac{32 G_\rmii{F}^2}{\pi^5_{ }}
 \bigl(\, g^2_{\alpha\rmii{L}} + g^2_{\alpha\rmii{R}} \,\bigr)
 \Bigl[\, 
   4 \bigl(\, 
 \overbrace{
 T_\gamma^9
 - 
 T_\nu^9  e^{\frac{2\mu_\nu}{T_\nu}}_{ }
 }^{\rm gain-loss}
 \,\bigr)
 + 
   7 
 \overbrace{
 T_\gamma^4 \, T_\nu^4 \, 
   \bigl(\, T^{ }_\gamma - T^{ }_\nu \,\bigr) e^{\frac{\mu_\nu}{T_\nu} }_{ }
 }^{\rm scattering}
 \,\Bigr]
 \;, \label{Q^MB} 
 \\[2mm] 
 J^\rmii{MB}_{ } 
 & 
 \equiv 
 & 
 \frac{32 G_\rmii{F}^2}{\pi^5_{ }}
 \bigl(\, g^2_{\alpha\rmii{L}} + g^2_{\alpha\rmii{R}} \,\bigr)
 \bigl(\, 
 \underbrace{
 T_\gamma^8
 -
 T_\nu^8  e^{\frac{2\mu_\nu}{T_\nu}}_{ }
 }_{\rm gain-loss}
 \,\bigr)
 \;, \label{J^MB}
\end{eqnarray}
\end{subequations}
where the physical origin of each term has been indicated. 
Scatterings do not transfer particle number while pair creations 
and annihilations (``gain$-$loss'') do. 
The values of the couplings used are collected
in \app\ref{se:couplings}. 

Now, if we go beyond the Maxwell-Boltzmann approximation, and
include $m^{ }_e \neq 0$ in the computation, the functional forms
of $Q$ and $J$ become more complicated, and cannot be written analytically. However, the result of any
computation can still be expanded to first order
around equilibrium. So, let us define
\begin{equation}
 \frac{T^{ }_\nu}{T^{ }_\gamma} 
 \; \equiv \; 
 1 + \epsilon^{ }_{\iT} + \rmO(\epsilon^2_{ })
 \;, \quad
 \frac{\mu^{ }_\nu}{T^{ }_\nu} 
 \; \equiv \; 
 \epsilon^{ }_\mu + \rmO(\epsilon^2_{ })
 \;. \label{lin_perts}
\end{equation}
Then, to facilitate the linearisation, we can express the 
full leading-order (LO) transfer rates as 
\ba
  \frac{ Q_{ }^\rmii{LO} }{ T_\gamma^9  G^2_\rmiii{F} } 
  &
  =
  & 
  \frac{1}{\pi^5}
  \biggl\{
  \bigl(\, g^2_{\alpha\rmii{L}} + g^2_{\alpha\rmii{R}} \,\bigr)
  \,
  \widehat{Q}_{1}^{ } 
  \bigg(
    \frac{T_\nu}{T_\gamma},
    \frac{\mu_\nu}{T_\nu};
    \frac{m_e}{T_\gamma}
  \bigg) 
  \; + \;
  g^{ }_{\alpha\rmii{L}}  g^{ }_{\alpha\rmii{R}}  
  \;
  \widehat{Q}_{2}^{ } 
  \bigg(
    \frac{T_\nu}{T_\gamma},
    \frac{\mu_\nu}{T_\nu};
    \frac{m_e}{T_\gamma}
  \bigg) 
  \biggr\}
  \; , \la{Q^LO}
  \\[2mm]
  \frac{ J_{ }^\rmii{LO} }{ T_\gamma^8  G^2_\rmiii{F} } 
  &
  =
  & 
  \frac{1}{\pi^5}
  \biggl\{
    \bigl(\, g^2_{\alpha\rmii{L}} + g^2_{\alpha\rmii{R}} \,\bigr)
    \;
  \widehat{J}_{1}^{ } 
  \bigg(
    \frac{T_\nu}{T_\gamma},
    \frac{\mu_\nu}{T_\nu};
    \frac{m_e}{T_\gamma}
  \bigg) 
  \; + \;
  g^{ }_{\alpha\rmii{L}}  g^{ }_{\alpha\rmii{R}}  
  \;
  \widehat{J}_{2}^{ } 
  \bigg(
    \frac{T_\nu}{T_\gamma},
    \frac{\mu_\nu}{T_\nu};
    \frac{m_e}{T_\gamma}
    \bigg) 
  \biggr\} 
  \; , \la{J^LO}
\ea
where $\widehat{Q}$ and $\widehat{J}$ are dimensionless functions of 
dimensionless variables. 
The terms proportional to  $g^{ }_{\alpha\rmii{L}}  g^{ }_{\alpha\rmii{R}}$
arise when $m_e \neq 0\,$ (cf.\ \eq\nr{res_MM}). 
The LO functions can be decomposed into  
``gain$-$loss'' and ``scattering'' parts, 
as done in \eqs\nr{Q^MB} and \nr{J^MB}.
After inserting \eq\nr{lin_perts} into $Q^\rmii{LO}$ and expanding, 
the leading terms in the expansion vanish 
due to the equilibrium condition. 
Thus we obtain, e.g.,
\ba
  \frac{ Q^\rmii{LO} }{ T_\gamma^9  G^2_\rmiii{F} } 
  &
  =
  & 
  \frac{1}{\pi^5}
  \biggr\{
  \bigl(\, g^2_{\alpha\rmii{L}} + g^2_{\alpha\rmii{R}} \,\bigr)
  \;
  \Bigl[\, 
  \epsilon^{ }_{\iT} \widehat{Q}^{(1,0)}_{1} 
  +
  \epsilon^{ }_\mu \widehat{Q}^{(0,1)}_{1} 
  \,\Bigr] 
  \; + \;
  g^{ }_{\alpha\rmii{L}} g^{ }_{\alpha\rmii{R}} 
  \,
  \Bigl[\,
  \epsilon^{ }_{\iT} \widehat{Q}^{(1,0)}_{2} 
  +
  \epsilon^{ }_\mu \widehat{Q}^{(0,1)}_{2} 
   \,\Bigr]
  \biggr\} 
  +
  {\cal O}
  \bigl( 
  \epsilon^{2}_{\iT}, \epsilon^{ }_{\iT} \epsilon^{ }_\mu, \epsilon^{2}_\mu 
  \bigr) 
  \; , \la{LO_lin_response}
\ea
where we introduced the following coefficients for $i = \{ 1,2 \}$, 
which are functions of $m_e/T_\gamma\,$ alone,
\ba
  \widehat{Q}^{(1,0)}_{i} 
  &
  \equiv
  &
  \partial^{ }_{\frac{\raise0.5ex\hbox{\scalebox{0.7}{$T_\nu$}}}
                     {\raise-0.5ex\hbox{\scalebox{0.7}{$T_\gamma$}}}}
  \widehat{Q}_{i}^{ } 
  \bigg(
    \frac{T_\nu}{T_\gamma},
    \frac{\mu_\nu}{T_\nu};
    \frac{m_e}{T_\gamma}
  \bigg) 
  \bigg|_{T_\nu = T_\gamma,\, \mu_\nu=0}
  \; , \la{Q10}
  \\[2mm]
  \widehat{Q}^{(0,1)}_{i} 
  &
  \equiv
  &
  \partial^{ }_{\frac{\raise0.5ex\hbox{\scalebox{0.7}{$\mu_\nu$}}}
                     {\raise-0.5ex\hbox{\scalebox{0.7}{$T_\nu$}}}}
  \widehat{Q}_{i}^{ } 
  \bigg(
    \frac{T_\nu}{T_\gamma},
    \frac{\mu_\nu}{T_\nu};
    \frac{m_e}{T_\gamma}
  \bigg) 
  \bigg|_{T_\nu = T_\gamma,\, \mu_\nu=0}
  \; . \la{Q01}
\ea
A similar expression can be obtained for $J_{ }^\rmii{LO}$ in \eq\nr{J^LO}, 
which then involves the coefficients $\widehat{J}^{\hspace*{0.4mm}(1,0)}_{i}\,$ 
and $\widehat{J}^{\hspace*{0.4mm}(0,1)}_{i}$. 
The same strategy can also be employed at higher orders in $\alpha_{\rm em}$,
though for the moment complete results are only available in 
the $m_e = 0$ limit~\cite{Jackson:2023zkl,Jackson:2024gtr}. 

In order to extend \eqs\nr{Q^MB} and \nr{J^MB} in the most general way, 
as allowed by the linearisation provided by \eq\nr{LO_lin_response}, 
we introduce $m_e$-dependent correction factors as 
\ba
  \bigl(\, g^2_{\alpha\rmii{L}} + g^2_{\alpha\rmii{R}} \,\bigr)
  \, 
  \bigl(
  T_\gamma^9 - T_\nu^9 e^{2\frac{\mu_\nu}{T_\nu}} 
  \bigr) 
  &
  \to
  &
  \Bigl[\,
  \bigl(\, g^2_{\alpha\rmii{L}} + g^2_{\alpha\rmii{R}} \,\bigr)
  \;
  f^{ }_{a1}\Big( \frac{m_e}{T_\gamma} \Big)
  \; + \;
  g^{ }_{\alpha\rmii{L}} g^{ }_{\alpha\rmii{R}} 
  \;
  f^{ }_{a2}\Big( \frac{m_e}{T_\gamma} \Big)
  \,\Bigr] \;
  \bigl(
  T_\gamma^9 - T_\nu^9 e^{2\frac{\mu_\nu}{T_\nu}} 
  \bigr) 
  \nn[2mm]
  &
  +
  &
  \Bigl[\,
  \bigl(\, g^2_{\alpha\rmii{L}} + g^2_{\alpha\rmii{R}} \,\bigr)
  \;
  f^{ }_{a3}\Big( \frac{m_e}{T_\gamma} \Big)
  \; + \;
  g^{ }_{\alpha\rmii{L}} g^{ }_{\alpha\rmii{R}} 
  \;
  f^{ }_{a4}\Big( \frac{m_e}{T_\gamma} \Big)
  \,\Bigr] \;
  \sqrt{T_\gamma^9 
  T_\nu^9}
  \bigl(
  e^{2\frac{\mu_\nu}{T_\nu}}  - 1
  \bigr) 
  \; , \la{newMB1}
  \\[4mm]
  \bigl(\, g^2_{\alpha\rmii{L}} + g^2_{\alpha\rmii{R}} \,\bigr)
  T_\gamma^4 \, T_\nu^4 \, 
  \bigl(\, T^{ }_\gamma - T^{ }_\nu \,\bigr) e^{\frac{\mu_\nu}{T_\nu} }_{ }
  &
  \to
  &
  \Bigl[\,
  \bigl(\, g^2_{\alpha\rmii{L}} + g^2_{\alpha\rmii{R}} \,\bigr)
  \;
  f^{ }_{s1}\Big( \frac{m_e}{T_\gamma} \Big)
  \; + \;
  g^{ }_{\alpha\rmii{L}} g^{ }_{\alpha\rmii{R}} 
  \;
  f^{ }_{s2}\Big( \frac{m_e}{T_\gamma} \Big)
  \,\Bigr] \;
  T_\gamma^4 \, T_\nu^4 \, 
  \bigl(\, T^{ }_\gamma - T^{ }_\nu \,\bigr) e^{\frac{\mu_\nu}{T_\nu} }_{ }
  \nn[2mm]
  &
  +
  &
  \Bigl[\,
  \bigl(\, g^2_{\alpha\rmii{L}} + g^2_{\alpha\rmii{R}} \,\bigr)
  \;
  f^{ }_{s3}\Big( \frac{m_e}{T_\gamma} \Big)
  \; + \;
  g^{ }_{\alpha\rmii{L}} g^{ }_{\alpha\rmii{R}} 
  \;
  f^{ }_{s4}\Big( \frac{m_e}{T_\gamma} \Big)
  \,\Bigr] \;
  \sqrt{
  T_\gamma^9
  T_\nu^9
  }
  \bigl(
  e^{\frac{\mu_\nu}{T_\nu}}  - 1
  \bigr) 
  \; , \la{newMB2}
  \\[4mm]
  \bigl(\, g^2_{\alpha\rmii{L}} + g^2_{\alpha\rmii{R}} \,\bigr)
  \, 
  \bigl(
  T_\gamma^8 - T_\nu^8 e^{2\frac{\mu_\nu}{T_\nu}} 
  \bigr) 
  &
  \to
  &
  \Bigl[\,
  \bigl(\, g^2_{\alpha\rmii{L}} + g^2_{\alpha\rmii{R}} \,\bigr)
  \;
  f^{ }_{n1}\Big( \frac{m_e}{T_\gamma} \Big)
  \; + \;
  g^{ }_{\alpha\rmii{L}} g^{ }_{\alpha\rmii{R}} 
  \;
  f^{ }_{n2}\Big( \frac{m_e}{T_\gamma} \Big)
  \,\Bigr] \;
  \bigl(
  T_\gamma^8 - T_\nu^8 e^{2\frac{\mu_\nu}{T_\nu}} 
  \bigr) 
  \nn[2mm]
  &
  +
  &
  \Bigl[\,
  \bigl(\, g^2_{\alpha\rmii{L}} + g^2_{\alpha\rmii{R}} \,\bigr)
  \;
  f^{ }_{n3}\Big( \frac{m_e}{T_\gamma} \Big)
  \; + \;
  g^{ }_{\alpha\rmii{L}} g^{ }_{\alpha\rmii{R}} 
  \;
  f^{ }_{n4}\Big( \frac{m_e}{T_\gamma} \Big)
  \,\Bigr] \;
  T_\gamma^4
  T_\nu^4
  \bigl(
  e^{2\frac{\mu_\nu}{T_\nu}}  - 1
  \bigr) 
  \; . \la{newMB3}
  \\[-.2cm]
  \nonumber
\ea
In the second lines of these equations, we chose 
to employ a temperature dependence symmetric 
in $T^{ }_\gamma$ and $T^{ }_\nu\,$ 
(this will be convenient shortly, 
when we consider also the transfers between the neutrino ensembles).

We can then match linear-response versions of \eqs\nr{newMB1}--\nr{newMB3}
to \eq\nr{LO_lin_response}, thereby determining the unknown functions. 
The results are given in \app\ref{app:match},  
and illustratated numerically in \fig\ref{fa_fs_fn_plot}.
The non-zero values for $m_e^{ }/T^{ }_\gamma = 0$ are 
$f_{a1}(0) = 0.884\,$, 
$f_{s1}(0) = 0.829\,$,  
$f_{n1}(0) = 0.829\,$, 
$f_{a3}(0) = 0.032\,$,
and 
$f_{n3}(0) = 0.041\,$.
Furthermore, $f_{s3}$ and $f_{s4}$
are identically zero for any $m_e\,$, 
indicating that 
there is no order $\epsilon_\nu$ term in the 
expansion of $Q^\rmi{LO}_{\rm scattering}\,$. 
This fact can be shown quite generally\footnote{%
 The proof relies on detailed balance, 
 as well as exchanging integration variables,  
 and it should be true even if $\mu^{ }_{\bar \nu} \neq \mu^{ }_\nu$.
}
by observing that  
$\widehat{Q}_i$ vanishes when $T_\nu=T_\gamma\,$, 
for any value of $\mu_\nu$ and $m_e\,$.

\begin{figure}[t]

\hspace*{-0.1cm}
\centerline{
  \includegraphics[scale=.54]{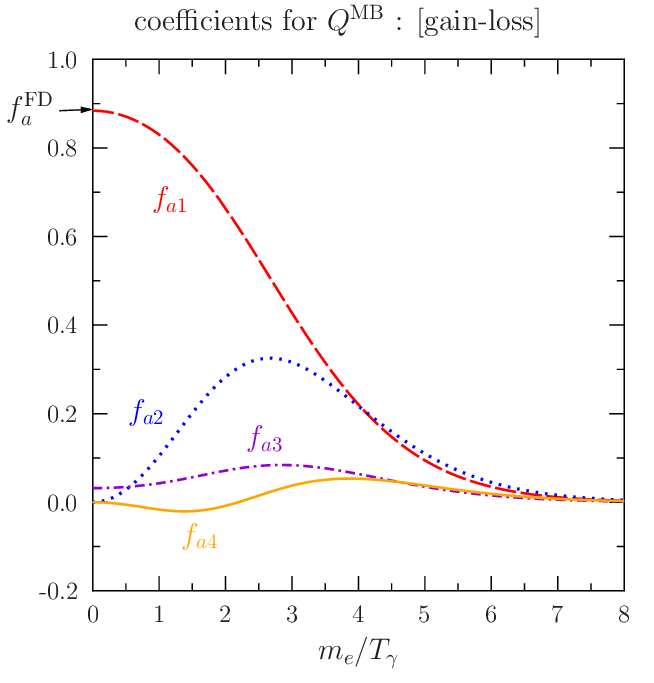}
  \hspace{.05cm}
  \includegraphics[scale=.54]{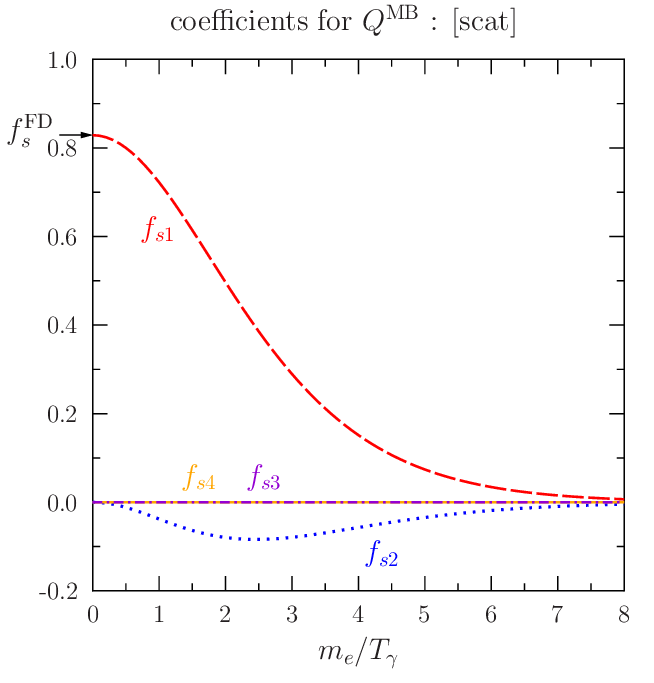}
  \hspace{.05cm}
  \includegraphics[scale=.54]{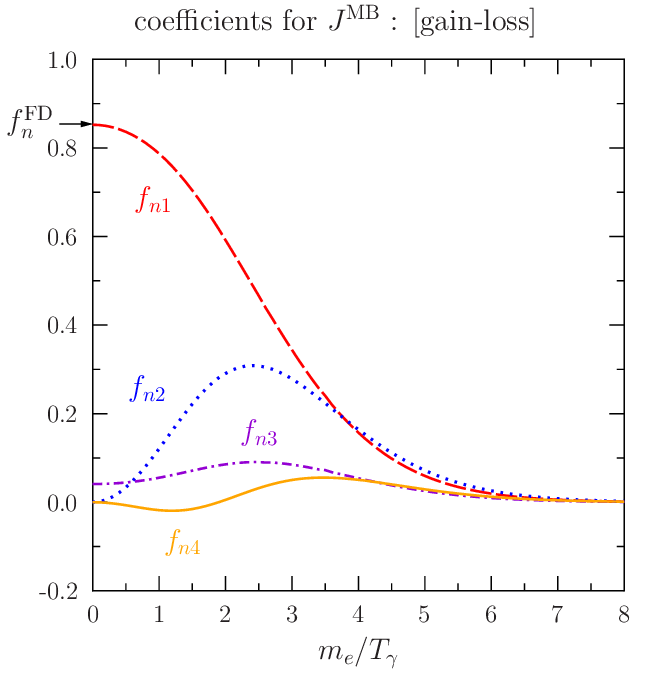}
}
\vspace{-0.5cm}
\caption[a]{\setstretch{1.3}
  The MB correction factors for the energy density 
  and number density transfer rates, 
  determined by matching to the linear response regime. 
  These correct \eqs\nr{Q^MB} and \nr{J^MB} 
  as laid out by the replacements in \eqs\nr{newMB1}--\nr{newMB3}.
}
\la{fa_fs_fn_plot}
\end{figure}

\vspace*{3mm}

We now return to the general matrix structure of the rate coefficients.
Supplementing \eqs\nr{newMB1}--\nr{newMB3} by the contributions of the 
other neutrino ensembles, we can write down full rates 
for a neutrino of a given flavour $\alpha\,$. 
In order to do so, we introduce the functions 
\ba
    F_a^{12}(T^{ }_\rmiii{A},T^{ }_\rmiii{B},\mu^{ }_\rmiii{A},\mu^{ }_\rmiii{B}) 
    & \;\equiv\; &
    T_\rmiii{A}^9 
    e^{2\frac{\mu^{ }_\rmiii{A}\vphantom{ | }}{T^{ }_\rmiii{B} \vphantom{ | } }}
    \; - \;
    T_\rmiii{B}^9 
    e^{2\frac{\mu^{ }_\rmiii{B} \vphantom{ | } }{T^{ }_\rmiii{B} \vphantom{ | } }}
    \; ,
    \\[2mm]
    F_a^{34}(T^{ }_\rmiii{A},T^{ }_\rmiii{B},\mu^{ }_\rmiii{A},\mu^{ }_\rmiii{B}) 
    & \;\equiv\; & 
    \big(T_\rmiii{A}^9 T_\rmiii{B}^9\big)_{ }^{\frac{1}{2}} 
    \,\big(
    e_{ }^{2\frac{\mu^{ }_\rmiii{B} \vphantom{ | } }{T^{ }_\rmiii{B} \vphantom{ | } }  } 
    \; - \; 
    e_{ }^{2\frac{\mu^{ }_\rmiii{A} \vphantom{ | } }{T^{ }_\rmiii{A} \vphantom{ | } }  }
    \big) 
    \; ,
    \\[2mm]
    F_s^{12}(T^{ }_\rmiii{A},T^{ }_\rmiii{B},\mu^{ }_\rmiii{A},\mu^{ }_\rmiii{B}) 
    & \;\equiv\; & 
    T_\rmiii{A}^4 T_\rmiii{B}^4(T^{ }_\rmiii{A}-T^{ }_\rmiii{B}) 
    e^{\frac{\mu^{ }_\rmiii{A} \vphantom{ | } }{T^{ }_\rmiii{A} \vphantom{ | } }}
    e^{\frac{\mu^{ }_\rmiii{B} \vphantom{ | } }{T^{ }_\rmiii{B} \vphantom{ | } }} 
    \; ,
    \\[2mm]
    F_n^{12}(T^{ }_\rmiii{A},T^{ }_\rmiii{B},\mu^{ }_\rmiii{A},\mu^{ }_\rmiii{B}) 
    & \;\equiv\; &
    T_\rmiii{A}^8 e^{2\frac{\mu^{ }_\rmiii{A} \vphantom{ | } }{T^{ }_\rmiii{A} \vphantom{ | } }}
    \; - \;
    T_\rmiii{B}^8 e^{2\frac{\mu^{ }_\rmiii{B} \vphantom{ | } }{T^{ }_\rmiii{B} \vphantom{ | } }} 
    \; ,
    \\[2mm]
    F_n^{34}(T^{ }_\rmiii{A},T^{ }_\rmiii{B},\mu^{ }_\rmiii{A},\mu^{ }_\rmiii{B}) 
    & \;\equiv\; & 
    T_\rmiii{A}^4 T_\rmiii{B}^4
    \,\big(e^{2\frac{\mu^{ }_\rmiii{B} \vphantom{ | } }{T^{ }_\rmiii{B} \vphantom{ | } }}
    \; - \;
    e^{2\frac{\mu^{ }_\rmiii{A} \vphantom{ | } }{T^{ }_\rmiii{A} \vphantom{ | } }}\big)
    \; .
\ea  
The subscript and superscript in $F$ refer to which 
mass-dependent correction the structure concerns
(cf.\ \eqs\nr{newMB1}--\nr{newMB3}). 
By construction, these functions are odd under the 
simultaneous exchange 
$T^{ }_\rmiii{A} \leftrightarrow T^{ }_\rmiii{B}$ and 
$\mu^{ }_\rmiii{A} \leftrightarrow \mu^{ }_\rmiii{B}\,$, 
ensuring that the partial 
rates satisfy $Q^{ }_{i \leftarrow j} = -\, Q^{ }_{j \leftarrow i}$ 
and vanish in equilibrium. 
The mutual exchanges among the 
QED and neutrino systems can then 
be written as
\ba
 Q^{ }_{\nu^{ }_{\alpha} \leftarrow \EMQED}
 & = &
 \frac{32 G_\rmii{F}^2}{\pi^5_{ }}
 \biggl\{\, 
   4  \times \biggl[
  \,
  \biggl(
  \bigl(\, g^2_{\alpha\rmii{L}} + g^2_{\alpha\rmii{R}} \,\bigr)
  \;
  f^{ }_{a1}\Big( \frac{m_e}{T_\gamma} \Big)
  \; + \;
  g^{ }_{\alpha\rmii{L}} g^{ }_{\alpha\rmii{R}} 
  \;
  f^{ }_{a2}\Big( \frac{m_e}{T_\gamma} \Big)
  \biggr)
  \;
  F_a^{12}(T_\gamma,T_{\nu_\alpha},0,\mu_{\nu_\alpha})
  \la{Q_matrix_gamma}
  \\ \nonumber
  & & 
  \hspace{17.5mm}
  +\,
  \biggl( \bigl(\, g^2_{\alpha\rmii{L}} + g^2_{\alpha\rmii{R}} \,\bigr)
  \;
  f^{ }_{a3}\Big( \frac{m_e}{T_\gamma} \Big)
  \; + \;
  g^{ }_{\alpha\rmii{L}} g^{ }_{\alpha\rmii{R}} 
  \;
  f^{ }_{a4}\Big( \frac{m_e}{T_\gamma} \Big)
  \biggr) \,
  F_a^{34}(T_\gamma,T_{\nu_\alpha},0,\mu_{\nu_\alpha})
  \,\biggr] \nn
  & & 
  \hspace{8.5mm}
  + 
  \hspace{1mm}
  7  \times  \biggl[\,
  \ \  \bigl(\, g^2_{\alpha\rmii{L}} + g^2_{\alpha\rmii{R}} \,\bigr)
  \;
  f^{ }_{s1}\Big( \frac{m_e}{T_\gamma} \Big)
  \; + \;
  g^{ }_{\alpha\rmii{L}} g^{ }_{\alpha\rmii{R}} 
  \;
  f^{ }_{s2}\Big( \frac{m_e}{T_\gamma} \Big)
  \,\biggr] \;
  F_s^{12}(T_\gamma,T_{\nu_\alpha},0,\mu_{\nu_\alpha})
  \, \biggr\}
  \;,
  \;\nn[3mm]
  Q^{ }_{\nu^{ }_{\alpha} \leftarrow \nu^{ }_{\beta}}
  & = & 
  \  \frac{8 G_\rmii{F}^2}{\pi^5_{ }}
  \, \biggl\{\, 
  4  \times
  \biggl[ \,
  f{ }_{a1}(0) \;F_a^{12}(T_{\nu_\beta},T_{\nu_\alpha},\mu_{\nu_\beta},\mu_{\nu_\alpha}) 
  \; + \; 
  f^{ }_{a3}(0) \; 
  F_a^{34}(T_{\nu_\beta},T_{\nu_\alpha},\mu_{\nu_\beta},\mu_{\nu_\alpha})
  \, \biggr]
  \la{Q_matrix_nu} \\
  & &  
  \hspace{8.5mm}
  + 
  \hspace{1mm}
  7 \, \times \, f^{ }_{s1}(0)
  \; F_s^{12}(T_{\nu_\beta},T_{\nu_\alpha},\mu_{\nu_\beta},\mu_{\nu_\alpha})  \, \biggr\} 
  \;, 
  \nn[3mm]
  J^{ }_{\nu_{\alpha} \leftarrow \EMQED}
  & = &
  \frac{32 G_\rmii{F}^2}{\pi^5_{ }}
  \biggl[
  \,
  \biggl(\bigl(\, g^2_{\alpha\rmii{L}} + g^2_{\alpha\rmii{R}} \,\bigr)
  \;
  f^{ }_{n1}\Big( \frac{m_e}{T_\gamma} \Big)
  \; + \;
  g^{ }_{\alpha\rmii{L}} g^{ }_{\alpha\rmii{R}} 
  \;
  f^{ }_{n2}\Big( \frac{m_e}{T_\gamma} \Big)
  \,\biggr)\;
  F_n^{12}(T_\gamma,T_{\nu_\alpha},0,\mu_{\nu_\alpha})
  \la{J_matrix_gamma} \\ \nonumber
  & & 
  \hspace{8mm}
  + \, \biggl( 
  \bigl(\, g^2_{\alpha\rmii{L}} + g^2_{\alpha\rmii{R}} \,\bigr)
  \;
  f^{ }_{n3}\Big( \frac{m_e}{T_\gamma} \Big)
  \; + \;
  g^{ }_{\alpha\rmii{L}} g^{ }_{\alpha\rmii{R}} 
  \;
  f^{ }_{n4}\Big( \frac{m_e}{T_\gamma} \Big)
  \,\biggr) \;
  F_n^{34}(T_\gamma,T_{\nu_\alpha},0,\mu_{\nu_\alpha})
  \,\biggr] 
  \;,
  \;\nonumber \\[3mm]
    J^{ }_{\nu^{ }_{\alpha} \leftarrow \nu^{ }_{\beta}}
  & = &
  \ 
  \frac{8 G_\rmii{F}^2}{\pi^5_{ }}
  \ 
  \Bigl[\, 
  f^{ }_{n1}(0) 
  \, F_n^{12}(T_{\nu_\beta},T_{\nu_\alpha},\mu_{\nu_\beta},\mu_{\nu_\alpha})
  \; + \; 
  f^{ }_{n3}(0) 
  \;
  F_n^{34}(T_{\nu_\beta},T_{\nu_\alpha},\mu_{\nu_\beta},\mu_{\nu_\alpha}) 
  \,  \Bigr]
  \;.
  \la{J_matrix_nu}
  \\[-1mm]
  \nonumber
\ea
The specific transfer rates introduced in 
\eqs\nr{e_i_cons} 
and \nr{n_i_cons}
are given in terms of these partial rates as
\ba
  Q^{ }_{\nu^{ }_{\alpha}}
  & = &
  Q^{ }_{\nu^{ }_{\alpha} \leftarrow \EMQED}
  \; + \;
  \sum_{\beta \neq \alpha}
  Q^{ }_{\nu^{ }_{\alpha} \leftarrow \nu^{ }_{\beta} }
  \;, \quad 
  Q^{ }_{\EMQED} 
  \; = \;
  \sum_\alpha Q^{ }_{\EMQED \leftarrow \nu^{ }_{\alpha}}
  \, , \\[3mm]
  J^{ }_{\nu^{ }_{\alpha}}
  & = &
  J^{ }_{\nu^{ }_{\alpha} \leftarrow \EMQED}
  \; + \;
  \sum_{\beta \neq \alpha}
  J^{ }_{\nu^{ }_{\alpha} \leftarrow \nu^{ }_{\beta} }
  \, .
\ea
We note that in this approach, the self-interactions of neutrinos of
flavour $\alpha$ do not play a role, even though in the physical world
they can redistribute momenta. 

The transfer rates close the system defined
by \eqs\eqref{eq:dTdt_dmudt_general}--\eqref{hubble}. 
While slightly lengthier than the $m_e\to 0$ Maxwell-Boltzmann 
approximations in eq.~\eqref{eqs:MB_rates}, they are still compact, and 
more accurate and flexible than before. 
As described in \se\ref{sec:data}, 
we provide an 
easy-to-use code and tables 
based on the above procedure
at \href{https://github.com/MiguelEA/nudec_BSM}{nudec\_BSM\_v2}. 

%
\addtocontents{toc}{\vspace{-0.5em}}
\section{Inclusion of dynamical oscillations in the momentum-averaged approach}
\la{ss:osc}
\addtocontents{toc}{\vspace{+0.0em}} 

It is possible to include dynamical neutrino oscillations 
in the momentum-averaged approach. 
Though the effect is not expected to be large 
in the SM~\cite{Mangano:2005cc,deSalas:2016ztq}, 
it is worth checking its magnitude within our approach, 
as it can become relevant for several BSM scenarios, 
as will be discussed in more detail in Part II~\cite{Escudero:2025mvt}. 

The system that we would like to solve is given by the Liouville equation, 
\begin{align}
  \bigl( \partial^{ }_t - H q\, \partial^{ }_q\bigr) 
  \varrho^{ }_\nu(t,\vec{q})
  \; = \;
  - i [{\cal H}, \varrho^{ }_\nu(t,\vec{q})] + j[\varrho^{ }_\nu] 
  \;,
  \la{liouville}
\end{align}
where the part $j[\varrho^{ }_\nu]$ accounts 
for the interaction rates~\cite{Sigl:1993ctk}.
In the absence of any asymmetry, the anti-neutrino
density matrix, $\varrho^{ }_{\bar\nu}\,$,
obeys the same evolution equation as $\varrho^{ }_\nu\,$. 
Considering ultrarelativistic neutrinos, 
and omitting a part proportional
to the unit matrix, 
the Hamiltonian $\mathcal{H}$ is given in the flavour basis by
\begin{align}
\label{eq:Hamiltonian}
  \mathcal{H} 
  \; = \; 
  U \frac{\Delta M^2}{2q} U^\dagger 
  + \mathcal{V} 
  \;, 
\end{align}
where $U$ is the PMNS matrix~\cite{ParticleDataGroup:2024cfk}, 
$\Delta M^2 = \mathrm{diag}(0,\Delta m_{21}^2, \Delta m_{31}^2)$ 
the neutrino mass matrix, 
and $\mathcal{V}$ the matter potential~\cite{Notzold:1987ik}. 
For the neutrino mixing angles and masses, 
we insert the latest global three-neutrino fit~\cite{Esteban:2024eli},
\ba
  (\sin^2\theta_{12}, \sin^2\theta_{23}, \sin^2\theta_{13})
  & \;=\; & 
  (0.308, 0.470, 0.02215) \;, \\[2mm] 
  (\Delta m_{21}^2, \Delta m_{31}^2)
  & \;=\; & 
  (7.49 \times 10^{-5}\,\mathrm{eV}^2, 2.514\times 10^{-3}\,\mathrm{eV}^2)
  \,,   \la{neutrino_params}
\ea
assuming normal hierarchy for the neutrino masses. 
We neglect a possible CP-violating phase in the leptonic sector, 
as it has been shown to have 
a negligible impact on $N_{\mathrm{eff}}$~\cite{Froustey:2020mcq}.

Following refs.~\cite{Domcke:2025lzg, Domcke:2025jiy}, 
the density matrix is generalized from 
\eqs\nr{varrho} and \nr{eq:f_nu} into
\be
 \varrho^{ }_\nu(t,\vec{q})
 \; \equiv \; 
 \frac{1}{e^{q/T^{ }_\nu}_{ } + 1} \times \mathcal{R}^{ }_\nu(t)
 \;, \la{factor}
\ee
where $\mathcal{R}^{ }_\nu$ retains the matrix structure and 
is independent of the momentum~$q$. The evolution equation
for $\mathcal{R}^{ }_\nu$ can be obtained by integrating \eq\nr{liouville} over $\vec{q}$; 
an explicit derivation can be found in ref.~\cite{Domcke:2025lzg}.

At temperatures sufficiently below neutrino decoupling, $T_\nu \leq 1.5\,\mathrm{MeV}$, 
neutrino flavour oscillations 
become increasingly fast. This implies that in the mass basis, $\mathcal{R}^{ }_\nu$ converges to a diagonal matrix. 
Physical observables can then be extracted in the same way as done from the solution of a full Liouville equation, following \eq\nr{f_nu_noneq}. 

For the numerical calculations, we use a {\tt C++} 
code, inspired by the publicly available $\texttt{COFLASY-C}$ code~\cite{Domcke:2025jiy}. 
Instead of using various approximations as in ref.~\cite{Domcke:2025lzg}, we include the full neutrino-neutrino and neutrino-electron collision terms reported in ref.~\cite{deSalas:2016ztq}. 
This allows us not only to include neutrino oscillations, 
but to have identical collision terms as in Fortepiano~\cite{Bennett:2020zkv}, 
and hence provide a fair comparison. 

\addtocontents{toc}{\vspace{-0.5em}} 
\section{QED corrections to energy density and pressure}
\label{sec:QED_corrections}
\addtocontents{toc}{\vspace{-0.5em}} 

Apart from the rates of energy and number density transfer between the QED and neutrino ensembles, 
an important ingredient for \eqs\nr{dot_T_gamma} and \nr{hubble} is given
by the QED energy density and pressure, $\e^{ }_\EMQED$ and $\p^{ }_\EMQED$.
The determination of these functions has a long history, with the 
neutrino decoupling codes usually making use of the expressions 
given in ref.~\cite{Bennett:2019ewm}. They include QED corrections up to 
$\rmO(e^3_{ })$, normally omitting 
a set of arguably small ``logarithmic'' corrections of $\rmO(e^2_{ })$. 

As a part of the current study, we have scrutinized the QED 
corrections to the energy density and pressure. Up to $\rmO(e^3_{ })$, 
our results agree with ref.~\cite{Bennett:2019ewm}, however we find 
representations which are substantially 
simpler than reported there. 
Given that
fewer special functions appear and unnecessary numerical 
cancellations can thereby be avoided, the simplified expressions
can help to speed up the numerical evaluation. Our detailed
results are given in \app\ref{sec:qed_eos}. 

We have also studied the influence of the terms of 
$\rmO(e^4_{ })$ and $\rmO(e^5_{ })$. These are fully known
only in the high-temperature limit, $T^{ }_\gamma \gg m^{ }_e$. They are 
conceptually important in the sense that in that 
domain, they are the leading corrections yielding a trace 
anomaly, i.e.\ $\e^{ }_\EMQED - 3\p^{ }_\EMQED = \rmO(e^4_{ }T^4_\gamma)$.
With the help of renormalization group invariance, we have been able to 
extend parts of the results to any temperature.

To be concrete, 
in the massless limit, the pressure is often expressed as 
\be
 \p^{ }_\EMQED = \frac{\pi^2_{ }T^4_{\gamma}}{45}
 \, 
 \biggl[ 
 c^{ }_0
 + e^2_{ }(\bmu) \, c^{ }_2
 + e^3_{ }(\bmu) \, c^{ }_3
 + e^4_{ }(\bmu) 
 \, \biggl(\, \tilde c_4^{ }
 \ln\frac{\bmu}{T^{ }_\gamma} + c^{ }_4 \,\biggr)
 + e^5_{ }(\bmu)
 \, \biggl(\, \tilde c_5^{ }
 \ln\frac{\bmu}{T^{ }_\gamma} + c^{ }_5 \,\biggr)
 + ...
 \biggr]
 \;. \la{def_p_cs}
\ee
Here $\bmu$ is the $\msbar$ scheme renormalization scale;
at this level of resolution, 
the running of~$e^2_{ }$ needs to be taken into account
(in contrast, $m^{ }_e$ 
is defined as the ``pole mass'', which does not run). At 1-loop
order, which is sufficient for our purposes, the running goes as 
\be
 \bmu \frac{{\rm d} e^2_{ }(\bmu)}{{\rm d}\bmu} 
 \; 
 = 
 \; 
 \frac{e^4_{ }(\bmu)}{6\pi^2_{ }}
 \; 
 \Rightarrow
 \; 
 e^2_{ }(\bmu) 
 \; 
 = 
 \; 
 \frac{6\pi^2_{ }}{\ln(\Lambda/\bmu)}
 \;, \la{running}
\ee
where $\Lambda$ is an integration constant. When no
argument is given, we assume $\bmu = m^{ }_e$, so that 
$
 e^2_{ } \equiv e^2_{ }(\bmu = m^{ }_e)
 \approx 4\pi/137
$.

As a physical quantity, the (temperature-dependent part of the) 
pressure must be independent
of $\bmu$. This implies that 
\be
 \bmu \frac{{\rm d} \p^{ }_\EMQED}{{\rm d}\bmu} 
 \; 
  =
 \;
 0
 \;
 \underset{\rmii{\nr{running}}}{
 \overset{\rmii{\nr{def_p_cs}}}{\Rightarrow}} 
 \; 
 \tilde c_4^{ } 
 \; 
 = 
 \; 
 -\frac{c^{ }_2}{6\pi^2_{ }}
 \;, \quad
 \tilde c_5^{ } 
 \; 
 = 
 \; 
 -\frac{c^{ }_3}{4\pi^2_{ }}
 \;.
 \la{c4p_c5p}
\ee
These relations are also valid in the massive case, and allow to extract
a part of the result in terms of 
the lower-order coefficients $c^{ }_2$ and $c^{ }_3$.\footnote{
   As discussed below \eq\nr{p^5}, 
   two coefficients remain in \eq\nr{def_p_cs}, 
   denoted by $c^{ }_4$ and $c^{ }_5$, 
   which are only known in the massless limit. Given that 
   only the massless values are known, these coefficients 
   do not go away at low $T^{ }_\gamma \ll m^{ }_e$,
   even though their mass-dependent
   generalizations would do so. However, this has no practical influence. 
 } 
The results are reported  in \app\ref{sec:qed_eos}.
In the end, the numerical effect that we find
from these corrections is clearly below the accuracy
of the momentum-averaged approach. 
The same is the case with the logarithmic corrections
of $\rmO(e^2_{ })$, 
though they grow in relative importance at low $T^{ }_\gamma$. 
As described in \se\ref{sec:data}, all corrections are included as 
tables (cf.\ \tabl\ref{tab:QED_pressure}) 
in \href{https://github.com/MiguelEA/nudec_BSM}{nudec\_BSM\_v2}, 
and for most of them we provide fast analytic representations 
(cf.\ \app\ref{qed_bessel}).

%
\section{Parametrization of the universe expansion in terms of $N^{ }_\rmi{eff}$ and $h^{ }_\rmi{eff}$}
\label{se:expansion}

The properties of the non-equilibrium neutrino ensembles, 
as determined by the solution of 
\eqs\nr{eq:dTdt_dmudt_general}--\nr{hubble}
or \nr{liouville}, 
dictate the 
expansion rate of the universe during the BBN and CMB epochs. Moreover, 
the overall amount of expansion that took place is relevant 
for the abundance of pre-existing cosmological relics, 
such as dark matter, baryon asymmetry, and primordial gravitational waves 
(cf.\ \se\ref{se:bsm}). These two characteristics, 
instantaneous expansion rate and expansion accrued, can be characterized
by the quantities that we call $N^{ }_\rmi{eff}$ and $h^{ }_\rmi{eff}$, 
respectively. 

If we consider the Hubble rate from \eq\nr{hubble}
at temperatures $T^{ }_m \ll T^{ }_\gamma \ll m^{ }_e$, where
$T^{ }_m \sim \mbox{eV}$ denotes the moment of matter-radiation
equality, the radiative QED corrections to 
$\e^{ }_\EMQED$ drop out, and we can write 
\be
  \e^{ }_\tot 
  \; \equiv \; 
  g^{ }_{*\e}\, \frac{\pi^2_{ }}{30} \, T^4_\gamma
  \quad \text{with} \quad
  g^{ }_{*\e} 
  \overset{T^{ }_m \;\ll\; T^{ }_\gamma \; \ll \; m^{ }_e}{\equiv}
  2 \, \biggl[ 1 + \frac{7}{8}\biggl( \frac{4}{11} \biggr)^{4/3}_{ }
  N^{ }_\rmi{eff} \biggr]
  \;. \la{def_Neff}
\ee
Therefore, $N^{ }_\rmi{eff}$ parametrizes the 
Hubble rate at keV scale temperatures. 
Put another way, 
\be
  N^{ }_\rmi{eff}
  \; \equiv \; 
  \frac{8}{7} 
  \biggl( \frac{11}{4} \biggr)^{4/3}_{ } 
  \biggl( \frac{\e^{ }_\tot - \e^{ }_\EMQED}{\e^{ }_\EMQED}\biggr)
  \bigg |^{ }_{T^{ }_m \;\ll\; T^{ }_\gamma \; \ll \; m^{ }_e}
  \;. \la{def_Neff_alt}
\ee 
In the Standard Model, and for the simplest case where one describes 
the neutrinos as a perfect fluid characterized by a single temperature,  
$N_{\rm eff}$ is given by
\begin{align}
  N_{\rm eff}^{ }|^{ }_\rmii{SM}
  \; \approx  \;
  3\, \left(\frac{11}{4}\right)^{4/3} 
  \left(\frac{T_{\nu}}{T_\gamma}\right)^{4}\,.
\end{align}

However, knowing the 
Hubble rate at $T^{ }_\gamma \ll m^{ }_e$
does not provide complete information about 
the previous expansion history.
We therefore also define another quantity, 
denoted by $h^{ }_\rmi{eff}$, which 
determines how the cosmological scale factor
evolves from some initial time ($a^{ }_\rmi{ini}$) till today ($a^{ }_\now$). 
Given that the influence of neutrino decoupling on this 
characteristic has received less attention than $N^{ }_\rmi{eff}$, 
let us provide some more background. 

The text-book logic for determining the evolution of the scale factor goes
via entropy conservation. {\em Assuming} thermal equilibrium, 
and combining Friedmann equations and thermodynamic identities, one finds
that $s a^3_{ } = $ constant, where $s$ denotes the entropy 
density (cf.\ the discussion below \eq\nr{1st_law}). 
Once neutrinos decouple, they are no longer in thermal equilibrium, 
and the logic needs to be rethought. However, 
one possibility is to {\em define} 
a non-equilibrium entropy such that 
it is effectively conserved, writing
\begin{equation}
  h^{ }_* \times (a T^{ }_\gamma)^3_{ }
 \; \equiv \; 
 \mbox{constant}
 \;, \quad
  s
 \; \equiv \; 
  h^{ }_*  \, 
 \frac{2 \pi^2_{ }T^3_{\gamma}}{45}
 \;. \label{conv_def}
\end{equation}
Then the evolution of the scale factor is given by 
\be
 \frac{a^{ }_\rmii{ini}}{a^{ }_\inow} 
 \; \overset{\rmii{\nr{conv_def}}}{=} \; 
 \biggl( \frac{h^{ }_{*,\inow}}
            {h^{ }_{*,\rmii{ini}}} \biggr)^{1/3}_{ }
 \frac{T^{ }_{\gamma,\inow}}{T^{ }_{\gamma,\rmii{ini}}}
 \;. \la{evo}
\ee
In terms of the variable $z$ determined via \eq\nr{e-folds},
we can compute $h^{ }_{*,\now}$ as 
\begin{equation}
 h^{ }_\rmi{eff} 
 \;
  \equiv
 \;
 h^{ }_{*,\now}
 \; 
 \underset{\rmii{\nr{evo}}}{
 \overset{\rmii{\nr{e-folds}}}{=}} 
 \; 
 \frac{ h^{ }_{*,\rmii{ini}} }{z^3_\inow}
 \;. \label{heff_res}
\end{equation}

After neutrino decoupling has completed,
the photon temperature redshifts with the universe expansion, 
so that \eq\nr{heff_res} becomes constant with 
respect to $T^{ }_{\gamma,\now}$. Therefore, 
we may evaluate $h^{ }_\rmi{eff}$ 
at the final moment of our evolution, 
$T^{ }_{\gamma,\now} \to T^{ }_{\gamma,\rmi{fin}}$.
In practice, we choose $T^{ }_{\gamma,\rmi{fin}} \approx 0.01$~MeV.
To ensure that particles like muons and pions, which we
do not include in our equation of state, 
matter very little, a suitable initial moment
is $T^{ }_{\gamma,\rmi{ini}} = 10\,$MeV. From
\eq\nr{heff}, we then obtain $h^{ }_{*,\rmi{ini}} = 10.736$.
With this initial value and a solution of \eq\nr{e-folds},
the value of $h^{ }_\rmi{eff}$ follows from \eq\nr{heff_res}.

It is appropriate to stress that the definition of a non-equilibrium
entropy coefficient via \eq\nr{heff_res} is {\em not} unique. 
Indeed there is another 
frequently used definition in the context of neutrino decoupling, 
which we recall in \app\ref{ss:params}; to make the distinction
clear, we refer to the latter with $g^{ }_{*s}$.

\addtocontents{toc}{\vspace{-0.5em}} 
\section{Main results and their comparison with the Liouville equation}
\addtocontents{toc}{\vspace{-0.5em}} 
\la{se:comparison}

As discussed above, in the momentum-averaged approach 
the neutrino decoupling process is described by simple ordinary 
differential equations for the electromagnetic temperature
(\eq\eqref{dot_T_gamma}), as well as the neutrino temperature(s) 
and effective chemical potentials (\eqs\eqref{eq:dTdt_dmudt_general}), 
all linked via the Hubble rate (\eq\nr{hubble}). 
This system of equations is 
parametrized by the neutrino interaction rates (\se\ref{ss:matching}) 
and by the energy density and pressure 
of the QED plasma (\se\ref{sec:QED_corrections}). 
With these ingredients, we obtain a fast solution  
($t^{ }_\rmii{CPU} \lesssim 1 \,{\rm s}$), 
which can be compared with the solution of 
the momentum-dependent system of $\mathcal{O}(100)$ 
coupled stiff integro-differential equations
obtained with Fortepiano~\cite{Bennett:2020zkv}. 
Here we summarize the results of the comparison, 
relegating the technical details to \app\ref{app:results}.

\begin{figure}[t]

\hspace*{-0.1cm}
\centerline{
  \includegraphics[scale=.56]{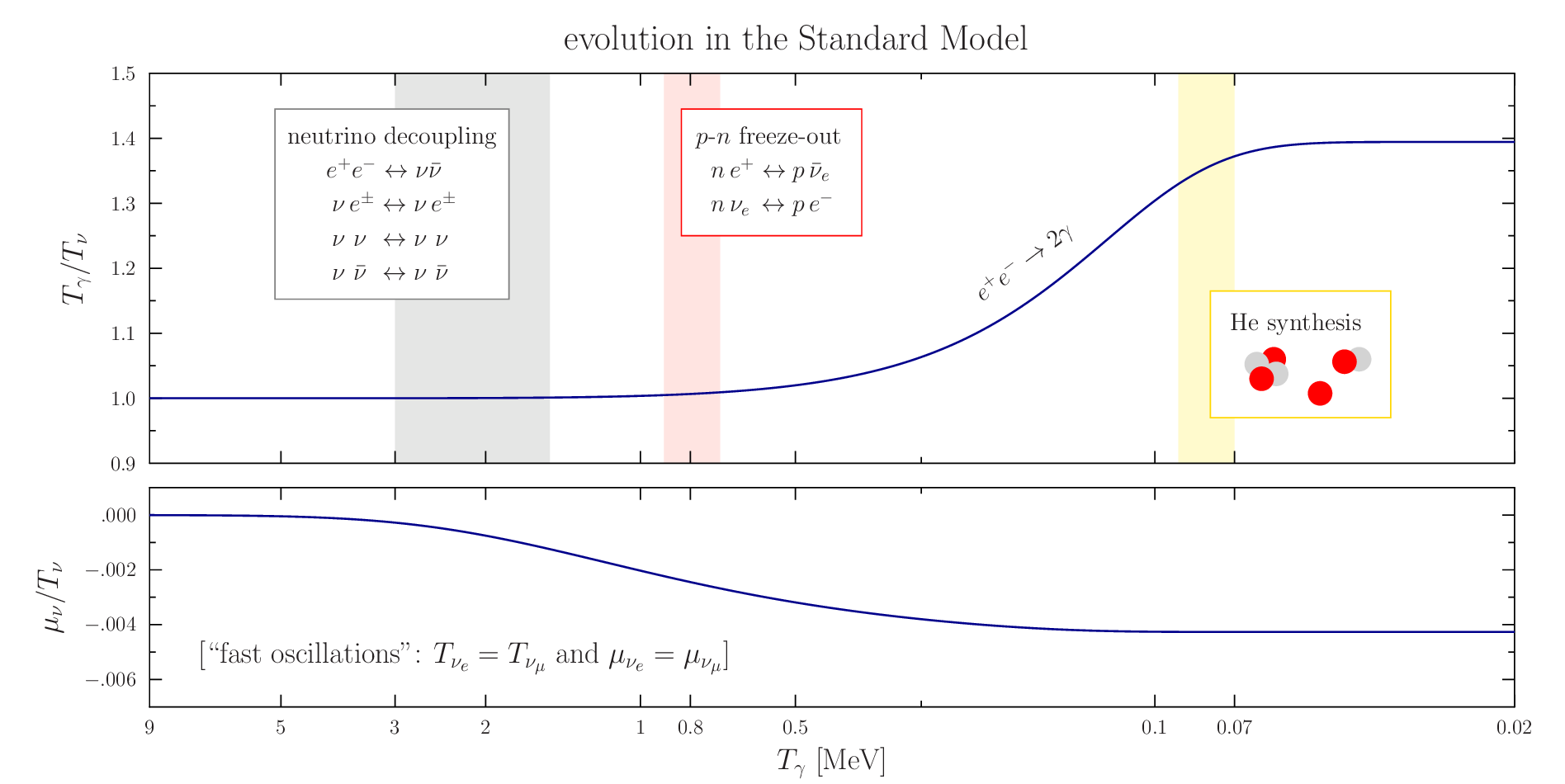}
}
\vspace{-0.5cm}
\caption[a]{\setstretch{1.3}
 Evolution of the neutrino temperature and (effective) 
  chemical potential in the Standard Model, 
  as a function of $T_\gamma^{ }$. 
  In this plot, the ``fast oscillation'' 
  case from \eq\nr{varrho} is assumed, 
  see \tabl\ref{tab:SM_summary} 
  for the corresponding cosmological parameters. 
  The increase in the ratio $T_\gamma^{ }/T_\nu^{ }$ 
  with time is chiefly due to 
  heating from $e^+_{ } e^-_{ } \to 2\gamma\,$, 
  while the 
  deviation in the parameter $\mu_\nu^{ }$ 
  from zero tracks the spectral distortions 
  in the neutrino momentum distribution. 
}
\la{evo_plot}
\end{figure}

Table~\ref{tab:SM_summary} shows the main results of our study. 
We tabulate 
the effective number of relativistic neutrinos,  
$N^{ }_\rmi{eff}$ (cf.\ \eq\nr{def_Neff_alt}); 
the corresponding total number of degrees of freedom 
contributing to the energy density, 
$g^{ }_{*\e}$ (cf.\ \eq\nr{def_Neff}); 
a definition of entropy degrees of freedom obtained from kinetic theory, 
$g^{ }_{*s}$ (cf.\ \eq\nr{s_v2}); 
a definition of entropy degrees of freedom parametrizing
the expansion of the universe, 
$h^{ }_\rmi{eff}$ (cf.\ \eq\nr{heff_res}); 
and a measure of the energy density that neutrinos carry 
in the late universe, 
$\sum m^{ }_\nu / [\Omega^{ }_\nu h^2_{ }\,\mbox{eV}]$ 
(cf.\ \eq\nr{Omega_nu}). 
In our results, the part with 
\textit{no oscillations} means that we treat $\nu_e$ and $\nu_{\mu/\tau}$ 
as separate fluids, while for 
the results \textit{with oscillations}, 
the approach of \eq\nr{varrho} assumes the 
oscillations to be so fast that 
all flavours are described by a joint averaged fluid. 

\textit{Accuracy of the approach for Standard Model characteristics. ---} 
The comparison between our results and those of Fortepiano 
indicates %
that the differences of all cosmological parameters 
are less than 0.05\% in the Standard Model, 
and can be reduced to below 0.03\% if chemical potentials are  
included in the neutrino distribution functions. 
If oscillations are omitted on both sides, 
chemical potentials increase the accuracy to 0.01\%. 
This is rather remarkable in view of the simplicity of our approach. 
Our $N^{ }_\rmi{eff}$ 
agrees very well with the value $N^{ }_\rmi{eff} \simeq 3.044\,$, 
obtained by refs.~\cite{Akita:2020szl,Froustey:2020mcq,Bennett:2020zkv}. 
We note that the numerical convergence of these references 
is $ 0.006\%$, which means we have agreement 
up to numerical errors in some cases. 
From table~\ref{tab:SM_summary}, we see that 
allowing for neutrino chemical potentials increases 
the precision most significantly for $\Omega^{ }_\nu h^2_{ }$, 
as chemical potentials affect more strongly the number density of neutrinos. 

As for the entropy density, we find a remarkable numerical agreement between 
the two definitions we have considered, $g^{ }_{*s}$ and $h^{ }_\rmi{eff}$
(cf.\ table~\ref{tab:SM_summary}). 
The final value,  $h^{ }_\rmi{eff} \approx 3.930$, 
differs noticeably from  that implied by  the latest 
PDG edition~\cite{ParticleDataGroup:2024cfk}.\footnote{%
  When citing a value for the current entropy density of 
  the universe, which is regularly used in the sense of \eq\nr{evo}, 
  ref.~\cite{ParticleDataGroup:2024cfk} 
  cites $h^{ }_{*,\inow} = 43/11 \approx 3.91$,
  which originates from essentially the same estimate
  as the naive $N^{ }_{\rm eff} = 3.0$.
} We believe that this number is worth taking into account, 
as the improvement is comparable to the current error on 
e.g.\ the dark matter 
or baryon asymmetry in the universe, 
and has an impact on the corresponding theoretical computations. 

In fig.~\ref{evo_plot} we show the thermodynamic evolution 
as a function of $T_\gamma$ for the case where 
the entire neutrino fluid is characterized 
by a common temperature and effective chemical potential. 
The evolution of $T_\gamma/T_\nu$ 
illustrates the usual behaviour: 
at high temperatures $T_\gamma\gtrsim 3\,{\rm MeV}$, 
photons and neutrinos remain in tight thermal equilibrium, 
followed by subsequent relative photon heating 
due to 
$e^+e^-\to 2\gamma$ annihilations. 
In the lower panel, we show $\mu_\nu^{ }/T_\nu^{ }$, 
highlighting when the spectral distortions 
to the neutrino distribution function develop. 

\begin{figure}[!t]
\hspace*{-0.1cm}
\centerline{
  \hspace{-.28cm}
  \includegraphics[scale=.54]{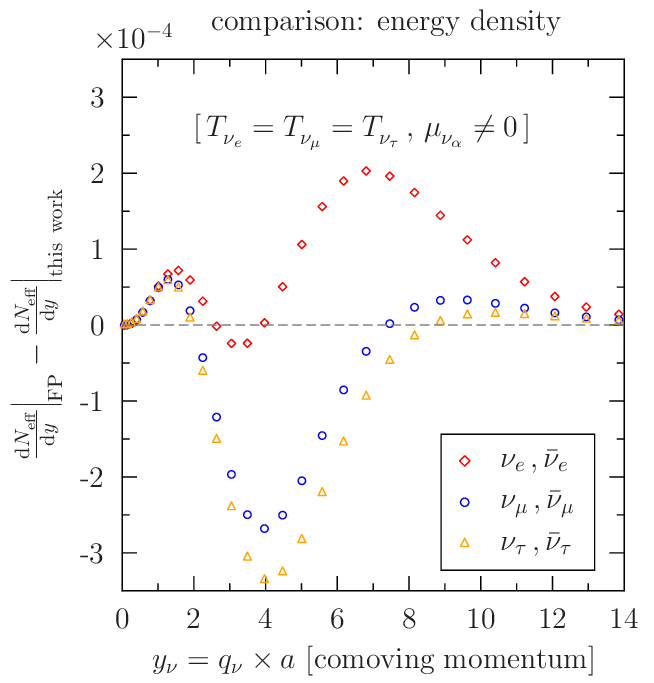}
  \hspace{.02cm}
  \includegraphics[scale=.54]{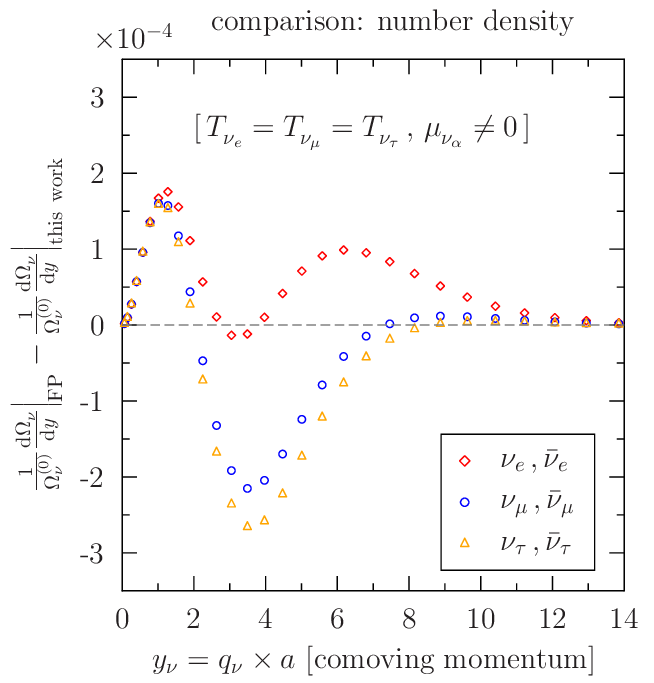}
  \hspace{.02cm}
  \includegraphics[scale=.54]{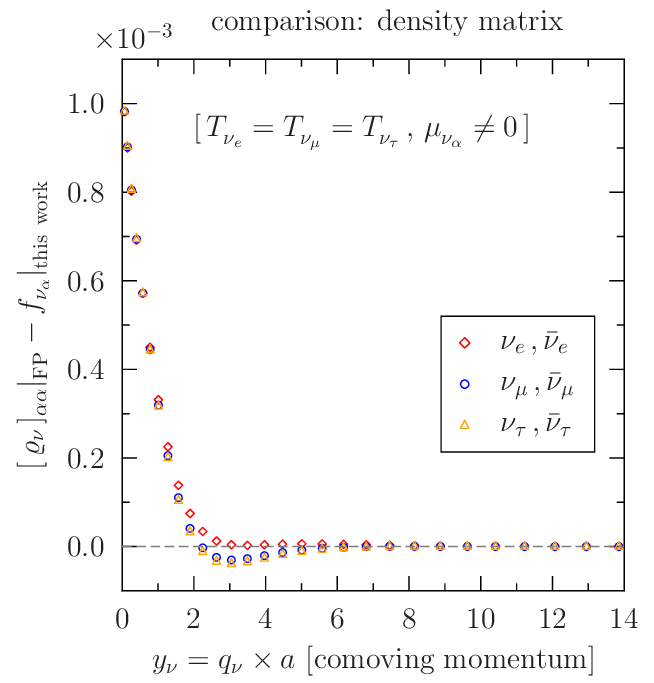}
}
\vspace{-0.5cm}
\caption[a]{\setstretch{1.3}
The difference of the (weighted) neutrino distribution function between our results 
and Fortepiano~\cite{Bennett:2020zkv}. 
The panels display the same information in different ways, 
moving from left to right: 
the differential energy density 
(${\rm d}N^{ }_\rmi{eff} \equiv \frac{120}{7\pi^4} 
(\frac{11}4)^{4/3} \frac{ y^3\, {\rm d}y}{z^4} f^{ }_{\nu_\alpha}$, 
cf.\ \eqs\nr{e_fit} and \nr{def_Neff_alt}); 
the differential number density
(${\rm d}\Omega^{ }_\nu / \Omega_\nu^{(0)} \equiv
 \frac{2}{3\zeta(3)} \frac{11}{4} \frac{ y^2 \, {\rm d}y }{z^3} f^{ }_{\nu_\alpha}\,$, 
cf.\ eqs.~\nr{Omega_nu}, \nr{n_fit}, and table~\ref{thermo_massless}); 
and the momentum distribution itself. 
On the $x$-axis, $a$ is normalized to $1/T_\nu$ at the initial temperature, i.e.\ 
$
 a \equiv a/(a T^{ }_\nu)^{ }_\rmii{ini}
$.
In the nomenclature of \tabl\ref{tab:SM_summary}, 
we show the comparison {\em with oscillations}, having included the neutrino chemical potentials. 
In the left and middle panels, the spectral distortions are $\lesssim 0.03\%$ across comoving neutrino momenta.
 }
\label{fig:comparison_FP_main}
\end{figure}

The differential neutrino distribution functions are 
shown in \fig\ref{fig:comparison_FP_main}. 
The determination of spectral distortions is 
a key goal of the Liouville approach 
(cf., e.g., \fig{3} of ref.~\cite{Akita:2020szl} 
or \fig{4} of ref.~\cite{Froustey:2020mcq}), 
however usually the comparison is against a Fermi distribution 
without a chemical potential. 
Given that neutrino temperatures and chemical potentials 
are ingredients of our momentum-averaged approach, 
it is natural to ask how large the distortions 
are relative to this generalized reference point. 
Our comparison shows that the differences are  
smaller than $\lesssim 0.03\%$. 
This is in concordance with the results of~ref.~\cite{Bond:2024ivb}, 
which highlight that a modified temperature 
and chemical potential can account for most of, 
but not all, the spectral distortions. 
In any case, from our approach we conclude that 
the spectral distortions of the Cosmic Neutrino Background 
are at the $\lesssim 0.03\%$ level. 
In the right panel of \fig\ref{fig:comparison_FP_main}, 
we also show the comparison directly with 
the distribution function, observing differences of up to $0.1\%$ 
in the infrared domain $q_\nu/T_\nu \lesssim 2.0$. 
However, this part is phase-space suppressed
in the integrated number and energy densities.

\textit{Treatment of neutrino oscillations. ---} 
In \tabl\ref{tab:SM_summary}, we include 
results from the approach described in \se\ref{ss:osc} 
that accounts for dynamical neutrino oscillations. 
We see that the agreement between that approach and Fortepiano 
is better than 0.012\% for any cosmological parameter, 
and the same holds for the neutrino distribution function. 
While this procedure is significantly more involved 
than just tracking temperatures and chemical potentials, 
the agreement is important, as it will allow us to scrutinize 
the impact of neutrino oscillations in BSM settings~\cite{Escudero:2025mvt}.

\textit{Higher-order QED effects. --- } Although not shown 
in \tabl\ref{tab:SM_summary}, we find the impact of the $\rmO(e^4_{ })$ and $\rmO(e^5_{ })$ 
corrections to the thermodynamics of the QED plasma to be quite small. In particular, 
they shift $N_{\rm eff}$ by $\lesssim 2 \times 10^{-5} $ as compared with 
the result considering corrections up to $\rmO(e^3_{ })$. 
This shift is actually somewhat smaller than the impact of 
the logarithmic $\rmO(e^2_{ })$ contribution, which shifts $N^{ }_\rmi{eff}$ 
by $\lesssim 3\times 10^{-5}$, and which is also not included by default in Fortepiano. 
Given their small numerical impact, we conclude that these corrections can be neglected.

\textit{On the accuracy of the collision terms. --- } 
We note that we have performed all calculations for \tabl\ref{tab:SM_summary} 
using the linear-response transfer rates shown in \fig\ref{fa_fs_fn_plot}. 
However, by solving the neutrino and QED plasma evolution using 
at each time step the full collision terms  (integrated over three dimensions), 
we obtain the very same results for $N_{\rm eff}$. This demonstrates the accuracy 
of our linear-response collision terms, and makes us confident that they will perform 
well also beyond the Standard Model.

%
\begin{table}[t]
\begin{center}
\begin{tabular}{l||l|l|l|l|l}
\hline\hline
\multicolumn{6}{c}{Neutrino Decoupling in the Standard Model: key parameters and observables}  \\ \hline
case/parameter      	                & \,\,\,$N^{ }_\rmi{eff}$   & $ \,\,\,\,\,g^{ }_{*\rho} $   & $\,\,\,\,g^{ }_{* s}$ & $\,\,\,\, h^{ }_\rmi{eff} $ &$\sum m_\nu/[\Omega_\nu h^2\,{\rm eV}]$   \\ \hline
{\em no oscillations:} & & & & &  \\ 
Fortepiano~\cite{Bennett:2020zkv}           & 3.0435 &  3.3824  &    3.9298  &  3.9296  &  93.129   \\
this work $\mu_\nu = 0$  & 3.0443  \, [0.028\hspace*{0.3mm}\%] \,\,&  3.3828  \,  [0.012\hspace*{0.3mm}\%]  &  3.9302  \,  [0.011\hspace*{0.3mm}\%] &  3.9301 \, [0.019\hspace*{0.3mm}\%] & 93.035 \,  [$-$0.10\hspace*{0.3mm}\%] \\
this work $\mu_\nu \neq 0$  & 3.0437  \, [0.006\hspace*{0.3mm}\%] &  3.3825 \, [0.003\hspace*{0.3mm}\%]  &  3.9299  \,  [0.002\hspace*{0.3mm}\%] & 3.9298 \, [0.010\hspace*{0.3mm}\%] &  93.127 \,  [$-$0.002\hspace*{0.3mm}\%] \\ \hline
{\em with oscillations:} $\,\,\,\,$ & & & & & \\
Fortepiano~\cite{Bennett:2020zkv}          & 3.0439  &  3.3826  &  3.9302  &  3.9298  &  93.119   \\
this work $\mu_\nu = 0$   & 3.0453 \, [0.044\hspace*{0.3mm}\%]  &  3.3832 \, [0.018\hspace*{0.3mm}\%]  &  3.9307 \,  [0.011\hspace*{0.3mm}\%]  & 3.9306 \, [0.025\hspace*{0.3mm}\%] & 93.013  \,  [$-$0.11\hspace*{0.3mm}\%] \\
this work $\mu_\nu \neq 0$& 3.0446 \, [0.020\hspace*{0.3mm}\%] &  3.3829 \,  [0.009\hspace*{0.3mm}\%] &  3.9303  \, [0.003\hspace*{0.3mm}\%] & 3.9302 \, [0.020\hspace*{0.3mm}\%]  & 93.108  \,  [$-$0.01\hspace*{0.3mm}\%]  \\ 
this work \se\ref{ss:osc} & 3.0443 \, [0.012\hspace*{0.3mm}\%] & 3.3827 \,  [0.004\hspace*{0.3mm}\%] &  3.9302  \, [$-$0.001\hspace*{0.3mm}\%] & 3.9301 \, [0.008\hspace*{0.3mm}\%] & 93.116  \,  [$-$0.003\hspace*{0.3mm}\%]  \\ 
\hline \hline
\end{tabular}
\end{center}

\vspace{-0.2cm}

\caption[a]{\setstretch{1.3}
Standard Model results as obtained in this work (\href{https://github.com/MiguelEA/nudec_BSM}{nudec\_BSM\_v2}) 
by solving for the time evolution of $T_\gamma^{ }$, $T_{\nu_e}^{ }$, $T_{\nu_\mu}^{ }$, 
effective neutrino chemical potentials, and the scale factor, $a$. We compare our results in terms 
of $N_\rmii{eff}^{ }$ (cf.\ \eq\nr{def_Neff_alt}), 
$g_{*\rho}^{ }$ (cf.\ \eq\nr{def_Neff}), 
$g_{* s}^{ }$ (cf.\ \eq\nr{s_v2}), 
$h_\rmii{eff}^{ }$ (cf.\ \eq\nr{heff_res}) and 
$\sum m_\nu^{ }/\Omega_\nu^{ } h^2_{ }$ (cf.\ \eq\nr{Omega_nu})
against the Liouville solution from Fortepiano with high accuracy settings~\cite{Bennett:2020zkv}. 
The results include the QED equation of state including $\rmO(e^3_{ })$,  
ignoring logarithmic corrections at $\rmO(e^2_{ })$, and
employing the low-energy constants from \eq\nr{g's}.  
``No oscillations'' and ``with oscillations'' refer to 
the assumptions in \eq\nr{varrho}, or to the more general 
framework of \se\ref{ss:osc}. 
Sec.~\ref{ss:osc} with $\Delta M^2 \to 0 $ in eq.~\eqref{eq:Hamiltonian} 
leads to essentially the same result as ``no oscillations'' with $\mu_\nu \neq 0$.
The neutrino masses and mixing angles are 
fixed according to \eq\nr{neutrino_params}.
}\label{tab:SM_summary}
\end{table}
%

\section{Outline of BSM extensions}
\la{se:bsm}
\addtocontents{toc}{\vspace{-0.5em}} 

The main motivation for the momentum-averaged approach is that it can 
straightforwardly incorporate other subsystems on top of the QED and neutrino ones. 
For concreteness, denoting such a species by $x$, \eqs\nr{e_i_cons} and 
\nr{n_i_cons} would then be supplemented by
\be
 \dot{\e}^{ }_x + 3 H (\e^{ }_x + \p^{ }_x) 
 \; = \; Q^{ }_x
\;, \quad
\dot{n}^{ }_x + 3 H n^{ }_x 
\; = \; J^{ }_x
\;.
 \la{e_x_cons}
\ee
The transfer rates include a contribution from the other
ensembles, e.g.\ 
$Q^{ }_x = \sum_i Q^{ }_{x\leftarrow i}$, with 
$
 i \in \{ \EMQED, \nu^{ }_e, \nu^{ }_\mu, \nu^{ }_\tau \}
$. 
Vice versa, the previous energy transfer rates, $Q^{ }_i$, 
need to be supplemented by the energy insertions from $x$, 
$Q^{ }_{i\leftarrow x}$, as dictated by detailed balance. 
In addition, the contribution from $x$ needs to 
be added to the Hubble rate, in \eq\nr{hubble}, and 
to the initial value of $h^{ }_*$, in \eq\nr{heff}.

In many BSM models, 
$Q^{ }_{x} = J^{ }_{x} = 0$
during the decoupling period,
implying that $x$ is {\em not} in equilibrium with the QED plasma. 
Then the dynamics of $x$
is easy to incorporate. 
For instance, for typical cold
dark matter candidates, with 
$\p^{ }_x \ll \e^{ }_x$, \eq\nr{e_x_cons}
amounts to 
$
 \dot{\e}^{ }_x + 3 H \e^{ }_x \approx 0
$, 
with the solution 
\be
 \e^{ }_{x,\now} 
 \; 
 \underset{\p^{ }_x \; \ll \; \e^{ }_x}{
 \overset{Q^{ }_x \; \approx \; 0}{\approx}} 
 \; 
 \e^{ }_{x,\rmi{ini}} 
 \biggl( \frac{a^{ }_\rmi{ini}}{a^{ }_\now} \biggr)^3_{ }
 \; 
 \underset{\rmii{\nr{heff_res}}}{
 \overset{\rmii{\nr{conv_def}}}{=}}  
 \; 
 \e^{ }_{x,\rmi{ini}}\,
 \frac{h^{ }_\rmii{eff}}
            {h^{ }_{*,\rmii{ini}}} 
          \biggl(  
 \frac{T^{ }_{\gamma,\inow}}{T^{ }_{\gamma,\rmii{ini}}}
          \biggr)^3_{ }
 \;. \la{dm}
\ee 
Similarly, for dark radiation, such as 
gravitational waves, with $\p^{ }_x \approx \e^{ }_x/3$, 
the initial energy density just gets diluted as
$(a^{ }_\rmi{ini}/a^{ }_\rmi{fin})^4_{ }$, and their
contribution to $N^{ }_\rmi{eff}$ can subsequently
be estimated from \eq\nr{def_Neff}.

The non-trivial case is if $Q^{ }_x \neq 0$, 
or if $x$ contributes significantly to $H$ during neutrino
decoupling. Such situations are straightforward to add to 
our framework, however the details and results are 
model-dependent. We return to well-motivated
examples in ref.~\cite{Escudero:2025mvt}. 

\section{Notes on numerical codes and data tables}
\label{sec:data}
\addtocontents{toc}{\vspace{-0.5em}} 

The \texttt{Mathematica} and \texttt{python} 
packages implementing our framework are available
as \href{https://github.com/MiguelEA/nudec_BSM}{nudec\_BSM\_v2}.
In addition to the numerical integrators, the repository
includes data tables for the functions plotted
in \fig\ref{fa_fs_fn_plot}, and an efficient evaluator
of the QED equation of state discussed in 
\se\ref{sec:QED_corrections}. 
The parameters are chosen as specified
in \app\ref{se:couplings}, and the initial
and final temperatures are by default set to 
\be
  T^{ }_{\gamma,\rmi{ini}} = 10\,\mbox{MeV}
  \;, \quad
  T^{ }_{\gamma,\rmi{fin}}
   \;\approx\; 0.01\,\mbox{MeV}
 \;. 
  \la{int_boundaries}
\ee

The main \texttt{Mathematica} and \texttt{python} module files 
are stored in the subdirectory {\tt BasicModules\_source/}. 
For the \texttt{python} implementation, 
some minimal examples of how to 
run the code are provided in 
{\tt test.py} and {\tt nudec.ipynb}. 
For the \texttt{Mathematica} implementation, 
several different Standard Model scenarios are 
provided in the notebook {\tt Neff\_SM.nb}.
When the test programs are run, they produce 
output files containing the solution. 
For reference, 
the results obtained with our default parameters can be found in the subdirectory
{\tt SM\_results/}.

The pre-computed input tables are in a subdirectory
called {\tt BasicModules\_data/}. They contain the rate coefficients
defined in \se\ref{ss:matching}, as well as the QED pressure and its first
and second temperature derivatives. For reference, we show examples of 
the rate coefficients in table~\ref{tab:MB_data}, 
and examples of the QED pressure in table~\ref{tab:QED_pressure}. We also remark
that, apart from downloading the QED pressure from tables, the code
contains an option for its fast evaluation in terms of Bessel functions, 
as explained in \app\ref{sec:Bessel}.

The \texttt{Mathematica} and \texttt{python} codes have identical modular 
structures, with the top layer solving the differential equations given 
in \eqs\nr{eq:dTdt_dmudt_general}--\nr{hubble}. It is straightforward to 
complement this system with further equations, as outlined in \se\ref{se:bsm},
or to modify the input  
that enters the Standard Model solver 
(notably the rate coefficients, the QED equation of state, or the fundamental
constants that we treat as fixed parameters). 

%
\begin{table}[t]
\begin{center}
\begin{tabular}{c||cccc|cccc|cccc}
\hline
 $m_e/T_\gamma$ & 
 ~~$f^{ }_{a1}$~~ & ~~$f^{ }_{a2}$~~ & ~~$f^{ }_{a3}$~~ & ~~$f^{ }_{a4}$~~ &   
 ~~$f^{ }_{s1}$~~ & ~~$f^{ }_{s2}$~~ & ~~$f^{ }_{s3}$~~ & ~~$f^{ }_{s4}$~~ &   
 ~~$f^{ }_{n1}$~~ & ~~$f^{ }_{n2}$~~ & ~~$f^{ }_{n3}$~~ & ~~$f^{ }_{n4}$~~ $\vphantom{\Big | }$ \\
\hline
\hline
 $0.0$ & 
 $0.884$ & $0.0$ & $0.0318$ & $0.0$ &   
 $0.829$ & $0.0$ & $0.0$ & $0.0$ &   
 $0.852$ & $0.0$ & $0.041$ & $0.0$ \\
 $\vdots$ & 
 $$ & $$ & $$ & $$ &   
 $$ & $$ & $$ & $$ &   
 $$ & $$ & $$ & $\vdots$ \\
 $1.0$ & 
 $0.830$ & $0.104$ & $0.0428$ & $-0.0165$ &   
 $0.722$ & $-0.0379$ & $0.0$ & $0.0$ &   
 $0.787$ & $0.121$ & $0.0553$ & $-0.0179$ \\
 $\vdots$ & 
 $$ & $$ & $$ & $$ &   
 $$ & $$ & $$ & $$ &   
 $$ & $$ & $$ & $\vdots$ \\
\hline
\hline
\end{tabular}
\end{center}\vspace{-0.5cm}
\caption[a]{\setstretch{1.3}
Examples of the rate coefficients shown in \fig\ref{fa_fs_fn_plot}, as they appear in the 
data file {\tt rate\_coefficients\_neutrinos.dat} 
that is available at \href{https://github.com/MiguelEA/nudec_BSM}{nudec\_BSM\_v2}.
}\label{tab:MB_data}
\end{table}
%

%
\begin{table}[t]
\begin{center}
\begin{tabular}{c||cccccc}
\hline
  $T_\gamma$~[MeV] & 
  ~~$p^{(0)}_\EMQED$~[MeV$^4$]~~ & 
  ~~$p^{(2)\msl\ln}_\EMQED$~[MeV$^4$]~~ & 
  ~~$p^{(2)\ln}_\EMQED$~[MeV$^4$]~~ & 
  ~~$p^{(3)}_\EMQED$~[MeV$^4$]~~ & 
  ~~$p^{(4)}_\EMQED$~[MeV$^4$]~~ & 
  ~~$p^{(5)}_\EMQED$~[MeV$^4$] $\vphantom{p^{(5)\big |}_\EMQED \Big | }$ \\
\hline
\hline
 $\vdots$ & 
 $$ &
 $$ & 
 $$ & 
 &
 $$ & 
 $\vdots$ \\
 $1.0$ & 
 $+\,5.83 \times 10^{-1}_{ }$ &
 $-\,1.33 \times 10^{-3}_{ }$ &
 $+\,1.77 \times 10^{-5}_{ }$ & 
 $+\,1.34 \times 10^{-4}_{ } $ &
 $-\,4.33 \times 10^{-6}_{ }$ &
 $-\,1.81 \times 10^{-7}_{ }$ \\
 $\vdots$ &  
 $$ &
 $$ & 
 $$ & 
 &
 $$ & 
 $\vdots$ \\
 $10.0$ & 
 $+\,6.03 \times 10^{+3}_{ }$ &
 $-\,1.59 \times 10^{+1}_{ }$ & 
 $+\,3.04 \times 10^{-3}_{ }$ &
 $+\,1.42 \times 10^{+0}_{ }$ & 
 $-\,1.02 \times 10^{-1}_{ }$ & 
 $+\,5.89 \times 10^{-3}_{ }$ \\
 $\vdots$ &
 $$ &
 $$ & 
 $$ & 
 &
 $$ & 
 $\vdots$ \\
\hline
\hline
\end{tabular}
\end{center}\vspace{-0.5cm}
\caption[a]{\setstretch{1.3}
Examples of contributions of various orders
to the QED pressure, as they appear in the 
data file {\tt QED\underline{~}p\underline{~}int.dat} 
that is available at \href{https://github.com/MiguelEA/nudec_BSM}{nudec\_BSM\_v2}.
We have here used $m_e = 0.51099895$~MeV and
$\alpha_{\rm em}(\bmu = m^{ }_e) = 1/137.035999084\,$. 
We supply similar tabulations for 
$\partial^{ }_{T_\gamma} p$ and $\partial^{2}_{T_\gamma} p$ 
in the files 
{\tt QED\underline{~}dp\underline{~}dT\underline{~}int.dat} 
and 
{\tt QED\underline{~}d2p\underline{~}dT2\underline{~}int.dat}, 
respectively.
}
\label{tab:QED_pressure}
\end{table}
%

\section{Conclusions}
\label{sec:conclusions}
\addtocontents{toc}{\vspace{-0.5em}} 

The main goal of this study has been to significantly improve on
the momentum-averaged approach to solve for neutrino decoupling, and thereby to 
accurately determine cosmological parameters like $N_{\rm eff}$ and $h_{\rm eff}$, see \eqs\eqref{def_Neff} and \eqref{heff_res}. 
In this context, direct outputs of our analysis are:
\begin{enumerate}
    \item Refined but still compact linear-response representations of energy and number density transfer rates that take into account all relevant neutrino-electron and neutrino-neutrino interactions (\se\ref{sec:Evolution}).  We have validated the new representations by solving for neutrino decoupling in the Standard Model both with them, and with the full leading-order rates, finding identical results. 
    
    \item An investigation of corrections to the QED energy density
    and pressure at $\rmO(e^4)$ and $\rmO(e^5)$, showing that these yield numerically small 
    contributions to $N_{\rm eff}$ and $h^{ }_\rmi{eff}$ 
    (\se\ref{sec:QED_corrections}).
    
    \item Verification that the accuracy of the momentum-averaged approach as compared with the Liouville equation
    is better than 0.04\% for any integrated cosmological parameter in the Standard Model (\se\ref{se:comparison}).
    Once we include dynamical neutrino oscillations,  
    the agreement is close to the numerical precision level.
    
    \item Determination of spectral distortions of the Cosmic Neutrino Background in the Standard
    Model, showing that compared with a distribution function containing an effective chemical
    potential, and including a phase-space factor as it appears in the number or energy
    density, they are $\lesssim 0.03\%$ for all comoving momenta (\se\ref{se:comparison}).

    \item Tabulation of key cosmological parameters for the Standard Model
    as a function of the photon temperature. We have also carefully defined and 
    computed the entropy-related parameter $h_{\rm eff} = 3.930$,
    improving upon the PDG value, which leads to $\sim 1\%$ shifts
    on primordial relic density abundances of dark matter or baryon asymmetry.  
\end{enumerate}

While we have improved upon many aspects of the momentum-averaged approach 
to solve for neutrino decoupling, it remains simple, numerically tractable, 
and computationally efficient. The corresponding 
\texttt{Mathematica} and \texttt{python} packages can be found at \gitlink $\,$ under the name \texttt{nudec\_BSM\_v2} 
(see \se\ref{sec:data} for a short description). An important aspect is that as described in 
\app\ref{sec:Bessel}, we have been able to speed up the calculations by a factor of $\sim 10$, 
making the codes suitable for fast cosmological inferences. Critically, the packages are portable 
and can easily be linked with BBN codes. Given that the framework is general, it can also be employed 
for other thermodynamic processes in the early universe (e.g.\ involving dark sectors). 

While we have focussed on the Standard Model in this paper, a key advantage of the approach, 
given its flexibility, is for studying BSM physics affecting neutrino decoupling. 
In fact, the present release
already incorporates the possibility to explore the consequences of time-varying fundamental constants, 
such as $m_e^{ }$, $s_\rmii{W}^2$, $G^{ }_\rmii{F}$, $G^{ }_\rmii{N}$, 
or $\alpha^{ }_\EMQED$.
Given the successful calibration of 
the accuracy of the momentum-averaged approach within the Standard Model, 
we will soon provide detailed investigations of well-motivated BSM extensions~\cite{Escudero:2025mvt}.


\vspace{0.5 cm}
\begin{center}
\textbf{Acknowledgments}
\end{center}

We thank Cara Giovanetti for helpful feedback about the implementation of our code in a BBN setup. 
M.E.\ and S.S.\ are grateful to Valerie Domcke and Mario Fernandez Navarro 
for their contribution to the development of neutrino oscillations 
in the momentum-averaged approach for neutrino decoupling. 
We acknowledge support from the DOE Topical Collaboration ``Nuclear Theory for New Physics,'' award No.~DE-SC0023663.  
S.S.\ is supported by the U.S. Department of Energy Office and by the Laboratory Directed Research and Development (LDRD) 
program of Los Alamos National Laboratory under project numbers 20230047DR and 20250164ER. 
Los Alamos National Laboratory is operated by Triad National Security, LLC, 
for the National Nuclear Security Administration of the U.S. Department of Energy (Contract No. 89233218CNA000001). 
G.J.\ is funded by the Agence Nationale de la Recherche (France), 
under grant ANR-22-CE31-0018 (AUTOTHERM).

\vspace{0.25cm}

\newpage

\appendix


\vspace{0.0cm}
\begin{center}
    \textbf{Appendices}
\end{center}

These appendices complement the main text by providing details aimed at
practitioners. In summary:
\begin{enumerate}

    \item[\ref{se:couplings}.]
     We specify our choices for the low-energy neutrino-electron interaction constants as well as the calculation of the relevant matrix element.

    \item[\ref{app:match}.]
    We show how the $f$ functions 
    in \eqs\nr{newMB1}--\nr{newMB3} can be matched onto the general linear-response coefficients from \eqs\nr{Q10} and \nr{Q01}.
   
    \item[\ref{app:results}.]
     We provide definitions of key cosmological parameters and present a detailed comparison between our results and those obtained by solving the Liouville equation in the Standard Model.

    \item[\ref{sec:qed_eos}.]
     We present the calculation of the equation of state of the electromagnetic plasma up to $\mathcal{O}(e^5)$. We provide more compact expressions for the contributions up to $\rmO(e^3)$ than those 
    in ref.~\cite{Bennett:2019ewm}. The $\rmO(e^4)$ and $\rmO(e^5)$ 
    contributions are new, but they turn out to be numerically negligible.

   \item[\ref{sec:Bessel}.]
   We show how to expand various integrals over the thermal distribution functions
   as series in Bessel functions, or as polylogarithms. 
   The corresponding speed-up in the code is a factor of $\sim 10$. All of these functions are defined in \texttt{Mathematica} and \texttt{python} at \href{https://github.com/MiguelEA/nudec_BSM}{nudec\_BSM\_v2}.  

\end{enumerate}

\addtocontents{toc}{\vspace{+0.5em}}  
\section{Couplings and matrix elements squared}
\la{se:couplings}
\addtocontents{toc}{\vspace{-0.5em}}  

The rate coefficients playing a role in the Standard Model 
originate from a Fermi vertex, which after a Fierz transformation 
can be expressed as  
\begin{eqnarray}
 \mathcal{L}^{ }_\iI
 &
 \supset
 & 
 - 2\sqrt{2} G^{ }_\iF\,
 \bar{\nu}^{ }_{\alpha} \gamma^{ }_\mu \aL \nu^{ }_{\alpha} 
 \nn[2mm]
 & \times & 
    \, \bar{\ell}^{ }_e  \gamma^\mu _{ }
           \Bigl\{\,
           \bigl[\,
             \delta^{ }_{\alpha,e} \, g^{ }_{e\iR}
           + 
             ( 1 - \delta^{ }_{\alpha,e} ) \, g^{ }_{\mu\iR}
           \,\bigr]\,\aR
       \; + \; 
           \bigl[\,
             \delta^{ }_{\alpha,e} \, g^{ }_{e\iL}
           + 
             ( 1 - \delta^{ }_{\alpha,e} ) \, g^{ }_{\mu\iL}
           \,\bigr]\,\aL
           \,\Bigr\}\, \ell^{ }_e 
 \;. \hspace*{5mm}
 \label{L_I_lr}
\end{eqnarray}
Here $\alpha$ is a flavour index, 
$\aL \equiv (1 - \gamma^{ }_5)/2$, 
$\aR \equiv (1 + \gamma^{ }_5)/2$, 
and $G^{ }_\iF = 1.1663787 \times 10^{-5}_{ }\, \mbox{GeV}^{-2}_{ }$
is the Fermi coupling. For the Wilson 
coefficients we employ the values~\cite{Erler:2013xha}
\be
 g^{ }_{e\iL} \;=\; 0.727 \;, \quad
 g^{ }_{e\iR} \;=\; 0.233 \;, \quad 
 g^{ }_{\mu\iL} \;=\; -0.273 \;, \quad
 g^{ }_{\mu\iR} \;=\; 0.233
 \;, \la{g's}
\ee
and for the QED parameters 
$m^{ }_e = 0.51099895\,$MeV and 
$\alpha^{ }_\rmi{em} = e^2_{ }/(4\pi) = 1/137.035999084$ as in Fortepiano. 

The vertex in \eq\nr{L_I_lr} gives rise
to pair creation, pair annihilation, and scattering processes.
The corresponding amplitudes can be 
obtained by crossing relations from each other, so it is sufficient
to choose one example, say
\begin{equation}
 \nu^{ }_e (\vec{k}^{ }_1,\s^{ }_1) 
 + 
 \bar\nu^{ }_e (\vec{k}^{ }_2,\s^{ }_2) 
 \; 
 \to 
 \; 
 \ell^{ }_e (\vec{p}^{ }_1,\t^{ }_1) 
 + 
 \bar\ell^{ }_e (\vec{p}^{ }_2,\t^{ }_2)  
 \;, 
\end{equation}
where $\s^{ }_i$, $\t^{ }_i$ are spin indices, 
and we work in the flavour basis. 
The matrix element can be written as 
\begin{equation}
 \M
 \;
  = 
 \;
 (\mbox{phase})\,
 2\sqrt{2}\, G^{ }_\iF \, 
 \bar v (\vec{k}^{ }_2,\s^{ }_2)
 \gamma^{ }_\mu \aL 
      u (\vec{k}^{ }_1,\s^{ }_1)
 \bar u (\vec{p}^{ }_1,\t^{ }_1)
 \gamma^\mu_{ }
 ( g^{ }_{e\iR}\aR + g^{ }_{e\iL}\aL )
      v (\vec{p}^{ }_2,\t^{ }_2)
 \;, 
\end{equation}
where $u$ and $v$ are the usual Dirac spinors. 
The matrix element squared reads 
\begin{eqnarray}
 |\M|^2_{ } & = & 
 8 G_\iF^2 \, 
 \bar v (\vec{k}^{ }_2,\s^{ }_2)
 \gamma^{ }_\mu \aL 
      u (\vec{k}^{ }_1,\s^{ }_1)
 \,
 \bar u (\vec{k}^{ }_1,\s^{ }_1)
 \gamma^{ }_\nu \aL
      v (\vec{k}^{ }_2,\s^{ }_2)
 \nn[2mm]
 & \times &
 \bar u (\vec{p}^{ }_1,\t^{ }_1)
 \gamma^\mu_{ }
 ( g^{ }_{e\iR}\aR + g^{ }_{e\iL}\aL )
      v (\vec{p}^{ }_2,\t^{ }_2)
 \,
 \bar v (\vec{p}^{ }_2,\t^{ }_2)
 \gamma^\nu_{ }
 ( g^{ }_{e\iR}\aR + g^{ }_{e\iL}\aL )
      u (\vec{p}^{ }_1,\t^{ }_1)
 \;. \hspace*{7mm} \label{MM}
\end{eqnarray}
We can sum over spins, and insert completeness relations. 
In addition, we can symmetrize the result in 
$\vec{p}^{ }_1 \leftrightarrow \vec{p}^{ }_2$, 
given that we assume the QED plasma to be charge-symmetric
(in the case of scatterings, this means that neutrinos can 
scatter both off electrons and off positrons). 
All in all this yields
\begin{eqnarray}
 \sum_{\s^{ }_i,\t^{ }_i} |\M|^2_{ }
 \; & 
 \overset{\rmii{symmetrize}}{
 \underset{\vec{p}^{ }_1\,\leftrightarrow\,\vec{p}^{ }_2}{\to}} 
 & \;
 16 G^2_\iF 
 \Bigl\{\,
 ( g^2_{e\iR} + g^2_{e\iL} )
 \bigl[\,
   (t - m_e^2)^2_{ } + (u - m_e^2)^2_{ } 
 \,\bigr]
 + 4 g^{ }_{e\iL}g^{ }_{e\iR} m_e^2 s
 \,\Bigr\}
 \;, \hspace*{6mm} \label{res_MM}
\end{eqnarray}
where $s$, $t$, $u$ are the usual Mandelstam variables. 
All other matrix elements squared can be obtained from 
\eq\nr{res_MM} by crossings.

\section{Parametrization of the rate coefficients}
\label{app:match}
\addtocontents{toc}{\vspace{-0.5em}} 

We specify here how the functions defined
in \eqs\nr{newMB1}--\nr{newMB3} can be matched onto 
the coefficients from \eqs\nr{Q10} and \nr{Q01}, so that 
the full electron mass dependence is retained in the 
linear response regime. 
All 12 functions can be related to the derivatives in 
\eqs\nr{Q10}--\nr{Q01}, by expanding the expressions 
for small $\epsilon^{ }_{\iT}$ and $\epsilon^{ }_\mu\,$. 
To be concrete,
\ba
  f^{ }_{a1}
  &=&
  \frac{-1}{1152} \, 
  \widehat{Q}_1^{(1,0)} \Big|_{{\rm gain-loss}}
  \, , 
  \hspace{2.2cm}
  f^{ }_{a2}
  \; = \; 
  \frac{-1}{1152} \, 
  \widehat{Q}_2^{(1,0)} \Big|_{{\rm gain-loss}}
  \, , \la{fa1_fa2}
  \\[2mm]
  f^{ }_{a3}
  &=&
  f^{ }_{a1} + \frac{1}{256} \, 
  \widehat{Q}_1^{(0,1)} \Big|_{{\rm gain-loss}}
  \, , 
  \hspace{1.5cm}
  f^{ }_{a4}
  \; = \; 
  f^{ }_{a2} +
  \frac{1}{256} \, 
  \widehat{Q}_2^{(0,1)} \Big|_{{\rm gain-loss}}
  \, , \la{fa3_fa4}
  \\[2mm]
  f^{ }_{s1}
  &=&
  \frac{-1}{224} \, 
  \widehat{Q}_1^{(1,0)} \Big|_{{\rm scattering}}
  \, , 
  \hspace{2.4cm}
  f^{ }_{s2}
  \; = \; 
  \frac{-1}{224} \, 
  \widehat{Q}_2^{(1,0)} \Big|_{{\rm scattering}}
  \, , \la{fs1_fs2}
  \\[2mm]
  f^{ }_{s3}
  &=&
  \frac{1}{224} \, 
  \widehat{Q}_1^{(0,1)} \Big|_{{\rm scattering}}
  \, , 
  \hspace{2.4cm}
  f^{ }_{s4}
  \; = \; 
  \frac{1}{224} \, 
  \widehat{Q}_2^{(0,1)} \Big|_{{\rm scattering}}
  \, , \la{fs3_fs4}
  \\[2mm]
  f^{ }_{n1}
  &=&
  \frac{-1}{256} \, 
  \widehat{J}_1^{\hspace*{0.3mm}(1,0)} \Big|_{{\rm gain-loss}}
  \, , 
  \hspace{2.5cm}
  f^{ }_{n2}
  \; = \; 
  \frac{-1}{256} \, 
  \widehat{J}_2^{\hspace*{0.3mm}(1,0)} \Big|_{{\rm gain-loss}}
  \, , \la{fn1_fn2}
  \\[2mm]
  f^{ }_{n3}
  &=&
  f^{ }_{n1} + \frac{1}{64} \, 
  \widehat{J}_1^{\hspace*{0.3mm}(0,1)} \Big|_{{\rm gain-loss}}
  \, , 
  \hspace{1.8cm}
  f^{ }_{n4}
  \; = \; 
  f^{ }_{n2} +
  \frac{1}{64} \, 
  \widehat{J}_2^{\hspace*{0.3mm}(0,1)} \Big|_{\rmi{gain-loss}}
  \, . \la{fn3_fn4}
\ea
The relevant partial derivatives can be evaluated numerically, 
making use of the spectral functions from 
\app{B} of ref.~\cite{Jackson:2024gtr} and 
implementing neutrino chemical potentials 
according to \app{A.3} of ref.~\cite{EscuderoAbenza:2020cmq}. 

Several of these functions have known limits for $m_e \to 0\,$. 
First, if electrons are taken to be massless, then 
$f^{ }_{x2}$ and $f^{ }_{x4}$ (for $x=a,s,n$) all vanish. 
And second, in this same limit, 
$f^{ }_{a1}(0) = f_a^\rmii{FD}\,$, 
$f^{ }_{s1}(0) = f_s^\rmii{FD}\,$, and 
$f^{ }_{n1}(0) = f_n^\rmii{FD}$, 
where the functions $f_x^\rmii{FD}$ were implemented in
ref.~\cite{EscuderoAbenza:2020cmq} 
to account for Fermi-Dirac statistics.

%
\section{Comparison in the Standard Model: momentum-averaged vs.\ Liouville solution}
\la{app:results}
\addtocontents{toc}{\vspace{-0.5em}} 

In this appendix we compare the results for neutrino decoupling between the momentum-averaged
equations from \se\ref{ss:momave}, and the Liouville equation as 
calculated in Fortepiano~\cite{Bennett:2020zkv}. 
In order to output a fair comparison we consider the same set of 
low-energy constants describing neutrino-electron interactions, 
specified in \eq\nr{g's}, and keep terms up to $\rmO(e^3_{ })$ 
in the QED energy density and pressure
(ignoring a logarithmic term of $\rmO(e^2_{ })$). 
Before proceeding to the comparison, 
we need to state how the relevant parameters are defined 
and how they can be extracted. 

%
\hiddenappsubsection[ss:params]{On different definitions of entropy density}

While we have given one definition of entropy density, 
parametrized by the coefficient $h^{ }_*$, 
in \eq\nr{conv_def}, let us start by elaborating on 
another definition. 
For a non-interacting fermionic fluid 
with two degrees of freedom, which is not in thermal equilibrium, an entropy density can be defined 
as (cf.,\ e.g.,\ \cite[eq.~(3.55)]{Bernstein:1988bw}  
or~\cite[eq.~(3.10)]{Sigl:1993ctk}) 
\begin{align}
 s^{ }_\nu \; \equiv \; -2\int_\vec{q} 
 \bigl[\, f^{ }_\nu \ln f^{ }_\nu 
 +(1-f^{ }_\nu)\ln(1-f^{ }_\nu) \,\bigr] 
 \;. \la{s_nu_alt}
\end{align}
Making use of the pressure from \eq\nr{e_p_n}, 
and inserting the ansatz from \eq\nr{eq:f_nu}, we get
\ba
 s^{ }_\nu & = & \frac{\p^{ }_\nu}{T^{ }_\nu}
 -2 \int_\vec{q} f^{ }_\nu \ln \frac{f^{ }_\nu}{1-f^{ }_\nu}
 \; \underset{\rmii{\nr{e_p_n}}}
 {\overset{\rmii{\nr{eq:f_nu}}}{=}}
 \frac{\e^{ }_\nu + p^{ }_\nu - \mu^{ }_\nu n^{ }_\nu}{T^{ }_\nu}
 \;. \la{s_nu_eq}
\ea
Summing over the photons and neutrino flavours, we define
the coefficient $g^{ }_{*s}$ via
\ba
 \sum_i s^{ }_i 
\; \equiv \; 
 \sum_i
 \frac{\e^{ }_i 
 + p^{ }_i
 - \mu^{ }_i 
   n^{ }_i }{T^{ }_i} 
 \; 
 \underset{\mu^{ }_\gamma \; = \; 0}{
 \overset{T^{ }_m\; \ll \; T^{ }_\gamma \; \ll \; m^{ }_e}{
 \equiv}} 
 \; 
 g_{* s}  \frac{2\pi^2}{45} T_\gamma^3
 \;, \la{g*s_def}
\ea
where $T^{ }_m \sim \mbox{eV}$ 
is the moment of matter-radiation equality.
Looking at changes of \eq\nr{s_nu_alt}, we get
\ba
 {\rm d}s^{ }_\nu 
 & \overset{\rmii{\nr{s_nu_alt}}}{=} & 
  -2 \int_\vec{q}
  \bigl[\, {\rm d} f^{ }_\nu \ln f^{ }_\nu + \cancel{{\rm d}f^{ }_\nu} 
 -{\rm d} f^{ }_\nu \ln(1-f^{ }_\nu) - \bcancel{{\rm d}f^{ }_\nu} \,\bigr] 
 \; = \; 
 -2 \int_\vec{q} {\rm d} f^{ }_\nu \ln \frac{f^{ }_\nu}{1-f^{ }_\nu}
 \; \underset{\rmii{\nr{e_p_n}}}
 {\overset{\rmii{\nr{eq:f_nu}}}{=}}
 \frac{{\rm d}\e^{ }_\nu - \mu^{ }_\nu {\rm d} n^{ }_\nu}{T^{ }_\nu}
 \;, \la{ds_nu_alt}
\ea
which has the form of the first law of thermodynamics, 
\be
 {\rm d}\e^{ }_\nu = T^{ }_\nu {\rm d}s^{ }_\nu + \mu^{ }_\nu {\rm d}n^{ }_\nu
 \;. \la{1st_law}
\ee
Inserting $\e^{ }_i + \p^{ }_i = T^{ }_i s^{ }_i + \mu^{ }_i n^{ }_i$ 
(cf.\ \eq\nr{g*s_def}) and 
\eq\nr{1st_law} into overall energy conservation, \eq\nr{e_i_cons}, 
we find
\be
 \sum_i \bigl[\,
  T^{ }_i \partial^{ }_t (s^{ }_i a^3_{ })
 + 
  \mu^{ }_i \partial^{ }_t (n^{ }_i a^3_{ })
 \,]
 \; = \; 0
 \;.
\ee
If we had a unique temperature ($T^{ }_i = T \;\forall i$), 
unique chemical potentials ($\mu^{ }_i = \mu \;\forall i$), 
and conserved particle numbers ($\partial^{ }_t (\sum_i n^{ }_i a^3) = 0$), 
the overall entropy defined in accordance with \eq\nr{g*s_def}, 
$\sum_i s^{ }_i a^3_{ }$, would be conserved. As none of these 
conditions is met, this is not guaranteed to be the case. Therefore, 
$g^{ }_{*s}$ could differ from $h^{ }_*$, with the latter having
been defined so as to 
make the entropy-like quantity in \eq\nr{conv_def} a constant of motion.

%
\hiddenappsubsection[ss:extract]{Extraction of cosmological parameters from a neutrino density matrix}

When solving the Liouville equation (cf.\ \eq\nr{liouville}), whether without approximation
on a momentum grid, or via a momentum-averaged ansatz as explained in \se\ref{ss:osc}, 
the output quantity 
is the neutrino density matrix, $\varrho^{ }_\nu$. The neutrino phase space distributions
can be obtained by  reading out the diagonal components
(cf.\ \eq\nr{varrho}), 
\be
 f^{ }_{\nu^{ }_\alpha}(t,q)
 \; \equiv \; 
 [\, \varrho^{ }_\nu(t,\vec{q}) \,]^{ }_{\alpha\alpha}
 \;, 
 \quad
  \alpha \; \in \; \{e,\mu,\tau \}
 \;. \la{f_nu_noneq}
\ee
Given the phase space distributions, energy density, number density, and 
pressure can be determined according to \eq\nr{e_p_n}. From 
the energy density, 
we then obtain $g^{ }_{*\e}$ and $N^{ }_\rmi{eff}$ 
in accordance with \eqs\nr{def_Neff} and \nr{def_Neff_alt}, respectively.
The entropy-related coefficient $h^{ }_\rmi{eff}$ can be obtained
from \eq\nr{heff_res}. 

The entropy defined according to \eq\nr{s_nu_alt} can 
similarly obtained from $f^{ }_{\nu^{ }_\alpha}$. 
To be quite concrete, we remark that 
Fortepiano employs the notation
\begin{align}
    x \;\equiv\; a m_e \,, \qquad y &\;\equiv\; a q \,, \qquad z \;\equiv\; a T_\gamma \,,
   \la{rescalings2}
\end{align}
where the units of the scale factor have
been chosen so that $a=1/T^{ }_\nu = 1/T^{ }_\gamma$ at the initial time, i.e.\ 
$
 a \equiv a/(a T^{ }_\gamma)^{ }_\rmi{ini}
$
(we show this explicitly in
\eqs\nr{e-folds} and \nr{rescalings}).
The time variable is effectively $x$, while $y$ represents 
the comoving momentum. 
With this in mind, we can write $g_{* s}$ as
\begin{align}
 g_{* s} & \underset{\rmii{\nr{g*s_def}}}{\overset{\rmii{\nr{s_nu_alt}}}{=}} 
 2 - \frac{45}{2\pi^2} \frac{2}{2\pi^2} \frac{1}{z^3} \sum_\alpha \int_0^{\infty} \! {\rm d}y\,y^2_{ }\, 
 \bigl[\, f_{\nu^{ }_\alpha} \ln f_{\nu^{ }_\alpha} 
+(1-f_{\nu^{ }_\alpha})\ln(1-f_{\nu^{ }_\alpha}) \,\bigr] 
 \qquad [\text{non-thermal distributions}]\,,
 \\
 g_{* s} & \underset{\rmii{\nr{g*s_def}}}{\overset{\rmii{\nr{s_nu_eq}}}{=}}
 2 +  \frac{45}{2\pi^2} \sum_{\alpha} \frac{\frac{4}{3}\rho_{\nu_\alpha}  - \mu_{\nu_\alpha} n_{\nu_\alpha} }{T_\gamma^3 T_{\nu_\alpha}}\qquad [\text{thermal distributions}]\,.
\la{s_v2}
\end{align}
In \eq\nr{s_v2}, a factor 2 already appears in the definitions of 
the thermodynamic functions according to \eq\nr{e_p_n}. 

In the late universe, neutrinos become non-relativistic. 
Then they contribute to the energy density as 
\begin{align}
\Omega_\nu h^2 
\;\equiv\;
\frac{\sum_\alpha m_{\nu^{ }_\alpha}  
 (n_{\nu^{ }_\alpha}+n_{\bar \nu^{ }_\alpha})}{\rho_c/h^2} \bigg|^{ }_{T^{ }_\gamma = T^{ }_\rmii{CMB}}
\overset{\rmii{fast~oscillations}}{\simeq}  \quad 
\frac{\sum_\nu m_{\nu} }{3} \frac{\sum_\alpha  (n_{\nu_\alpha}+n_{\bar \nu_\alpha})}{\rho_c/h^2}
\bigg|^{ }_{T^{ }_\gamma = T^{ }_\rmii{CMB}}
\;, \la{Omega_nu_pre}
\end{align}
where we have assumed that oscillations have equilibrated the number densities. If this 
assumption is not valid, we should view \eq\nr{Omega_nu_pre} 
as a definition of what we mean by $\sum_\nu m^{ }_\nu$.
The critical energy density is defined as $H_0^2 \equiv 8\pi \rho_c /(3\mpl^2) $, 
and  hence $\rho_c/h^2 \equiv  3 \mpl^2/(8\pi) \times (100 {\rm km/s/Mpc})^2  
 = 8.0959\times 10^{-11} \,{\rm eV}^4$. 
After neutrino decoupling, 
$n^{ }_{\nu^{ }_\alpha}$ scales as $a^{-3}_{ }$, 
which in turn scales as $T_\gamma^3$. 
It is therefore convenient to rewrite \eq\nr{Omega_nu_pre} as
\begin{align}
 \Omega_\nu^{ } h^2  
 \quad
 \overset{T^{ }_\rmiii{CMB} \; \le \; T^{ }_\gamma \; \le \; 0.01\,\text{MeV} }{\simeq} 
 \quad
 \frac{\sum_\nu m_{\nu} }{3} \frac{\sum_\alpha  (n_{\nu^{ }_\alpha}+n_{\bar \nu^{ }_\alpha})}{T_\gamma^3} 
 \frac{T_\rmii{CMB}^3}{\rho_c/h^2}
 \,.
 \la{Omega_nu}
\end{align}
For the CMB temperature today 
we take $T_\rmii{CMB} = 2.7255\,{\rm K } = 2.34865 \times 10^{-4} \mathrm{eV}$~\cite{2009ApJ...707..916F}.

Given a Fortepiano output for \eq\nr{f_nu_noneq},
effective temperatures and chemical potentials are extracted through 
a matching between the integrated energy and number densities.
In particular, at a given photon temperature, parametrized through $z$
(cf.\ \eq\nr{rescalings2}), we equate
\begin{align}
    \frac{\e_\nu }{\e_\gamma} &\;=\; \frac{1}{z^4}\frac{1}{2\pi^2/{30}} \frac{2}{2\pi^2} \int_0^{\infty} \! {\rm d}y\, y^3_{ } f_\nu(y)
    \hspace*{-3cm} & 
    & \;\overset{\rmii{\nr{e_plus_massless}}}{=}\; 
    -2\,\frac{3 T_\nu^4 \text{Li}_4\left(-e^{\mu_\nu/T_\nu}\right)}{\pi ^2}\bigg/\frac{2\pi^2}{30} T_{\gamma}^4
    \,,
    \la{e_fit}\\[2mm]
    \frac{n_\nu }{T_\gamma^3} &\;=\; \frac{1}{z^3} \frac{2}{2\pi^2} \int_0^{\infty} \! {\rm d}y\, y^2_{ }\, f_\nu(y) 
    \hspace*{-3cm} &
    & \;\overset{\rmii{\nr{n_plus_massless}}}{=}\; 
    -2\,\frac{T_\nu^3 \text{Li}_3\left(-e^{\mu_\nu/T_\nu}\right)}{\pi ^2}\bigg/ T_{\gamma}^3\,. \la{n_fit}
\end{align}
The last two cases correspond to a Fermi-Dirac neutrino distribution function, 
and allow us to extract numerically the dimensionless ratios
$T^{ }_\nu/T^{ }_\gamma$ and $\mu^{ }_\nu/T^{ }_\nu$.

%
\hiddenappsubsection[ss:compa]{Comparison of numerical results between Fortepiano and this work}

In \tabl\ref{tab:SM_summary} on p.~\pageref{tab:SM_summary},
we have highlighted the comparison 
between our calculations and the results from Fortepiano~\cite{Bennett:2020zkv}
for integrated quantities. 
In Fortepiano we have used high-accuracy settings, with the solver precision $10^{-7}$ and 
the Gauss-Laguerre method with $N_y = 35$ and $y_{\rm max} = 20$ for the collision terms.
The observables were extracted as described in \app\ref{ss:extract}. 

From \tabl\ref{tab:SM_summary} we see that describing neutrinos with just their temperature 
leads to a result for $N^{ }_\rmi{eff}$ that differs by 
$0.04\%$ from Fortepiano. This difference is reduced by a factor of two 
when neutrinos are allowed to have effective chemical potentials. Similar trends are seen
in the other quantities.
The influence of a neutrino chemical potential is most clear in the energy density parameter of neutrinos today
($\sum m^{ }_\nu/[\Omega^{ }_\nu h^2\,\mbox{eV}]$), where the precision increases by a factor of $\sim 10$.

Table~\ref{tab:SM_summary_extended} 
on p.~\pageref{tab:SM_summary_extended}
shows a similar comparison, but including more information. 
In particular, the neutrino temperatures 
and effective chemical potentials are indicated. 
The chemical potentials turn out to be quite small, $\sim 5\times 10^{-3} T^{ }_\nu$, 
but nevertheless they increase the precision significantly. 
For the Fortepiano's rows and for our approach
based on \eq\nr{factor}, the temperature and chemical potentials 
are obtained from a matching between the energy density and number density 
to a Fermi-Dirac distribution, 
as described by \eqs\nr{e_fit} and \nr{n_fit}. 
Our neutrino temperatures are closer to
each other than those of Fortepiano, an effect
that we associate with the universal momentum
dependence assumed by \eq\nr{factor}, however this
has little practical influence on physical observables.
Finally, the last two columns highlight the increase of the neutrino temperature compared to a purely decoupled neutrino, 
which would have $a T^{ }_\nu|_{T^{ }_\gamma \ll m^{ }_e}/(a T^{ }_{\nu})|_{T^{ }_\gamma \gg m^{ }_e} = 1$. 
This factor enters the relationship between physical and comoving momenta,
$q/T^{ }_\nu = y / (a T^{ }_\nu)$, cf.\ \eq\nr{rescalings2}. 

\begin{figure}[!t]
\centering
\begin{tabular}{ccc}
\hspace{-0.cm}\includegraphics[width=0.35\textwidth]{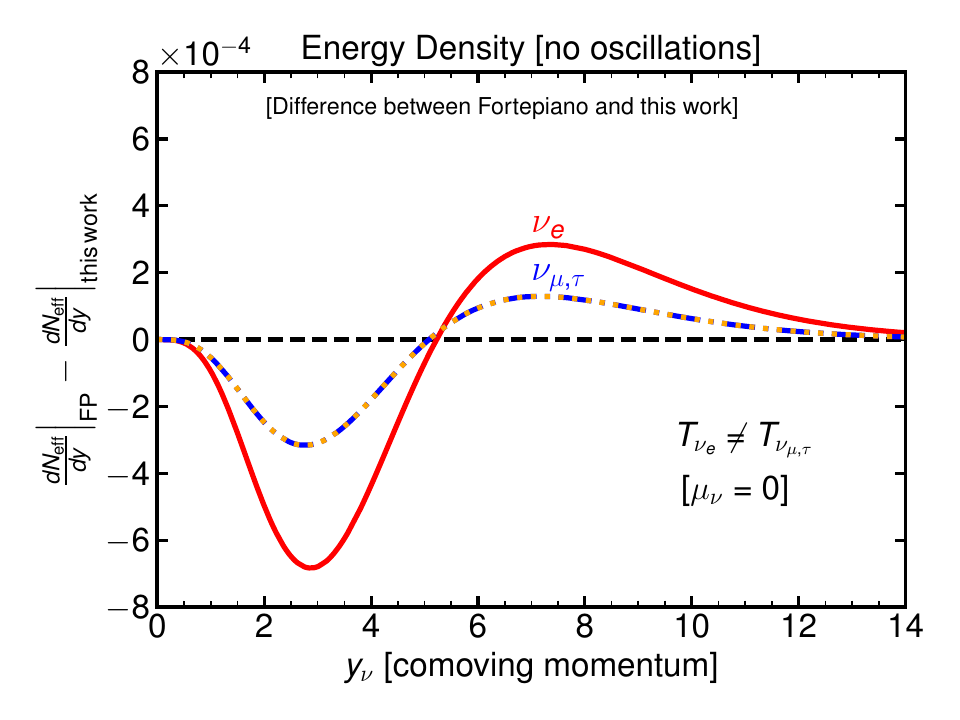} & \includegraphics[width=0.35\textwidth]{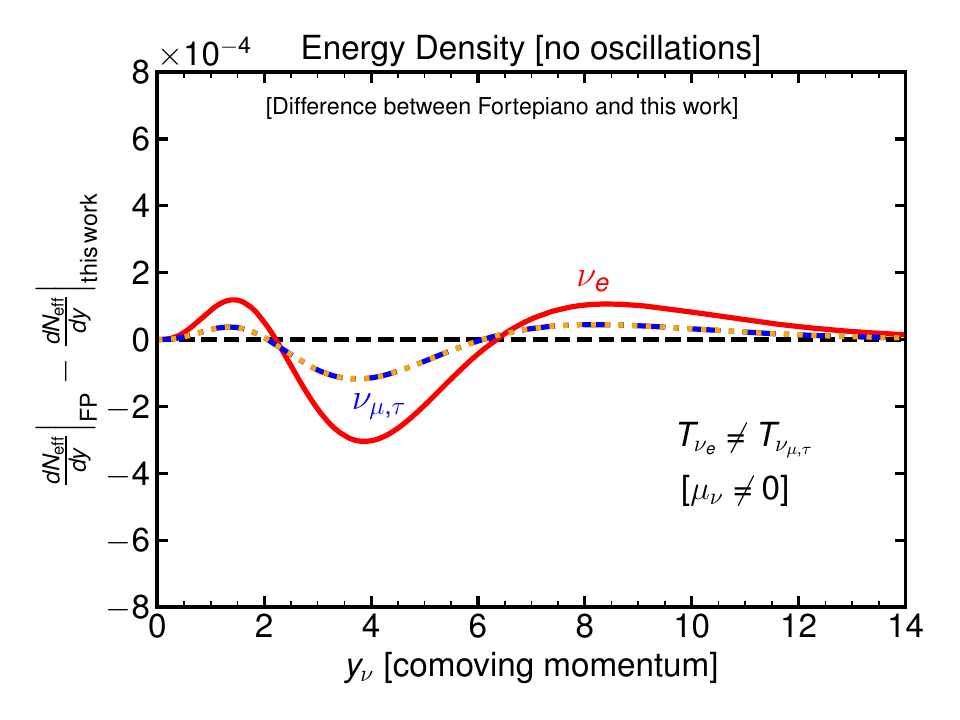} \\
\hspace{-0.cm}\includegraphics[width=0.35\textwidth]{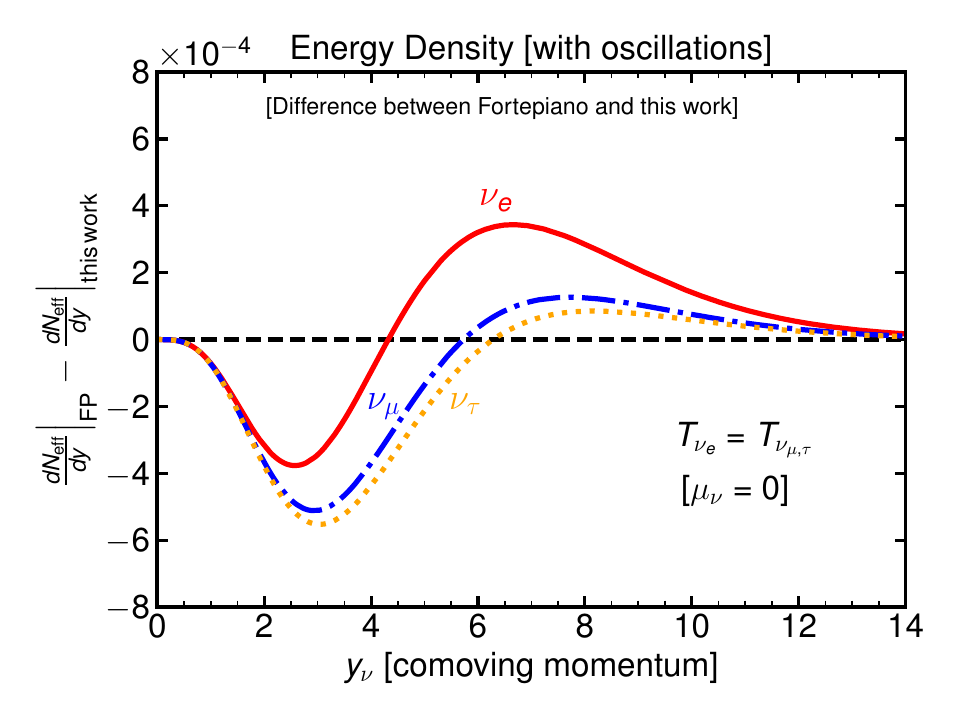}  & \hspace{-0.cm}\includegraphics[width=0.35\textwidth]{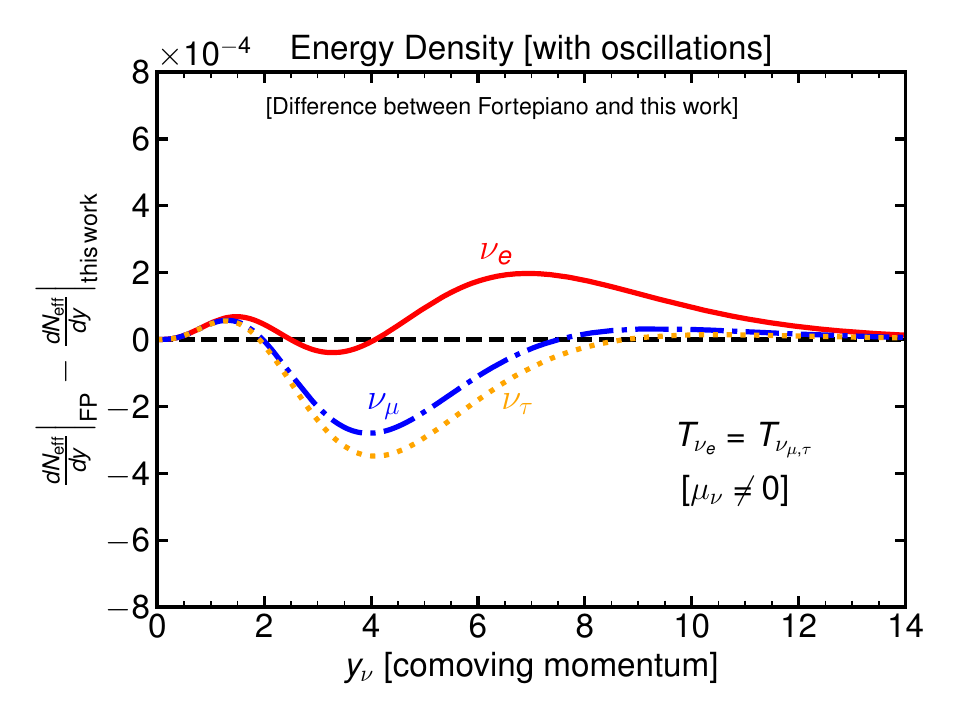}
\\
\hspace{-0.cm}\includegraphics[width=0.35\textwidth]{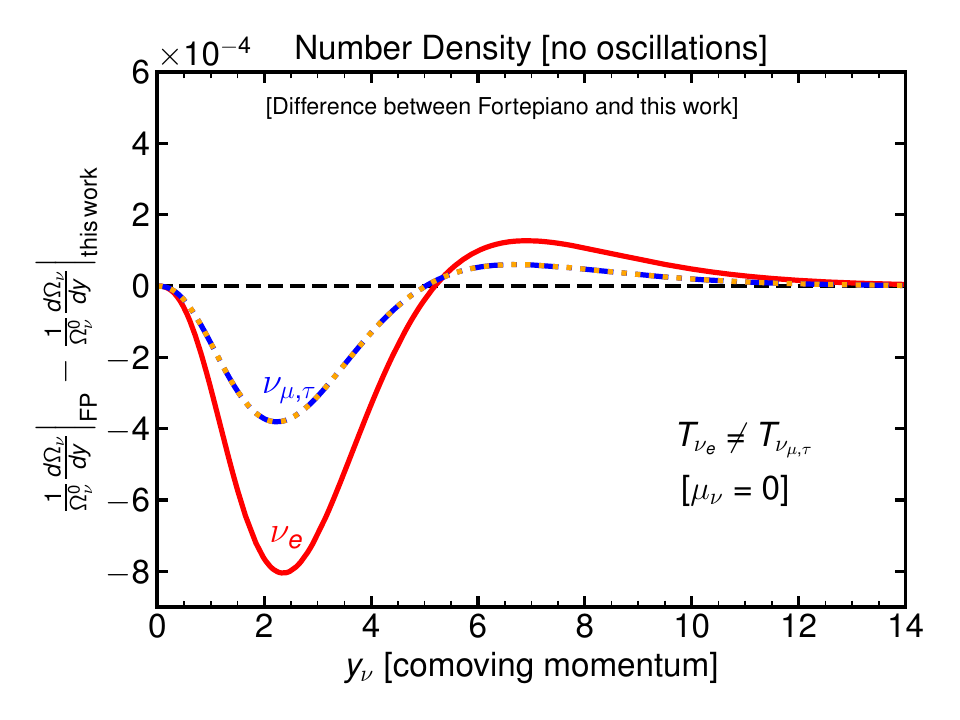} & \includegraphics[width=0.35\textwidth]{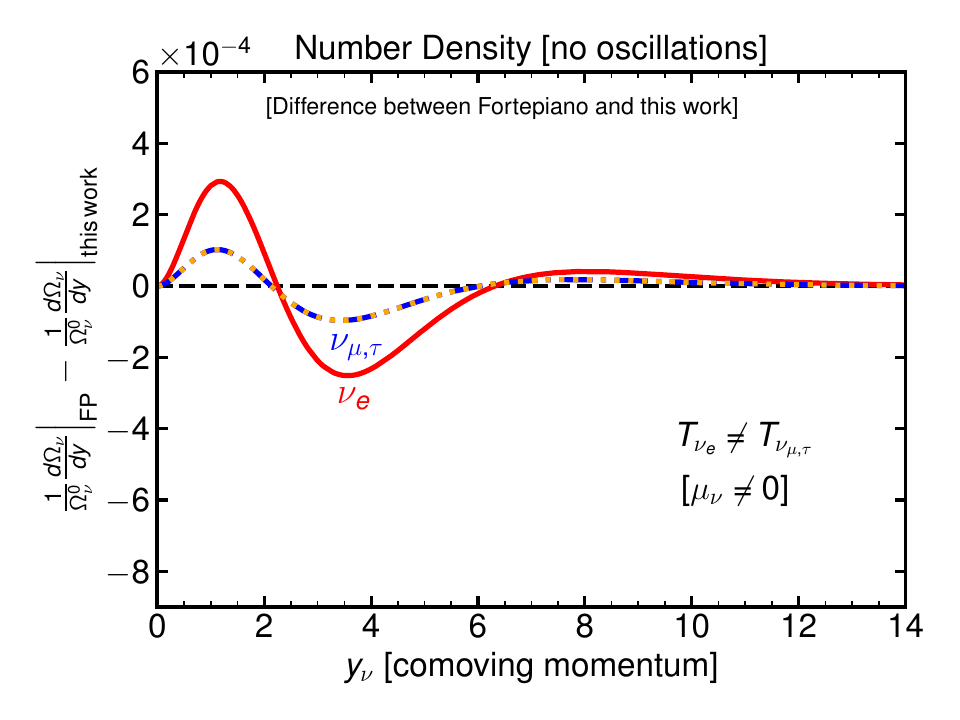} \\
\hspace{-0.cm}\includegraphics[width=0.35\textwidth]{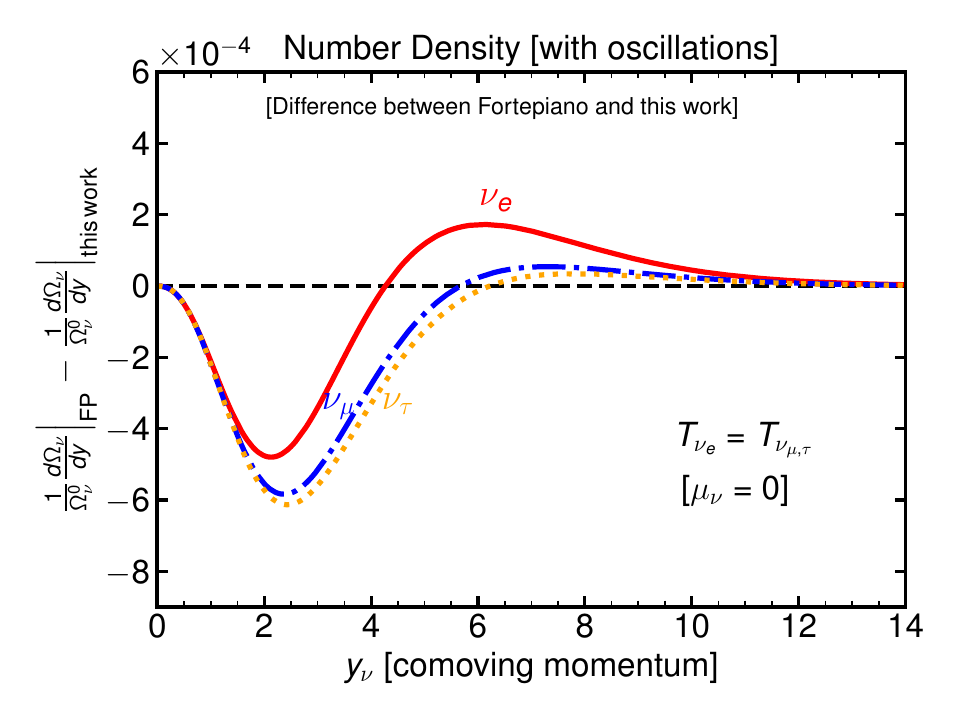}  & \hspace{-0.cm}\includegraphics[width=0.35\textwidth]{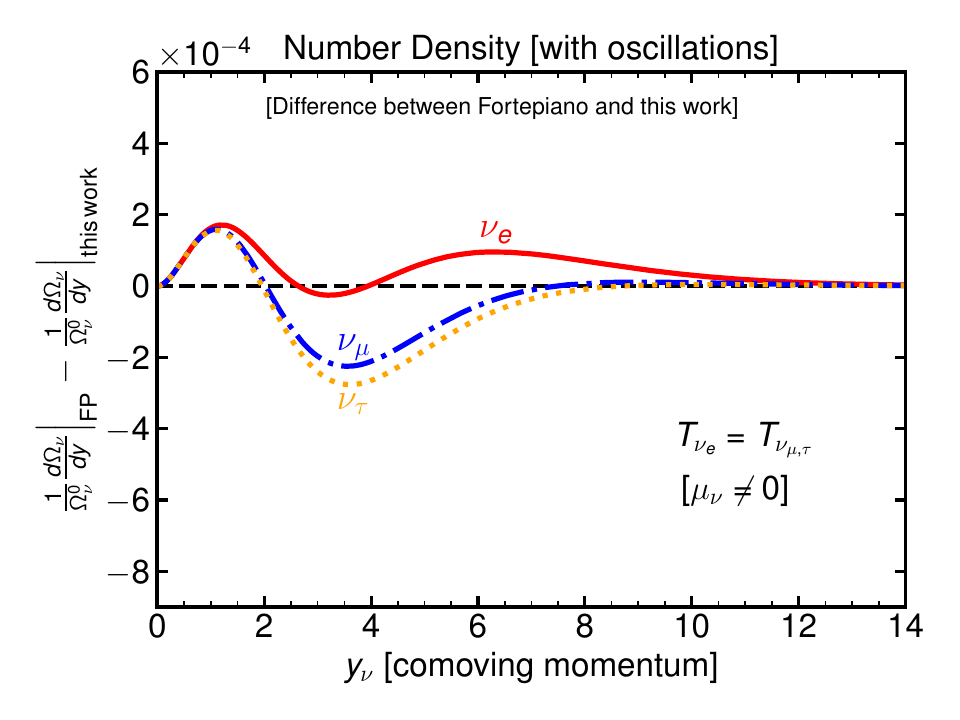}
\\
\end{tabular}
\caption[a]{\setstretch{1.3}
Comparison of the neutrino distribution function between our results and 
those obtained in Fortepiano~\cite{Bennett:2020zkv}. The four upper panels show the relative difference 
in terms of the neutrino energy density and the four lower ones for the number density. We can see that 
the spectral distortions are quite small, $< 8\times 10^{-4}$. The left panels correspond to solutions 
with $\mu_\nu = 0$, and the right ones with $\mu_\nu \neq 0$. Including the chemical potentials 
brings the differences to $< 4\times 10^{-4}$. 
 }
\label{fig:comparison_FP}
\end{figure}

In \fig\ref{fig:comparison_FP} we highlight the comparison between our results and 
those from Fortepiano at the level of the neutrino momentum distribution as a function of $y$. 
The four top panels show the comparison for the differential energy density 
and the four lower ones for the differential number density. Overall, the spectral distortions
are small, $< 8\times 10^{-4}$ in all cases. 
Including the chemical potentials is important in reducing the spectral distortions 
both for the number and energy densities, as seen by a comparison 
of the left ($\mu_\nu^{ } = 0$) and right ($\mu_\nu^{ } \neq 0 $) panels.

All in all, the agreement between the two approaches is excellent both at the level of integrated quantities, 
such as $N^{ }_\rmi{eff}$, $h^{ }_\rmi{eff}$ or $\Omega_\nu h^2$, 
but also for the neutrino distribution function, particularly
once we include effective chemical potentials.

\begin{table}[t]
\begin{center}
\begin{tabular}{l||c|c|c|c|c|c|c}
\hline\hline
\multicolumn{8}{c}{Neutrino Decoupling in the Standard Model: key parameters and observables extended}  \\ \hline
case/parameter      	                
& $\,\,T_\gamma/T_{\nu_e} \,\,$ & $ \,\, T_\gamma/T_{\nu_\mu} \,\,$ &$T_{\nu_e}/\mu_{\nu_e}$ &$T_{\nu_\mu}/\mu_{\nu_\mu}$  
& $(a T_{\nu_e})^{ }_\rmi{fin}/(a T_{\nu_e})^{ }_\rmi{ini} $ 
& $(a T_{\nu_\mu})^{ }_\rmi{fin}/(a T_{\nu_\mu})^{ }_\rmi{ini} $ 
& $(a T^{ }_\gamma)^{ }_\rmi{fin}/(a T^{ }_\gamma)^{ }_\rmi{ini}$
\\ \hline
{\em no oscillations:} & & & & & & & \\
Fortepiano~\cite{Bennett:2020zkv} 
& 1.3927  &  1.3957  &  $-$159  &  $-$348  & -  &  -  &  1.3980 \\
this work $\mu_\nu = 0$ 
&  1.3946  &  1.3965  &  -  &  -  &  1.0024  &  
1.0010 & 1.3979 \\
this work $\mu_\nu \neq 0$
&  1.3925  &  1.3956  &  $-$152  &  $-$341  &  1.0039  &  1.0017 & 1.3979 \\ \hline
{\em with oscillations:} & & & & & & & \\
Fortepiano~\cite{Bennett:2020zkv} 
& 1.3939  &  1.3948  &  $-$202  &  $-$283  &  -  &  -  &  1.3980 \\
this work $\mu_\nu = 0$ 
&  1.3958  &  1.3958  &  -  &  -  &  1.0015  &  1.0015 & 1.3978 \\
this work $\mu_\nu \neq 0$ 
& 1.3944  &  1.3944  &  $-$234  &  $-$234  &  1.0025  &  1.0025  &  1.3979 \\
this work \se\ref{ss:osc} &  1.3945 & 1.3944 & $-$297 & $-$215 & 1.0024 & 1.0025 & 1.3979 \\ 
\hline
\hline
\end{tabular}
\end{center}

\vspace{-0.2cm}

\caption[a]{\setstretch{1.3}
Standard Model results as obtained in this work (\href{https://github.com/MiguelEA/nudec_BSM}{nudec\_BSM\_v2}),
by solving for the time evolution of $T^{ }_\gamma$, $T^{ }_{\nu^{ }_e}$, $T^{ }_{\nu^{ }_\mu}$, 
effective neutrino chemical potentials, and the scale factor, $a$. 
We compare our results against the Liouville solution from Fortepiano with high accuracy settings~\cite{Bennett:2020zkv}. 
The results include the QED equation of state at $\rmO(e^3_{ })$,  ignoring logarithmic corrections at $\rmO(e^2_{ })$, and
employing the low-energy constants from \eq\nr{g's}.  
``No oscillations'' and ``with oscillations'' refer to 
the assumptions in \eq\nr{varrho}, or to the more general 
framework of \se\ref{ss:osc}.
For the results with Fortepiano, $T_\nu$ and $\mu_\nu$ are not actual temperatures/chemical potentials, 
but are rather obtained from a fit to the neutrino energy and number densities 
given by a Fermi-Dirac distribution function, as described by \eqs\nr{e_fit} and \nr{n_fit}. 
}
\label{tab:SM_summary_extended}
\end{table}

%
\section{QED equation of state}
\label{sec:qed_eos}
\addtocontents{toc}{\vspace{-0.5em}} 

The influence of the QED equation-of-state 
on neutrino decoupling has been extensively discussed
in ref.~\cite{Bennett:2019ewm}. In this appendix,
we comment on ref.~\cite{Bennett:2019ewm} in two respects: 
(i)~the contributions of $\rmO(e^2_{ })$ and $\rmO(e^3_{ })$ 
are rewritten with fewer special functions; 
(ii)~the contributions of $\rmO(e^4_{ })$ and $\rmO(e^5_{ })$
are worked out in an approximate form, pointing out
a new parametric enhancement.

%
\hiddenappsubsection{General framework and notation}

Given that we compare with ref.~\cite{Bennett:2019ewm} at several point, 
we start by summarizing their notation. 
The evolution equation considered 
in ref.~\cite{Bennett:2019ewm}
is \eq\nr{e_i_cons} summed over all species, 
\be
 \dot{\e}^{ }_\tot + 3 H (\e^{ }_\tot + \p^{ }_\tot) 
 \;
 \overset{\rmii{\nr{e_i_cons}}}{=}
 \;
 0
 \;,  \la{de_1}
\ee
where 
the subscript $(...)^{ }_\tot$ refers to the total 
energy density~($\e^{ }_\tot$)
and pressure~($\p^{ }_\tot$). 
If we now rescale $\e^{ }_\tot$ and $\p^{ }_\tot$
by powers of the scale factor, $a$,
and replace $t$ through $a$ as the time-like
coordinate, \eq\nr{de_1} turns into
\be
 \frac{{\rm d} (\e^{ }_\tot a^4_{ })}{{\rm d}a}
 \; = \; 
 \frac{ \e^{ }_\tot a^4_{ } - 3 \p^{ }_\tot a^4_{ } }{a}
 \;. \la{de_2}
\ee
Furthermore, we split the energy density and pressure into QED and 
neutrino parts, 
\be
 \e^{ }_\tot \; = \; \e + \e^{ }_\nu
 \;, \quad \e^{ }_\nu = \sum_\alpha \e^{ }_{\nu^{ }_\alpha}
 \;, \quad
 \p^{ }_\tot \; = \; \p + \p^{ }_\nu 
 \;, \quad \p^{ }_\nu = \sum_\alpha \p^{ }_{\nu^{ }_\alpha}
 \;, \la{splitup}
\ee
where $\e \equiv \e^{ }_\EMQED$ and $\p \equiv \p^{ }_\EMQED$
in our notation, and define rescaled variables as 
\be
 x \; \equiv \; \frac{a\, m^{ }_e }{ 
 (a^{ }  T^{ }_\gamma)^{ }_\rmii{ini} }
 \;, \quad
 z \; \overset{\rmii{\nr{e-folds}}}{\equiv} \;
 \frac{a\, T^{ }_\gamma}{
 (a^{ }  T^{ }_\gamma)^{ }_\rmii{ini}}
 \;, \quad
 \tau \; \equiv \; \frac{m^{ }_e}{T^{ }_\gamma } \; = \; \frac{x}{z}
 \;, \quad
 \bar\e \; \equiv \; \frac{\e \, a^4_{ }}{ 
  (a^{ }  T^{ }_\gamma)^{4}_\rmii{ini} }
 \;. \la{rescalings}
\ee
Inserting these rescalings as well as the splitup 
from \eq\nr{splitup} into \eq\nr{de_2}, it turns into
\be
 \frac{{\rm d} \bar \e}{{\rm d}x}
 \; 
 = 
 \;
 \frac{\bar \e - 3 \bar \p}{ x }
 + 
 \frac{\bar \e^{ }_\nu - 3 \bar \p^{ }_\nu}{ x }
 -  
 \frac{{\rm d} \bar \e^{ }_\nu}{{\rm d}x}
 \;. \la{de_dx}
\ee
Subsequently, we write 
${\rm d}/{\rm d}x = \partial/\partial x 
+ ({\rm d}z/{\rm d}x)\, \partial/\partial z$
on the left-hand side, and then solve for 
$ {\rm d}z/{\rm d}x $, rescaling also the numerator 
and denominator by the common factor $1/(2z^3_{ })$,  
\be
 \frac{{\rm d}z}{{\rm d}x}
 \;
 = 
 \; 
 \frac{ \displaystyle \frac{1}{2 z^3_{ }}
 \biggl[\, 
 \frac{\bar \e - 3 \bar \p}{ x }
  - 
 \frac{\partial \bar \e }{\partial x}
 + 
 \frac{\bar \e^{ }_\nu - 3 \bar \p^{ }_\nu}{ x }
  - 
 \frac{{\rm d} \bar \e^{ }_\nu} {{\rm d}x}
 \,\biggr]
 }{ \displaystyle \frac{1}{2 z^3_{ }}
 \biggl[\,
 \frac{\partial \bar{\e}}{\partial z}
 \,\biggr]
 }
 \;
 \underset{\bar{\e}^{ }_\nu - 3 \bar{\p}^{ }_\nu \vphantom{\big |} 
 \; 
 = 
 \; 0}{\equiv}
 \;
 \frac{\displaystyle
 \sum_n G^{(n)}_1
 -
 \frac{1}{2 z^3_{ }} \frac{{\rm d} \bar \e^{ }_\nu} {{\rm d}x}
 }{\displaystyle
 \sum_n G^{(n)}_2
 }
 \;. \la{dz_dx}
\ee
Here 
the QED energy density and pressure
have been expanded in a perturbative series, 
$
 \bar \e = \sum_n \bar\e^{(n)}_{ }
$,
$
 \bar \p = \sum_n \bar\p^{(n)}_{ }
$,
and we have defined 
\ba
 G_1^{(n)}
 & \equiv & 
 \frac{1}{2 z^3_{ }} 
 \biggl[\,
 \frac{\bar \e^{(n)}_{ } - 3 \bar \p^{(n)}_{ }}{ x }
  - 
 \frac{\partial \bar \e^{(n)}_{ } }{\partial x}
 \,\biggr]
 \;, \la{G_1} \\[2mm]
 G_2^{(n)}
 & \equiv & 
 \frac{1}{2 z^3_{ }} 
 \biggl[\,
 \frac{\partial \bar \e^{(n)}_{ } }{\partial z}
 \,\biggr]
 \;.  \la{G_2}
\ea
We note that the factor $1/(2z^3_{ })$ 
in \eq\nr{dz_dx} is missing in front of
$-{\rm d}\bar\e^{ }_\nu/{\rm d}x$
in \eq(2.15) of ref.~\cite{Bennett:2019ewm}.

In order to determine the functions $G^{(n)}_{1,2}$
from \eqs\nr{G_1} and \nr{G_2}, the starting
point is the pressure (minus the free energy density), from which 
all other quantities can be derived. Let us write it as
\be
 \p^{(n)}_{ } 
 \; \equiv \; 
 T^4_{ \gamma }\,\tilde\p^{(n)}_{ }(\tau)
 \;, 
 \quad
 \tau 
 \; 
 \overset{\rmii{\nr{rescalings}}}{=}
 \; 
 \frac{m^{ }_e}{T^{ }_\gamma}
 \;. \la{def_tilde_p}
\ee
Making use 
of $\partial^{ }_{T^{ }_\gamma} = -(\tau/T^{ }_\gamma)\partial^{ }_\tau$, 
the entropy density ($s = {\rm d}\p/{\rm d}T^{ }_\gamma$), 
the energy density ($\e = T^{ }_\gamma\, s - \p $), 
and the ``trace anomaly'' ($\e-3\p$)
can be obtained as 
\be
 s 
 \; = \;
 T^3_{\gamma}\, \bigl(\, 4\tilde\p - \tau\, \tilde\p\hspace*{0.3mm}' \,\bigr)
 \;, \quad
 \e  
 \; 
 = 
 \; 
 T^4_{\gamma}\, \bigl(\, 3\tilde\p - \tau\, \tilde\p\hspace*{0.3mm}' \,\bigr)
 \;, \quad
 \e - 3 \p 
 \; 
 = 
 \; 
 T^4_{\gamma}\, \bigl(\, - \tau\, \tilde\p\hspace*{0.3mm}' \,\bigr)
 \;. \la{thermodynamics}
\ee
The rescaled functions from 
\eqs\nr{rescalings} and \nr{de_dx} then take the forms 
\be
 \bar\e  
 \; 
 \underset{\rmii{\nr{rescalings}}}{
 \overset{\rmii{\nr{thermodynamics}}}{=}} 
 \; 
 z^4_{ }\, \bigl(\, 3\tilde\p - \tau\, \tilde\p\hspace*{0.3mm}' \,\bigr)
 \;, \quad
 \bar\e - 3 \bar\p 
 \; 
 \underset{\rmii{\nr{rescalings}}}{
 \overset{\rmii{\nr{thermodynamics}}}{=}} 
 \; 
 z^4_{ }\, \bigl(\, - \hspace*{0.3mm} \tau\, \tilde\p\hspace*{0.3mm}' \,\bigr)
 \;. \la{bar_e}
\ee
Inserting \eq\nr{bar_e}
into \eqs\nr{G_1} and \nr{G_2}, and making use of 
\be
 \partial^{ }_x
 \;
 \overset{\rmii{\nr{rescalings}}}{=} 
 \; 
 \frac{\partial^{ }_\tau}{z}
 \;, \quad
 \partial^{ }_z
 \;
 \overset{\rmii{\nr{rescalings}}}{=} 
 \; 
 - \frac{\tau \, \partial^{ }_\tau}{z}
 \;, 
\ee
finally yields
\be
 \boxed{
 G^{ }_1 
 \;
 \underset{\rmii{\nr{bar_e}}}{
 \overset{\rmii{\nr{G_1}}}{=}}
 \; 
 \frac{\tau\hspace*{0.3mm}\tilde\p\hspace*{0.3mm}'' - 
 3 \tilde\p\hspace*{0.3mm}'}{2}
 \;, \quad
 G^{ }_2 
 \; 
 \underset{\rmii{\nr{bar_e}}}{
 \overset{\rmii{\nr{G_2}}}{=}}
 \;
 \frac{ \tau^2_{ } \hspace*{0.3mm}\tilde\p\hspace*{0.3mm}'' }{2}
 -
 3 \tau \tilde\p\hspace*{0.3mm}'
 +  
 6\tilde\p
 }
 \;. \la{res_G1_G2}
\ee
These relations will be used for determining
$G^{ }_1$ and $G^{ }_2$ in the following. We can
also give physically more transparent meanings to 
$G^{ }_1$ and $G^{ }_2$ by taking temperature derivatives directly from the unrescaled pressure,
yielding 
\be
 G^{ }_1
 \;
 \underset{\rmii{\nr{res_G1_G2}}}
 {\overset{\rmii{\nr{def_tilde_p}}}{=}}
 \; 
 \frac{1}{2\tau} 
 \Biggl(\, 
 \frac{\partial^2_{T^{ }_\gamma}  
  \p^{ }_{ } }{T^2_\gamma}
 - 
 \frac{3 \partial^{ }_{T^{ }_\gamma}   
  \p^{ }_{ } }{T^3_\gamma}
 \,\Biggr)\, 
 \;, \quad
 G^{ }_2
 \; 
 \underset{\rmii{\nr{res_G1_G2}}}
 {\overset{\rmii{\nr{def_tilde_p}}}{=}}
 \; 
 \frac{\partial^2_{T^{ }_\gamma}  
  \p^{ }_{ } }{2 T^2_\gamma}
 \;. \la{G1_G2_proper}
\ee

We note that the entropy density $s$, from \eq\nr{thermodynamics}, 
can also be written in terms of $G^{ }_1$ and $G^{ }_2$. Specifically,
adopting the re-parametrization in terms of $h^{ }_*$ from 
\eq\nr{conv_def}, and adding the contribution of three kinetically
equilibrated neutrinos, we find
\be
 \boxed{
 h^{ }_* 
 \;
 \underset{\rmii{\nr{conv_def}}}{
 \overset{\rmii{in~equilibrium}}{=}}
 \; 
 \frac{15}{\pi^2_{ }}
 \,\bigl(\, 
   G^{ }_2 - \tau\hspace*{0.3mm} G^{ }_1
 \,\bigr)
 \; + \; \frac{21}{4}
 }
 \;. \la{heff}
\ee

%
\hiddenappsubsection{Special functions and relations between them}

In order to present $G^{ }_{1}(\tau)$ and $G^{ }_{2}(\tau)$, 
five special functions have been defined
in ref.~\cite{Bennett:2019ewm}, 
\ba
 j(\tau) 
 & 
 \overset{\rmii{(4.33)~of~\cite{Bennett:2019ewm}}}{\equiv}
 & 
 \frac{1}{\pi^2_{ }}
 \int_0^\infty \! {\rm d}\omega \, 
 \frac{\exp\bigl(\,\sqrt{\omega^2_{ } + \tau^2_{ }} \,\bigr)}
 {\bigl[ \exp\bigl(\,\sqrt{\omega^2_{ } + \tau^2_{ }} \,\bigr)
    + 1 \bigr]^2_{ }}
 \;, \la{jtau} \\[2mm]
 J(\tau) 
 & 
 \overset{\rmii{(2.16)~of~\cite{Bennett:2019ewm}}}{\equiv}
 & 
 \frac{1}{\pi^2_{ }}
 \int_0^\infty \! {\rm d}\omega \, \omega^2_{ } \, 
 \frac{\exp\bigl(\,\sqrt{\omega^2_{ } + \tau^2_{ }} \,\bigr)}
 {\bigl[ \exp\bigl(\,\sqrt{\omega^2_{ } + \tau^2_{ }} \,\bigr)
    + 1 \bigr]^2_{ }}
 \;, \la{Jtau} \\[2mm]
 Y(\tau) 
 &
 \overset{\rmii{(2.16)~of~\cite{Bennett:2019ewm}}}{\equiv}
 & 
 \frac{1}{\pi^2_{ }}
 \int_0^\infty \! {\rm d}\omega \, \omega^4_{ } \, 
 \frac{\exp\bigl(\,\sqrt{\omega^2_{ } + \tau^2_{ }} \,\bigr)}
 {\bigl[ \exp\bigl(\,\sqrt{\omega^2_{ } + \tau^2_{ }} \,\bigr)
    + 1 \bigr]^2_{ }}
 \;, \la{Ytau} \\[2mm]
 k(\tau) 
 &
 \overset{\rmii{(4.33)~of~\cite{Bennett:2019ewm}}}{\equiv}
 & 
 \frac{1}{\pi^2_{ }}
 \int_0^\infty \! {\rm d}\omega \, 
 \frac{ 1 }{ \sqrt{\omega^2_{ } + \tau^2_{ }} } \, 
 \frac{1}
 { \exp\bigl(\,\sqrt{\omega^2_{ } + \tau^2_{ }} \,\bigr)
    + 1  }
 \;, \la{ktau} \\[2mm]
 K(\tau) 
 &
 \overset{\rmii{(4.15)~of~\cite{Bennett:2019ewm}}}{\equiv}
 & 
 \frac{1}{\pi^2_{ }}
 \int_0^\infty \! {\rm d}\omega \, 
 \frac{ \omega^2_{ } }{ \sqrt{\omega^2_{ } + \tau^2_{ }} } \, 
 \frac{1}
 { \exp\bigl(\,\sqrt{\omega^2_{ } + \tau^2_{ }} \,\bigr)
    + 1  }
 \;. \la{Ktau} 
\ea
Furthermore, the derivatives $j'$, $J'$, $Y'$ and $K'$ were treated
as independent functions, meaning that in total~9 special
functions appear in the results of ref.~\cite{Bennett:2019ewm}. 
For completeness
and ``symmetry'', we here define yet another function, 
\be
 Z(\tau) 
 \;
 \equiv
 \;
 \frac{1}{\pi^2_{ }}
 \int_0^\infty \! {\rm d}\omega \, 
 \frac{ \omega^4_{ } }{ \sqrt{\omega^2_{ } + \tau^2_{ }} } \, 
 \frac{1}
 { \exp\bigl(\,\sqrt{\omega^2_{ } + \tau^2_{ }} \,\bigr)
    + 1  }
 \;, \la{Ztau} 
\ee
but it is again not independent (see below),
and does not appear in the final results. 

By making use of partial integrations and algebraic
relations, the following identities 
can be established between the functions defined: 
\ba
 &&
 J = 2 K + \tau^2_{ } k
 \;, \quad
 Y = 4 Z + 3 \tau^2 K
 \;, \la{ibp1} \\[2mm]
 &&
 Z' = - 3 \tau K
 \;, \quad
 Y' = -3 \tau J
 \;, \quad
 K' = -\tau k 
 \;, \quad
 J' = - \tau j
 \;, \quad
 k' = -\frac{j}{\tau}
 \;. \la{ibp2}
\ea
While some of these are easy to verify, by making use of 
$
 \omega \partial^{ }_\tau f(\sqrt{\omega^2_{ } + \tau^2_{ }})
 = 
 \tau \partial^{ }_\omega f(\sqrt{\omega^2_{ } + \tau^2_{ }})
$, 
others may not be obvious. 
For \eq\nr{ibp1}, we write 
$ \omega / \sqrt{\omega^2_{ }+ \tau^2_{ }} 
 = \partial^{ }_\omega \sqrt{\omega^2_{ }+ \tau^2_{ }}
$
in the integral representation of $K$ or $Z$; 
carry out a partial integration; and then write
$ \sqrt{\omega^2_{ }+ \tau^2_{ }} = 
(\omega^2_{ }+ \tau^2_{ })/\sqrt{\omega^2_{ }+ \tau^2_{ }}$
in one of the resulting terms. 
For the last relation
in \eq\nr{ibp2}, we first show that 
$
 K' + \tau^2_{ }k' = -\tau(k + j)
$, 
and then subtract the simpler identity $K' = -\tau k$.
All in all, 
\eqs\nr{ibp1} and \nr{ibp2} permit for 
a simplification of some of the expressions 
in ref.~\cite{Bennett:2019ewm} (see below). 

%
\hiddenappsubsection{Contribution of $\rmO(e^0_{ })$}

For completeness, let us start at leading order. The pressure, 
normalized according to \eq\nr{def_tilde_p}, reads
\be
 \tilde\p^{(0)}_{ }(\tau)
 \; = \; 
 \frac{\pi^2_{ }}{45} 
 + \frac{2Z(\tau)}{3}
 \;. \la{p^0}
\ee
Inserting into \eq\nr{res_G1_G2} and making use of \eqs\nr{ibp1}
and \nr{ibp2}, we find 
\be
 G^{(0)}_1(\tau) 
 \;
 \underset{\rmii{\nr{ibp1},\nr{ibp2}}}{
 \overset{\rmii{\nr{p^0},\nr{res_G1_G2}}}{=}}
 \;
 \tau J(\tau)
 \;,\quad
 G^{(0)}_2(\tau)
 \;
 \underset{\rmii{\nr{ibp1},\nr{ibp2}}}{
 \overset{\rmii{\nr{p^0},\nr{res_G1_G2}}}{=}}
 \;
 \tau^2_{ } J(\tau) + Y(\tau) + \frac{ 2\pi^2_{ }}{ 15 }
 \;. \la{res_G1_G2^0}
\ee
These agree with \eq(2.15) of ref.~\cite{Bennett:2019ewm}. 

%
\hiddenappsubsection{Contribution of $\rmO(e^2_{ })$}

For the contributions of $\rmO(e^2_{ })$, we omit all temperature-independent
terms from the pressure, 
and for the moment also one non-factorizable contribution, which is referred to as 
a ``logarithmic'' term in ref.~\cite{Bennett:2019ewm}
(cf.\ \eq\nr{ln_p^2}). 
The remainder reads
\be
 \p^{(2)\msl\ln}_{ }
 \; 
 =
 \; 
 - 2 e^2_{ } 
 \biggl[\,
 \frac{T^2_{ \gamma }}{6} + 
 \int \! \frac{{\rm d}^3_{ }\vec{q}}{(2\pi)^3_{ }}
 \, \frac{\nF^{ }(\epsilon^{ }_q)}{\epsilon^{ }_q}
 \,\biggr]
 \int \! \frac{{\rm d}^3_{ }\vec{r}}{(2\pi)^3_{ }}
 \, \frac{\nF^{ }(\epsilon^{ }_r)}{\epsilon^{ }_r}
 \;, \la{orig_p^2}
\ee
where $q \equiv |\vec{q}|$, 
$\epsilon^{ }_q \equiv \sqrt{q^2 + m_e^2}$, 
and $\nF^{ }(x) \equiv 1/[\exp(x/T^{ }_\gamma) + 1]$ is
the Fermi distribution. Eq.~\nr{orig_p^2} agrees with 
\eq(4.7) of ref.~\cite{Bennett:2019ewm}.
Normalized like 
\eq\nr{def_tilde_p}, and making use of the special 
function $K$ from \eq\nr{Ktau}, it can be expressed as 
\be
 \tilde\p^{(2)\msl\ln}_{ }(\tau)
 \; = \; 
 - \frac{e^2}{2}
 \biggl[\, 
 \frac{1}{3} + K(\tau)
 \,\biggr] \, K(\tau) 
 \;. \la{p^2}
\ee
Inserting into \eq\nr{res_G1_G2} and making use of \eqs\nr{ibp1}
and \nr{ibp2}, we find 
\ba
 G^{(2)\msl\ln}_1(\tau) 
 &
 \underset{\rmii{\nr{ibp1},\nr{ibp2}}}{
 \overset{\rmii{\nr{p^2},\nr{res_G1_G2}}}{=}}
 &
 -
 \frac{e^2_{ }\tau}{12}
 \Bigl\{\,
       j(\tau) \bigl[ 1 + 6 K(\tau) \bigr]
 + 2\, k(\tau) \bigl[ 1 + 3 J(\tau) \bigr]
 \,\Bigr\}
 \;,
 \la{res_G1^2}
 \\[2mm]
 G^{(2)\msl\ln}_2(\tau)
 &
 \underset{\rmii{\nr{ibp1},\nr{ibp2}}}{
 \overset{\rmii{\nr{p^2},\nr{res_G1_G2}}}{=}}
 &
 -\frac{e^2_{ }}{12}
 \Bigl\{\,
  3 J(\tau) 
  + 
  2 
  \bigl[ J(\tau) + K(\tau) \bigr]
  \bigl[ 1 + 3 J(\tau) \bigr]
  + 
  \tau^2_{ } j (\tau)
  \bigl[ 1 + 6 K(\tau) \bigr]
 \,\Bigr\}
 \;. \nn \la{res_G2^2}
\ea
If we employ \eqs\nr{ibp1} and \nr{ibp2} in the more complicated
\eqs(4.13) and (4.14) of ref.~\cite{Bennett:2019ewm}, or evaluate the results
numerically, we find perfect agreement. 

For completeness, let us also write down  
the normally omitted logarithmic term, 
\be
 \p^{(2)\ln}_{ }
 \; 
 =
 \; 
 \frac{ e^2_{ } m_e^2 }{4 \pi^4_{ }} 
 \int_0^\infty \! \dd q^{ }
 \int_0^\infty \! \dd r^{ }
 \, \frac{q\, r\, \nF^{ }(\epsilon^{ }_q) \nF^{ }(\epsilon^{ }_r) }
 {\epsilon^{ }_q\, \epsilon^{ }_r}
 \, \ln\bigg| \frac{q+r}{q-r} \bigg|
 \;. \la{ln_p^2}
\ee
The temperature derivative of this expression agrees with the 
entropy density given in \eq(4.17) of ref.~\cite{Bennett:2019ewm}.

We remark that 
normalizing \eq\nr{ln_p^2}
like in \eq\nr{def_tilde_p}, and differentiating with 
respect to $\tau$, as needed for computing 
$G^{(2)\ln}_1$ and $G^{(2)\ln}_2$ from \eq\nr{res_G1_G2},
leads to complicated expressions.  
In analogy with \eqs\nr{ibp1} and \nr{ibp2}, there are 
relations between seemingly different integral representations,
which allow to simplify the outcomes. 
Probably the most compact formulae can be obtained
by taking temperature derivatives of \eq\nr{ln_p^2}, and 
extracting then $G_1^{(2)\ln}$ and $G_2^{(2)\ln}$ 
from \eq\nr{G1_G2_proper}.
Introducing the functions
\ba
 \mathcal{A}(\omega^{ }_1,\omega^{ }_2)
 & \;\equiv\; & 
 \frac{1}{
 \bigl( e^{ \sqrt{\omega_1^2 + \tau^2_{ }} }_{ } + 1 \bigr)
 \bigl( e^{ \sqrt{\omega_2^2 + \tau^2_{ }} }_{ } + 1 \bigr)
 \sqrt{\omega_1^2 + \tau^2_{ }}
 \sqrt{\omega_2^2 + \tau^2_{ }}
 }
 \;, \la{calA} \\[2mm]
 \mathcal{B}(\omega^{ }_1,\omega^{ }_2)
 & \;\equiv\; & 
 \frac{e^{ \sqrt{\omega_2^2 + \tau^2_{ }} }_{ }}{
 \bigl( e^{ \sqrt{\omega_1^2 + \tau^2_{ }} }_{ } + 1 \bigr)
 \bigl( e^{ \sqrt{\omega_2^2 + \tau^2_{ }} }_{ } + 1 \bigr)^2_{ }
 \sqrt{\omega_1^2 + \tau^2_{ }}
 }
 \;, \la{calB} \\[2mm]
 \mathcal{C}(\omega^{ }_1,\omega^{ }_2)
 & \;\equiv\; & 
 \frac{e^{ \sqrt{\omega_2^2 + \tau^2_{ }} }_{ }}{
 \bigl( e^{ \sqrt{\omega_1^2 + \tau^2_{ }} }_{ } + 1 \bigr)
 \bigl( e^{ \sqrt{\omega_2^2 + \tau^2_{ }} }_{ } + 1 \bigr)^2_{ }
 }
 \Biggl[\, 
 \frac{e^{ \sqrt{\omega_1^2 + \tau^2_{ }} }_{ }}{
  e^{ \sqrt{\omega_1^2 + \tau^2_{ }} }_{ } + 1  } 
 + 
 \frac{ \sqrt{\omega_2^2 + \tau^2_{ }} }{ \sqrt{\omega_1^2 + \tau^2_{ }} }
 \frac{e^{ \sqrt{\omega_2^2 + \tau^2_{ }} }_{ } - 1}{
  e^{ \sqrt{\omega_2^2 + \tau^2_{ }} }_{ } + 1  } 
 \,\Biggr]
 \;, \la{calC}
\ea
we thereby obtain
\ba
 \tilde\p^{(2)\ln}_{ }(\tau) 
 & 
 \; = \; 
 & 
 \frac{e^2_{ } \tau^2_{ }}{4\pi^4_{ }}
 \int_0^\infty \! {\rm d}\omega^{ }_1 
 \int_0^{\omega^{ }_1} \! {\rm d}\omega^{ }_2
 \, \omega^{ }_1 \omega^{ }_2 
 \, \ln \bigg| \frac{\omega^{ }_1 + \omega^{ }_2}
                    {\omega^{ }_1 - \omega^{ }_2} \bigg|
 \Bigl[\,
  \mathcal{A}( \omega^{ }_1 , \omega^{ }_2 ) 
 + 
  ( \omega^{ }_1 \leftrightarrow \omega^{ }_2 )
 \,\Bigr]
 \;, 
 \la{p2ln} 
 \\[2mm]
 G^{(2)\ln}_{1}(\tau)
 & 
 \; = \; 
 & 
 \frac{e^2_{ } \tau^{}_{ }}{4\pi^4_{ }}
 \int_0^\infty \! {\rm d}\omega^{ }_1 
 \int_0^{\omega^{ }_1} \! {\rm d}\omega^{ }_2
 \, \omega^{ }_1 \omega^{ }_2 
 \, \ln \bigg| \frac{\omega^{ }_1 + \omega^{ }_2}
                    {\omega^{ }_1 - \omega^{ }_2} \bigg|
 \Bigl[\,
  \mathcal{C}( \omega^{ }_1 , \omega^{ }_2 ) 
 - 
 5\,\mathcal{B}( \omega^{ }_1 , \omega^{ }_2 ) 
 + 
  ( \omega^{ }_1 \leftrightarrow \omega^{ }_2 )
 \,\Bigr]
 \;, 
 \la{G12ln} 
 \\[2mm]
 G^{(2)\ln}_{2}(\tau)
 & 
 \; = \; 
 & 
 \frac{e^2_{ } \tau^{2}_{ }}{4\pi^4_{ }}
 \int_0^\infty \! {\rm d}\omega^{ }_1 
 \int_0^{\omega^{ }_1} \! {\rm d}\omega^{ }_2
 \, \omega^{ }_1 \omega^{ }_2 
 \, \ln \bigg| \frac{\omega^{ }_1 + \omega^{ }_2}
                    {\omega^{ }_1 - \omega^{ }_2} \bigg|
 \Bigl[\,
  \mathcal{C}( \omega^{ }_1 , \omega^{ }_2 ) 
 - 
 2\,\mathcal{B}( \omega^{ }_1 , \omega^{ }_2 ) 
 + 
  ( \omega^{ }_1 \leftrightarrow \omega^{ }_2 )
 \,\Bigr]
 \;. 
 \la{G22ln} 
\ea

%
\hiddenappsubsection{Contribution of $\rmO(e^3_{ })$}

The contribution of $\rmO(e^3_{ })$ originates from the ``Debye mass'', 
\be
 \p^{(3)}_{ }
 \; = \; 
 \frac{m_\rmiii{D}^3 T^{ }_\gamma}{12\pi}
 \;, \quad
  m_\rmii{D}^2 
 \; = \; 
 4 e^2_{ }\int \! \frac{{\rm d}^3_{ }\vec{q}}{(2\pi)^3_{ }}
 \, \frac{\nF^{ }(\epsilon^{ }_q)}{\epsilon^{ }_q}
 \biggl(\, 2 + \frac{m_e^2}{q^2_{ }} \,\biggr)
 \;, 
\ee
where we use the same notation as in \eq\nr{orig_p^2}.
Expressed in terms of 
\eq\nr{def_tilde_p}, and employing the special 
functions $k$ and $K$ from \eqs\nr{ktau} and \nr{Ktau}, 
respectively, as well as \eq\nr{ibp1} to combine them, yields
\be
 \tilde\p^{(3)}_{ }(\tau)
 \; = \; 
 \frac{e^3_{ }}{3\sqrt{2}\pi} \bigl[\, J(\tau) \,\bigr]^{3/2}_{ }
 \;. \la{p^3}
\ee
Inserting into \eq\nr{res_G1_G2} and making use of \eqs\nr{ibp1}
and \nr{ibp2}, we find 
\ba
 G^{(3)}_1(\tau) 
 &
 \underset{\rmii{\nr{ibp1},\nr{ibp2}}}{
 \overset{\rmii{\nr{p^3},\nr{res_G1_G2}}}{=}}
 &
 \frac{e^3_{ }\tau}{4\pi}
 \sqrt{\frac{ J(\tau) }{2}}
 \biggl[\,
  2 j(\tau) - \tau^{ }_{ }j'(\tau)
  + \frac{\tau^2_{ }j^2_{ }(\tau)}{2 J(\tau)} 
 \,\biggr]
 \;,
 \la{res_G1^3}
 \\[2mm]
 G^{(3)}_2(\tau)
 &
 \underset{\rmii{\nr{ibp1},\nr{ibp2}}}{
 \overset{\rmii{\nr{p^3},\nr{res_G1_G2}}}{=}}
 &
 \frac{e^3_{ }}{4\pi}
 \sqrt{\frac{ J(\tau) }{2}}
 \biggl[\,
  8 J(\tau) 
  + 5 \tau^2_{ } j(\tau) - \tau^3_{ }j'(\tau)
  + \frac{\tau^4_{ }j^2_{ }(\tau)}{2 J(\tau)} 
 \,\biggr]
 \;. \la{res_G2^3}
\ea

Equations \nr{res_G1^3} and \nr{res_G2^3} can be compared with 
\eq(4.32) of ref.~\cite{Bennett:2019ewm}. Our results are simpler; 
however, making use of \eqs\nr{ibp1} and \nr{ibp2} in the expressions
of ref.~\cite{Bennett:2019ewm}, or checking numerically, 
it can be verified that 
the values are the same. 

%
\hiddenappsubsection{Contributions of $\rmO(e^4_{ })$ and $\rmO(e^5_{ })$}

We finally discuss the contributions of $\rmO(e^4_{ })$ and $\rmO(e^5_{ })$
that were {\em not} included in ref.~\cite{Bennett:2019ewm}. 
These terms have been fully
determined only in the massless limit
($\tau = m^{ }_e/T^{ }_\gamma\to 0$), and we start by recalling those 
results. However, we can subsequently upgrade parts of the expressions to the
massive situation, by making use of renormalization group invariance,
as discussed around \eqs\nr{def_p_cs}--\nr{c4p_c5p}. 

The finite coefficients $c^{ }_4$ and $c^{ }_5$ were first determined
in ref.~\cite{Parwani:1994xi}, 
however $c^{ }_4$ was computed only numerically, and the
renormalization scale was fixed in a non-standard way. An analytic computation
of $c^{ }_4$ can be found in ref.~\cite{Arnold:1994eb}, 
\ba
 c^{ }_4 
 & 
 \overset{\rmii{(5.1)~of~\cite{Arnold:1994eb}}}{=} 
 & 
 -\,\frac{5}{(4\pi)^4_{ }}
 \biggl[\,
  \frac{20}{3} \ln(4\pi)
 + \frac{8}{3}\frac{\zeta'(-3)}{\zeta(-3)} 
 - \frac{16}{3}\frac{\zeta'(-1)}{\zeta(-1)} 
 - 4 \gamma^{ }_\rmiii{E} 
 - \frac{319}{12}
 + \frac{208 \ln 2}{5}
 \,\biggr]
 \nn[2mm] 
 & 
 \overset{\rmii{(5.3)~of~\cite{Arnold:1994eb}}}{\approx} 
 & 
 -0.00159412
 \;. \la{res_c4}
\ea
An independent computation of $c^{ }_5$, with a general
renormalization scale so that the non-logarithmic part
can be easily isolated, can be found in ref.~\cite{Zhai:1995ac}, 
\ba
 c^{ }_5 
 & 
 \overset{\rmii{(4.3)~of~\cite{Zhai:1995ac}}}{=} 
 & 
 \,\frac{320}{\sqrt{3}(4\pi)^5_{ }}
 \biggl[\,
  \ln(4\pi)
 - \gamma^{ }_\rmiii{E} 
 - \frac{7}{4}
 - 2 \ln 2
 \,\biggr]
 \;
 \approx 
 \;
 -0.000697166
 \;. \la{res_c5}
\ea
The coefficients $c^{ }_4$ and $c^{ }_5$ will be needed
in \eqs\nr{res_G2^4} and \nr{res_G2^5}, respectively.

\vspace*{3mm}

We now move on to the massive case. The pressure can still be written
like in \eq\nr{def_p_cs}, however the coefficients $c_n^{ }$ become
functions of $\tau = m^{ }_e/T^{ }_\gamma$. We fix the renormalization
scale as $\bmu = m^{ }_e$, whereby the logarithms turn into $\ln\tau$. 
Fixing the values of $\tilde c_4^{ }$ and $\tilde c_5^{ }$
according to \eq\nr{c4p_c5p}, and going over to the normalization 
in \eq\nr{def_tilde_p}, we then find
\ba
 \tilde\p^{(4)}_{ }(\tau)
 & 
 \underset{\rmii{\nr{c4p_c5p}}}{
 \overset{\rmii{\nr{def_p_cs},\nr{def_tilde_p}}}{\approx}}
 & 
 - \frac{e^2_{ }\tilde\p^{(2)\msl\ln}_{ }(\tau) \ln\tau}{6\pi^2_{ }}
 + \frac{c^{ }_4\, e^4_{ } \pi^2_{ }}{45}
 \;, \la{p^4} 
 \\[2mm]
 \tilde\p^{(5)}_{ }(\tau)
 & 
 \underset{\rmii{\nr{c4p_c5p}}}{
 \overset{\rmii{\nr{def_p_cs},\nr{def_tilde_p}}}{\approx}}
 & 
 - \frac{e^2_{ }\tilde\p^{(3)}_{ }(\tau) \ln\tau}{4\pi^2_{ }}
 + \frac{c^{ }_5\, e^5_{ } \pi^2_{ }}{45}
 \;, \la{p^5} 
\ea
where $\tilde\p^{(2)}_{ }$ and $\tilde\p^{(3)}_{ }$
are given by \eqs\nr{p^2} and \nr{p^3}, respectively. 
While the logarithmic terms are fixed by renormalization
group invariance, the $\tau$-dependence of 
$c^{ }_4$ and $c^{ }_5$ is unknown, so we treat them as constants, 
as is appropriate for the massless limit. 

Subsequently, we can insert \eqs\nr{p^4} and \nr{p^5} into
\eq\nr{res_G1_G2}, in order to estimate the functions $G^{(4,5)}_{1,2}$.
At $\rmO(e^4_{ })$ we find 
\ba
 G^{(4)}_1(\tau)
 &
 \underset{\rmii{\nr{res_G1_G2}}}{
 \overset{\rmii{\nr{p^4}}}{\approx}}
 & 
  -\, \frac{e^2_{ } G^{(2)\msl\ln}_1 (\tau) \ln \tau}{6\pi^2_{ }}
 \; - \, 
 \frac{e^4_{ }}{36 \pi^2_{ }}
 \frac{  J(\tau) + 6 \bigl[ J(\tau) - K(\tau) \bigr] K(\tau) }{\tau}
 \;, \la{res_G1^4}
 \\[2mm]
 G^{(4)}_2(\tau)
 &
 \underset{\rmii{\nr{res_G1_G2}}}{
 \overset{\rmii{\nr{p^4}}}{\approx}}
 & 
 -\, \frac{e^2_{ } G^{(2)\msl\ln}_2 (\tau) \ln \tau}{6\pi^2_{ }}
 + \frac{2 c^{ }_4\, e^4_{ } \pi^2_{ }}{15}
 \; - \, 
 \frac{e^4_{ }}{72\pi^2_{ }}
 \Bigl\{
   3 K(\tau) \bigl[ 1 - K(\tau) \bigr] 
 + 2 J(\tau) \bigl[ 1 + 6 K(\tau) \bigr] 
 \Bigr\}
 \;, \la{res_G2^4}
\ea
where $G^{(2)\msl\ln}_1$ and $G^{(2)\msl\ln}_2$
are given by \eqs\nr{res_G1^2} and \nr{res_G2^2}, respectively. 
An analogous computation at $\rmO(e^5_{ })$ produces
\ba
 G^{(5)}_1(\tau)
 &
 \underset{\rmii{\nr{res_G1_G2}}}{
 \overset{\rmii{\nr{p^5}}}{\approx}}
 & 
 -\, \frac{e^2_{ } G^{(3)}_1 (\tau) \ln \tau}{4\pi^2_{ }}
 \; + \, 
 \frac{e^5_{ }}{24\pi^3_{ }}
 \sqrt{\frac{J(\tau)}{2}}
 \frac{ 4 J(\tau) + 3 \tau^2_{ }j(\tau)}{\tau}
 \;, \la{res_G1^5} \\[2mm]
 G^{(5)}_2(\tau)
 &
 \underset{\rmii{\nr{res_G1_G2}}}{
 \overset{\rmii{\nr{p^5}}}{\approx}}
 & 
 -\, \frac{e^2_{ } G^{(3)}_2 (\tau) \ln \tau}{4\pi^2_{ }}
 + \frac{2 c^{ }_5\, e^5_{ } \pi^2_{ }}{15}
 \; + \, 
 \frac{e^5_{ }}{24\pi^3_{ }}
 \sqrt{\frac{J(\tau)}{2}}
 \Bigl\{ 7 J(\tau) + 3 \tau^2_{ }j(\tau) \Bigr\}
 \;,
 \hspace*{8mm} \la{res_G2^5}
\ea
where $G^{(3)}_1$ and $G^{(3)}_2$
are given by \eqs\nr{res_G1^3} and \nr{res_G2^3}, respectively. 

\vspace*{3mm}

We note from \eqs\nr{res_G1^4} and \nr{res_G1^5} that 
$G^{ }_1$ obtains a $1/\tau$-enhancement at 
$\tau = m^{ }_e/T^{ }_\gamma \ll 1$
at $\rmO(e^4_{ })$ and $\rmO(e^5_{ })$. 
This is not a coincidence. It can be seen from \eq\nr{def_p_cs} that 
in the limit $\tau\to 0$, when the $c^{ }_n$'s are constants, 
$\tilde c_4^{ }$ and $\tilde c_5^{ }$ 
are the first coefficients contributing
to $\tilde\p\hspace*{0.3mm}'$, and therefore 
to the trace anomaly, cf.\ \eq\nr{thermodynamics}.
However, despite the small-$\tau$ enhancement of $G^{(4)}_1$ and $G^{(5)}_1$, 
in the end we find {\em no} substantial
numerical effect on $N^{ }_\rmi{eff}$ or $h^{ }_\rmi{eff}$
from them.

%
\section{Bessel and polylogarithmic representations of thermodynamic quantities}
\label{sec:Bessel}
\addtocontents{toc}{\vspace{-1.0em}} 

One novelty of our numerical packages is the option to evaluate thermodynamic formulae as a series in modified Bessel functions. In this appendix we document the relevant formulae and mention their possible limitations. 

%
\hiddenappsubsection{Neutrino and BSM sectors}

Considering non-interacting bosons ($-$) or fermions ($+$) with $g$ internal degrees of freedom, 
their number density, energy density and pressure are written as (cf.\ \eq\eqref{e_p_n})
\begin{subequations}
\begin{align}
    n^{ }_\pm &\;\overset{\rmii{\nr{e_p_n}}}{=}\; \frac{g}{2\pi^2} \int_{m}^{\infty}  {\rm d}E \,  E \sqrt{E^2-m^2} f_{\pm}
    \,, \label{eq:n_rho_p_1} \\[2mm]
    \e^{ }_\pm &\;\overset{\rmii{\nr{e_p_n}}}{=}\; \frac{g}{2\pi^2} \int_{m}^{\infty} {\rm d}E \,  E^2 \sqrt{E^2-m^2} f_{\pm}
    \,, \la{e_pm_def} \\[2mm]
    \p^{ }_\pm &\;\overset{\rmii{\nr{e_p_n}}}{=}\;\frac{g}{6\pi^2} \int_{m}^{\infty}  {\rm d}E \, (E^2-m^2)^{3/2} f_{\pm}
    \,. \label{eq:n_rho_p_3}
\end{align}
\end{subequations}
We include bosons, as they appear in BSM constructions. 
The distribution functions have the Bose or Fermi shape, 
\begin{align}
    f^{ }_\pm \;=\; \frac{1}{e^{(E-\mu)/T}_{ }\pm 1} \quad _{-\,\,\![{\rm bosons}]}^{+\,[{\rm fermions}]}\,. \la{f_pm}
\end{align}
For $m>0$ the integrals in \eqs\eqref{eq:n_rho_p_1}--\nr{eq:n_rho_p_3} 
are not analytically solvable, however they can be written as a sum of Bessel functions~\cite{1939isss.book.....C}. If $\mu < m$, so that $\mu < E$, we can write down a convergent 
geometric series,\footnote{%
 If $\mu > m$, there is an energy domain in which we need to expand in $e^{(E-\mu)/T}_{ }$ instead, leading to the 
 generalized representation
 \be
  f^{ }_\pm \;=\; 
  \theta(E-\mu)\,\sum_{n=0}^{\rm \infty} (\mp1)^{n}_{ } [e^{\mu/T}]^{n+1}_{ } [e^{-E/T}]^{n+1}_{ }
 - 
  \theta(\mu-E)\,\sum_{n=0}^{\rm \infty} (\mp1)^{n+1}_{ } [e^{-\mu/T}]^{n}_{ } [e^{E/T}]^{n}_{ }
  \;. \la{series_parts}
 \ee
 } 
\begin{align}
    f^{ }_\pm 
    \;=\;
    \frac{e^{(\mu - E)/T}_{ }}{1 \pm e^{(\mu - E)/T}_{ }} 
    \; \overset{\mu\;<\;E}{=} \; 
    \sum_{n=0}^{\rm \infty} (\mp1)^{n}_{ } [e^{\mu/T}]^{n+1}_{ } [e^{-E/T}]^{n+1}_{ }
    \,. \la{series_exact}
\end{align}
In usual scenarios 
without particle-antiparticle asymmetries, chemical potentials tend to be negative, 
so we can employ \eq\nr{series_exact}. 
For massless fermions, results applying to both signs can be 
derived from \eq\nr{series_parts}, and are given 
in \eqs\nr{n_plus_massless}--\nr{p_plus_massless}.

Recalling now representations of the generalized Bessel function $K^{ }_\nu$,  
\begin{subequations} \la{def_Knu}
\begin{align}
  K_\nu^{ }(z)
 & \;
 \underset{\re \nu \;>\; -\tfrac{1}{2} }{
 \overset{\re z \; >\; 0 \vphantom{\big | } }{=}}
 \; 
 \frac{\sqrt{\pi}\,\left(\tfrac{z}{2}\right)^{\nu}}{\Gamma\!\left(\nu+\tfrac{1}{2}\right)}
\int_{1}^{\infty} \! {\rm d}t\, e^{-z t}\,\bigl(t^{2}-1\bigr)^{\nu-\tfrac{1}{2}}
 \;, 
 \\[2mm] 
   K_{\nu+1}^{ }(z)
 & \;
 \underset{\re \nu \;>\; -\tfrac{1}{2} }{
 \overset{\re z \; >\; 0 \vphantom{\big | } }{=}}
 \; 
 \frac{\sqrt{\pi}\,\left(\tfrac{z}{2}\right)^{\nu}}{\Gamma\!\left(\nu+\tfrac{1}{2}\right)}
 \int_{1}^{\infty} \! {\rm d}t\, t \, e^{-z t} \, \bigl(t^{2}-1\bigr)^{\nu-\tfrac{1}{2}}
\;, 
\end{align}
\end{subequations}
we can rewrite all the thermodynamic integrals in terms of sums of Bessel functions,
\begin{subequations}
\begin{align}
    n_\pm &\;\underset{\rmii{\nr{series_exact},\nr{def_Knu}}}{\overset{\rmii{\nr{eq:n_rho_p_1}}}{=}}\; 
    \frac{g}{2\pi^2} \sum_{n=0}^\infty \frac{m^2 T}{n+1} \, (\mp 1)^n \, e^{\frac{\mu(n+1)}{T}} \, K_2\left[\frac{m(n+1)}{T}\right] 
     \;, \la{n_bessel} \\[2mm]
    \e_\pm &\;\underset{\rmii{\nr{series_exact},\nr{def_Knu}}}{\overset{\rmii{\nr{e_pm_def}}}{=}}\; \frac{g}{2\pi^2} \sum_{n=0}^\infty \frac{m^2 T}{(n+1)^2} \, (\mp 1)^n \, e^{\frac{\mu(n+1)}{T}}
    \left\{ m(n+1) K_1\left[\frac{m(n+1)}{T}\right] + 3T K_2\left[\frac{m(n+1)}{T}\right] \right\}
     \;, \la{e_bessel} \\[2mm]
    \p_\pm &\;\underset{\rmii{\nr{series_exact},\nr{def_Knu}}}{\overset{\rmii{\nr{eq:n_rho_p_3}}}{=}}\; \frac{g}{2\pi^2} \sum_{n=0}^\infty \frac{m^2 T^2}{(n+1)^2} \, (\mp 1)^n \, e^{\frac{\mu(n+1)}{T}} \, K_2\left[\frac{m(n+1)}{T}\right]
    \;. \la{p_bessel}
\end{align}
\end{subequations}
For \eqs\nr{T dot} and \nr{mu dot} we also need derivatives with respect to $T$ 
and $\mu$, which can be obtained with the recursion relations
$
 -z K'_\nu = \nu K^{ }_\nu + z K^{ }_{\nu-1}
$
and
$
 K^{ }_{\nu+1} = K^{ }_{\nu-1} + (2\nu/z) K^{ }_\nu
$, as
\begin{subequations}
\begin{align}
    \frac{\partial n_\pm}{\partial T} &\;=\; \frac{g}{2\pi^2} 
    \sum_{n=0}^\infty \frac{m^2}{T} \, (\mp 1)^n \, e^{\frac{\mu(n+1)}{T}} 
    \left\{ 
        m K_1\left[\frac{m(n+1)}{T}\right] 
        + \biggl( \frac{3T}{n+1} - \mu\biggr)
           K_2\left[\frac{m(n+1)}{T}\right] 
    \right\}\;, \\[2mm]
  \frac{\partial \e_\pm}{\partial T} &\;=\; \frac{g}{2\pi^2}
    \sum_{n=0}^\infty \frac{m}{(n+1)^2 T} \, (\mp 1)^n \, e^{\frac{\mu(n+1)}{T}} \biggl\{ 
        m \left[ m^2(n+1)^2 + 3T\bigl( 4T - \mu (n+1) \bigr) \right] K_0\left[\frac{m(n+1)}{T}\right] \nn
        &\quad + \frac{1}{n+1} \left[
         m^2(n+1)^2\bigl( 5T - \mu (n+1) \bigr) 
         + 6 T^2\bigl( 4 T - \mu (n+1) \bigr) \right] K_1\left[\frac{m(n+1)}{T}\right] 
    \biggr\}
    \;, \\[2mm] 
   \frac{\partial n_\pm}{\partial\mu} &\;=\; \frac{g}{2\pi^2} \sum_{n=0}^\infty m^2 \, (\mp 1)^n \, e^{\frac{\mu(n+1)}{T}} \, K_2\left[\frac{m(n+1)}{T}\right]
   \;, \\[2mm]
    \frac{\partial \e_\pm}{\partial\mu} &\;=\; \frac{g}{2\pi^2} \sum_{n=0}^\infty \frac{m^2}{n+1} \, (\mp 1)^n \, e^{\frac{\mu(n+1)}{T}} 
    \left\{ m(n+1) K_1\left[\frac{m(n+1)}{T}\right] + 3T K_2\left[\frac{m(n+1)}{T}\right] \right\}
    \;. \la{de_dmu_bessel}
\end{align}
\end{subequations}
The Maxwell-Boltzmann limit is given by the $n = 0$ term. 
In practice, we set $n_{\rm max} = 30$ by default in the code.

In the massless limit, applicable to neutrinos, the results can be given in terms 
of polylogarithms. For $|z| < 1$,  a polylogarithm is defined through 
\be
 \mbox{Li}^{ }_s(z) 
 \; \overset{|z|\;<\; 1}{=} \; \sum_{n=1}^{\infty} \frac{z^n_{ }}{n^s_{ }}
 \la{Li_def}
 \;, 
\ee
but it can be analytically continued outside of this domain. To obtain 
convergent sum representations, we
use \eq\nr{series_parts} as a starting point, leading to 
\begin{subequations}
\ba
 n^{ }_+ &\; \underset{\rmii{\nr{series_parts},\nr{Li_def}}}{\overset{\rmii{\nr{eq:n_rho_p_1}}}{=}} \;& 
 \frac{g}{\pi^2_{ }}
 \biggl[ - T^3_{ } \,\mbox{Li}^{ }_3 \Bigl(  - e^{- \frac{|\mu|}{T}}_{ }\Bigr)
+ \theta(\mu)\,\frac{\,\mu\, (\mu^2_{  } + \pi^2_{ }T^2_{ })}{6} \biggr]
  \nn[2mm]
 &\; \underset{\rmii{continuation}}{\overset{\rmii{analytic}}{=}} \;&
 \frac{g}{\pi^2_{ }}
  \biggl[ - T^3_{ } \,\mbox{Li}^{ }_3 \Bigl(  - e^{\frac{\mu}{T}}_{ }\Bigr) \biggr]
\;, \la{n_plus_massless} \\[2mm]
 \e^{ }_+ &\; \underset{\rmii{\nr{series_parts},\nr{Li_def}}}{\overset{\rmii{\nr{e_pm_def}}}{=}} \; &
 \frac{g}{\pi^2_{ }}
 \biggl[ 
   3 T^4_{ } \mbox{sign}(\mu) \,\mbox{Li}^{ }_4 \Bigl(  - e^{- \frac{|\mu|}{T}}_{ }\Bigr)
+ \theta(\mu)\biggl( 
  \frac{\mu^4_{  }}{8} + \frac{(\mu \pi T)^2_{ }}{4}
   + \frac{7 (\pi T)^4_{ }}{120} \biggr)
 \biggr]
   \nn[2mm]
  &\; \underset{\rmii{continuation}}{\overset{\rmii{analytic}}{=}} \;&
 \frac{g}{\pi^2_{ }}
  \biggl[ - 3 T^4_{ } \,\mbox{Li}^{ }_4 \Bigl(  - e^{\frac{\mu}{T}}_{ }\Bigr) \biggr]
 \;, \la{e_plus_massless} \\[2mm]
  \p^{ }_+ &\; \underset{\rmii{\nr{eq:n_rho_p_3}}}{\overset{\rmii{\nr{e_pm_def}}}{=}} \;& 
 \frac{\e^{ }_+}{3} \;.
  \la{p_plus_massless}
\ea
\end{subequations}
Making use of $\partial^{ }_z \mbox{Li}^{ }_s(-e^z_{ }) = \mbox{Li}^{ }_{s-1}(-e^z_{ })$, 
one can easily find partial temperature and chemical potential derivatives. Moreover,
for massless bosons, we note from \eq\nr{f_pm} that
$
 f^{ }_+ |^{ }_{\mu\to \mu + i \pi T} = -f^{ }_{-}
$, 
though then a branch cut opens at $\mu = 0$, and the representations
are only valid for $\mu < 0$. The resulting expressions are given
in \tabl\ref{thermo_massless} on p.~\pageref{thermo_massless}.

%
\begin{table}[t]
\begin{center}

\begin{equation*}
\def\arraystretch{1.25}
\begin{array}{cccc}
\toprule
      \multicolumn{4}{c@{}}{ \text{massless thermodynamics}, \,\, m = 0 \, ,\,\, x \equiv e^{\mu/T} }   \\[2mm]
    \toprule
  \,\,\, \,\,\,\,   \text{quantity} \,\,\,\, \,\,\,
       &\,\,\,\,    \text{Fermi-Dirac} \,\,\,\, 
       & \,\,\,\,   \text{Bose-Einstein}\;(\mu < 0) \,\,\,\, 
       &\,\,\,   \text{Maxwell-Boltzmann}\,\,\, 
   \\[2mm]
     \toprule
        n   & -g\frac{T^3}{\pi^2}\, \text{Li}_3\left(-x\right)
            &  g\frac{T^3}{\pi ^2}  \text{Li}_3\left(x\right)
            & g\frac{T^3}{\pi ^2} \, x\\[2mm]
        \e   &  -g\frac{3T^4}{\pi ^2}  \, \text{Li}_4\left(-x\right) 
               &   g\frac{3 T^4 }{\pi ^2} \text{Li}_4\left(x\right)
               & g \frac{3 T^4 }{\pi ^2} \, x\\[2mm]
        \p   &  \rho/3 & \rho/3  & \rho/3 \\[2mm]
  \partial n/\partial T   & g\frac{T \left(\mu  \text{Li}_2\left(-x\right)-3 T \text{Li}_3\left(-x\right)\right)}{\pi ^2}
                          & g\frac{T \left(3 T \text{Li}_3\left(x\right)-\mu  \text{Li}_2\left(x\right)\right)}{\pi ^2}
                          & g\frac{T  (3 T-\mu )}{\pi ^2} x\\[3mm]
  \partial \e/\partial T  
     & g\frac{3 T^2\left(\mu  \text{Li}_3\left(-x\right)-4 T \text{Li}_4\left(-x\right)\right)}{\pi ^2}   
     &  g\frac{3 T^2 \left(4 T \text{Li}_4\left(x\right)-\mu  \text{Li}_3\left(x\right)\right)}{\pi ^2}
     &  g\frac{3 T^2  (4 T-\mu )}{\pi ^2} x\\[2mm]
  \partial n/\partial \mu   
     & -g\frac{T^2 }{\pi^2} \text{Li}_2\left(-x\right)
     & g\frac{T^2 }{\pi^2} \text{Li}_2\left(x\right) 
     & g\frac{T^2 }{\pi^2} x \\[2mm]
   \partial \e/\partial \mu   
     & -g\frac{3 T^3 }{\pi^2} \text{Li}_3\left(-x\right)  
     & g\frac{3 T^3 }{\pi^2}\text{Li}_3\left(x\right) 
     & g\frac{3 T^3 }{\pi^2} x \tabularnewline
     \bottomrule 
\end{array}
\end{equation*}
\begin{equation*} 
\def\arraystretch{1.25}
\begin{array}{cccc}
\toprule
      \multicolumn{4}{c@{}}{ \text{massless thermodynamics}, \,\, m = 0 \, ,\,\, \mu = 0  }   \\[2mm]
    \toprule
  \,\,\, \,\,\,\,   \text{quantity} \,\,\,\, \,\,\,
  &\,\,\,\,         \text{Fermi-Dirac} \,\,\,\, 
  & \,\,\,\,        \text{Bose-Einstein} \,\,\,\, 
  &\,\,\, \text{Maxwell-Boltzmann}\,\,\, \\[2mm]
     \toprule
        n   & g \frac{3}{4} \frac{\zeta(3)}{\pi ^2}  T^3
            & g\frac{\zeta(3)}{\pi ^2} T^3 
            & g\frac{T^3}{\pi ^2} \\[3mm]
        \e   &  g\frac{7}{8} \frac{\pi^2}{30}  T^4 
             &  g\frac{\pi^2}{30}  T^4   
             & g \frac{3  T^4 }{\pi ^2}  \\[2mm]
        \p   &  \rho/3 & \rho/3  & \rho/3 \\[2mm]
       \partial n/\partial T    & 3\hspace*{0.3mm}n/T   &  3\hspace*{0.3mm}n/T      &  3\hspace*{0.3mm}n/T
       \\[2mm]
       \partial \e/\partial T   & 4\hspace*{0.3mm}\e/T  &  4\hspace*{0.3mm}\e/T     &  4\hspace*{0.3mm}\e/T \\
            \bottomrule 
\end{array}
\end{equation*}

\caption[a]{\setstretch{1.3}
Summary of results contained in, or derivable
from \eqs\nr{n_plus_massless}--\nr{p_plus_massless}. The bosonic 
expressions originate from the analytic continuation 
$
 f^{ }_+ |^{ }_{\mu\to \mu + i \pi T} = -f^{ }_{-}
$, 
but are only valid for $\mu < 0$. 
The Maxwell-Boltzmann limits
were obtained from \eqs\nr{n_bessel}--\nr{de_dmu_bessel}, 
by restricting to $n=0$ and employing
$
 K^{ }_\nu(z) \approx
 2^{\nu-1}_{ }\Gamma(\nu) z^{-\nu}_{ }
$ 
for $|z| \ll 1$.
In the lower panel, 
$\zeta (3)  = \text{Li}^{ }_3(1)=-(4/3) \text{Li}^{ }_3(-1) \approx 1.20206$ and 
$\pi^4_{ }/90 = \text{Li}^{ }_4(1)=-(8/7) \text{Li}^{ }_4(-1)$. 
}
\label{thermo_massless}
\end{center}
\end{table}
%

%
\hiddenappsubsection{Accuracy \& code speed-up}

We have checked the accuracy of the series 
representations in \eqs\nr{n_bessel}--\nr{p_bessel}, 
after restricting to $n^{ }_\rmi{max} = 30$. Even though the series correspond
to low-temperature expansions, their convergent nature guarantees that they can also be applied
for $T > m$. Restricting to $T < 100m$, the given $n^{ }_\rmi{max}$ yields the error
($- = $ bosons, $+ = $ fermions)
\begin{align}
    \delta n^{ }_{-} < 5\times 10^{-4}\,,\quad
    \delta n^{ }_{+} < 10^{-5}\,,\quad
    \delta \e^{ }_{-} < 10^{-5}\,,\quad
    \delta \e^{ }_{+} < 10^{-6}\,,\quad
    \delta \p^{ }_{-} < 10^{-5}\,,\quad
    \delta \p^{ }_{+} < 10^{-6}\,.
\end{align}
Inserting $\mu < 0$ accelerates the convergence. On the other hand, for $\mu > m > 0 $, the series are not convergent, 
but for fermions they still work reasonably well, 
provided that the chemical potentials are not too large. 
For massless fermions, we can use \eqs\nr{n_plus_massless}--\nr{p_plus_massless} for either sign. 

The speed-up of the code 
from using series expansions rather than integrating the functions numerically each time
they are called, 
is a factor of 12 in \texttt{Mathematica}, and similarly in \texttt{python}. 
We have seen this both in the Standard Model, but also in the case of a $\phi \leftrightarrow \nu\bar{\nu}$ BSM scenario. 

\pagebreak

%
\hiddenappsubsection{QED sector}
\la{qed_bessel}

As described in~\app\ref{sec:qed_eos}, six functions are needed in order to describe QED corrections 
to the electromagnetic energy density and pressure of the universe (provided that we neglect the
logarithmic term of $\rmO(e^2_{ }$)). They can be written as sums over Bessel functions similarly
to \eqs\nr{n_bessel}--\nr{de_dmu_bessel}, making use \eq\nr{series_exact} as well as the related
expansion
\begin{align}
        \frac{e^x_{ }}{(e^x_{ }+1)^2_{ }}
        \;=\; \sum_{n=0}^{\infty} (-1)^n (n+1)  e^{-x (n+1)}_{ } \,.
\end{align}
Inserting also the recursion relation $K^{ }_1(z) + z K_1'(z) = - z K^{ }_0(z)$, we find
\ba
 j(\tau) 
 & 
 \overset{\rmii{\nr{jtau}}}{=}
 & 
\frac{1}{\pi^2}\sum_{n=0}^{\infty } (-1)^n (n+1)\, \tau\,  K_1[(n+1) \tau ]
 \;,  \la{jtauB} \\[2mm]
  j'(\tau) 
 & 
 \overset{\rmii{\nr{jtauB}}}{=}
 & 
  \frac{1}{\pi^2}\sum_{n=0}^{\infty }(-1)^{n+1} (n+1)^2 \tau  K_0[(n+1) \tau ]
 \;,  \la{djtauB} \\[2mm]
 J(\tau) 
 & 
 \overset{\rmii{\nr{Jtau}}}{=}
 & 
 \frac{1}{\pi^2}\sum_{n=0}^{\infty } (-1)^n \tau ^2 K_2[(n+1) \tau ]
 \;, \la{JtauB} \\[2mm]
 Y(\tau) 
 &
 \overset{\rmii{\nr{Ytau}}}{=}
 & 
\frac{1}{\pi^2}\sum_{n=0}^{\infty} \frac{3 (-1)^n \tau ^3 K_3[(n+1) \tau ]}{n+1}
 \;, \la{YtauB} \\[2mm]
 k(\tau) 
 &
 \overset{\rmii{\nr{ktau}}}{=}
 & 
\frac{1}{\pi^2}\sum_{n=0}^{\infty} (-1)^n K_0[(n+1) \tau ]
 \;, \la{ktauB} \\[2mm]
 K(\tau) 
 &
 \overset{\rmii{\nr{Ktau}}}{=}
 & 
 \frac{1}{\pi^2}\sum_{n=0}^{\infty} \frac{(-1)^n \tau  K_1[(n+1) \tau ]}{n+1}
 \;, \la{KtauB}  \\[2mm]
 Z(\tau) 
 \;
 & \overset{\rmii{\nr{Ztau}}}{=} &
 \;
 \frac{1}{\pi^2}\sum_{n=0}^{\infty} \frac{3 (-1)^n \tau ^2 K_2[(n+1) \tau ]}{(n+1)^2}
 \;. \la{ZtauB} 
\ea
For most of these we can use $n^{ }_\rmi{max}=30$ if we restrict to $\tau > 0.01$
(i.e.\ $T^{ }_\gamma < 100 m^{ }_e$),
and then the speed-up is a factor of $\sim 10$ compared with
an exact integration. 
However, the functions $j(\tau)$ and $k(\tau)$ require
$n^{ }_\rmi{max} = 1000$ for a comparable precision, and  
$j'(\tau)$ requires $n^{ }_\rmi{max} = 3000$. As $\tau$
increases, convergence accelerates, though only   
for $\tau > 0.5$ is $n^{ }_\rmi{max} = 30$ sufficient for $j'(\tau)$. 
Otherwise, we can use small-$\tau$ expansions instead,
in particular
$j(\tau) \approx [1 + 7 \zeta'(-2) \tau^2_{ }]/(2\pi^2_{ })$,
$j'(\tau) \approx 7 \zeta'(-2) \tau / \pi^2$, 
$J(\tau) \approx [ 2 \pi^2_{ } - 3 \tau^2]/(12\pi^2_ { })$,
$Y(\tau) \approx [14 \pi^2_{ } - 15 \tau^2_{ }]/60$,
$k(\tau) \approx [\ln(\pi/\tau)-\gamma^{ }_\rmiii{E}]/(2\pi^2_{ })$,
$K(\tau) \approx [\pi^2 + 3\tau^2_{ } \ln(\tau)]/(12\pi^2)$,
$Z(\tau) \approx [7\pi^2_{ } - 15 \tau^2_{ }]/120$.

\addtocontents{toc}{\vspace{+0.5em}}  
\bibliography{biblio}

\end{document}